\documentclass[12pt]{iopart}

\usepackage{graphicx} 
\usepackage{epsfig}
\usepackage{dcolumn} 
\usepackage{bm} 
\usepackage{iopams}
\usepackage{setstack}

\begin{document}

\title[Shot noise in spin-orbit-coupled nanostructures]{What can we learn about the dynamics of transported spins by measuring shot noise in spin-orbit-coupled
nanostructures?}
\author{Branislav K Nikoli\' c and Ralitsa L Dragomirova}

\address{Department of Physics and Astronomy, University
of Delaware, Newark, DE 19716, USA}
\ead{bnikolic@physics.udel.edu}

\begin{abstract}
We review recent studies of the shot noise of spin-polarized charge currents and pure spin currents in multiterminal
semiconductor nanostructures, while focusing on the effects brought by the intrinsic Rashba spin-orbit (SO) coupling and/or
extrinsic SO scattering off impurities in two-dimensional electron gas (2DEG) based devices. By generalizing the scattering theory of quantum shot noise to include the full spin-density matrix of electrons injected from a spin-filtering electrode, we show how decoherence and dephasing in the course of spin precession can lead to substantial enhancement of the Fano factor (noise-to-current ratio) of spin-polarized charge currents. These processes are suppressed by decreasing the width of the diffusive Rashba wire, so that purely electrical measurement of the shot noise in a ferromagnet$|$SO-coupled-diffusive-wire$|$paramagnet setup can
quantify the degree of quantum coherence of transported spin through a remarkable one-to-one correspondence between the purity of the spin state
and the Fano factor. In four-terminal SO-coupled nanostructures, injection of unpolarized charge current through the longitudinal
leads is responsible not only for the pure spin Hall current in the transverse leads, but also for nonequilibrium random time-dependent current
fluctuations. The analysis of the shot noise of transverse pure spin Hall current and zero charge current, or transverse spin current and non-zero
charge Hall current, driven by unpolarized or spin-polarized injected longitudinal charge current, respectively, reveals a unique experimental tool to
differentiate between the intrinsic Rashba and extrinsic SO mechanisms underlying the spin Hall effect in 2DEG devices. When the intrinsic
mechanisms responsible for spin precession start to dominate the spin Hall effect, they also enhance the shot noise of transverse spin and charge transport in 
multiterminal geometries. Finally, we discuss the shot noise of transverse spin and zero charge currents in the quantum-interference-driven spin Hall effect in ballistic four-terminal Aharonov-Casher rings realized using high-mobility 2DEG with the Rashba SO coupling. The modulation of the Rashba coupling by the gate electrode imprints the oscillatory signature of constructive and destructive spin interference around the ring on both the spin and charge shot noise, which differ from the corresponding oscillations of the spin Hall conductance, thereby revealing quantum-interference-driven temporal correlations between spin-resolved charge currents of opposite spins.
\end{abstract}

\maketitle

\section{Introduction}\label{sec:intro}

Over the past two decades, the exploration of the shot noise accompanying charge currents in mesoscopic conductors has become one of the major tools for gathering information about microscopic mechanisms of transport and temporal correlations  between charge carriers which cannot be extracted from traditional measurements of time-averaged quantities~\cite{Blanter2000,Blanter2005,Beenakker2003}. Such nonequilibrium time-dependent fluctuations arise due to the discreetness of the electrical  charge, persist down to zero temperature, and  require stochasticity induced by either quantum-mechanical~\cite{Schomerus2005}  backscattering of charge carriers (as in mesoscopic and nanoscopic devices) or by random injection process (as in the textbook example of a Schottky vacuum tube where cathode emits electrons randomly and independently).

Theoretical description of the shot noise is typically formulated in terms of current fluctuations in a conductor with a non-fluctuating bias voltage applied between
the contacts~\cite{Blanter2000}. For zero bias voltage $V=0$, or in macroscopic systems where electrons thermalize in a short time to remain near equilibrium even
under finite $V$, one observes thermal noise. This equilibrium noise vanishes at $T=0$ and, being directly related to the conductance through the fluctuation-dissipation theorem, does not give any new information~\cite{Blanter2000,Beenakker2003}. The situation changes when time it takes for an electron to equilibrate becomes comparable to the time of flight through the conductor, which can be achieved by reducing the size of the system or by lowering the temperature. In this limit, nonequilibrium effects become essential and the relevant energy scale for the noise is set by the bias voltage $eV$ rather than the temperature $k_B T$~\cite{Nagaev1995}. At low frequencies, the  nonequilibrium current noise is dominated by time-dependent conductance fluctuations (arising from the random motion of impurities), termed ``1/f noise'' because of the characteristic frequency dependence of their spectral density which is quadratic function of the mean (i.e., time-averaged) current $I$. At higher frequencies, two principal signatures signifying the shot noise emerge---noise spectral density linearly depends on current while being  frequency-independent.

Macroscopic metallic conductors typically exhibit thermal noise, but no shot noise---in wires of length $L$ longer than the temperature-dependent inelastic electron-phonon scattering length $L_{\rm e-ph}$, the shot noise power is expected to be reduced by a factor  $(L_{\rm e-ph}/L)^{p}$ ($p > 0$)~\cite{Nagaev1995,Naveh1998,Huard2007} when electron-phonon interaction are able to efficiently drain extra energy from the electron subsystem to bring it closer to local thermal equilibrium~\cite{Steinbach1996}. While this leads to {\em a priori} assumption of vanishing nonequilibrium noise~\cite{Blanter2000,Blanter2005,Beenakker2003} in macroscopic metallic samples, which is typically confirmed experimentally~\cite{Steinbach1996}, finite shot noise can be encountered in specific devices that are much longer than $L_{\rm e-ph}$~\cite{Naveh1998,Gomila2004}. In contrast, inelastic electron-electron scattering, which persists to much lower temperatures than electron-phonon scattering, does not suppress shot noise, but slightly enhances the noise power~\cite{Blanter2000,Beenakker2003,Nagaev1995}. The low sensitivity of the shot noise to electron-electron scattering is due to its inability to drain the external-electric-field-supplied energy from the electron subsystem, so that shot noise may be considered as a direct result of such deviation from equilibrium~\cite{Nagaev1995}.

The zero-frequency shot noise spectral density $S=2FeI$ of conventional unpolarized  charge current in two-terminal {\em non-interacting} conductors reaches the maximum value $F=1$ (the Poissonian limit) when transport is determined by uncorrelated stochastic processes. This is the situation encountered in modern tunnel junctions or vacuum tubes explored in the early 1900s where the mean occupation of a state is so small that the Pauli principle is inoperative. On the other hand, correlations among electrons reduce the noise $F<1$, where the dominant source of correlations is the Pauli principle preventing double occupancy of an electronic state. While Coulomb repulsion is another source of correlations, in a metal it is strongly screened and ineffective. Nevertheless, electron-electron interactions in specific setups (e.g., involving various regimes under the Coulomb blockade condition~\cite{Onac2006}, as reviewed in Refs.~\cite{Blanter2005,Barnas2008}) can lead to experimental observation of the super-Poissonian $F>1$ shot noise~\cite{Onac2006}.

For some of the basic types of two-terminal nanostructures, the Fano factor characterizing the transport of non-interacting quasiparticles assumes universal values~\cite{Blanter2000} (i.e., independent of the details of the system, such as impurity distribution, band structure, and shape of the conductor):
$F = 1/2$ for a symmetric double barrier; $F=1/3$ for a diffusive wire; $F=1/2$ for a dirty interface; $F= 1/4$ for a symmetric ballistic chaotic cavity; and $F = 0$
for a ballistic conductor (e.g., quantum point contact in the plateau regime of its quantized conductance). These sub-Poissonian results have been confirmed experimentally~\cite{Heny1999,Oberholzer2001} and derived theoretically by various approaches~\cite{Blanter2000,Beenakker1997}---they are considered to be semiclassical in nature (in the sense that they can be reproduced via approaches based on the Boltzmann equation with Langevin random forces~\cite{Nagaev1995}) where quantum mechanics enters through the calculation of transmission eigenvalues or the Fermi statistics of electrons. The notion of noise can be generalized to multiterminal conductors~\cite{Sukhorukov1999} where temporal correlations calculated between currents in different terminals are always negative due to electrons obeying the Fermi statistics~\cite{Blanter2000}.

Interestingly enough, the mature field of the shot noise of non-interacting particles has been revived very recently by the studies of ballistic transport through evanescent (i.e., exponentially decaying) modes in two-terminal graphene nanoribbons where the scattering theory predicts $F=1/3$~\cite{Beenakker2008}. This surprising result, which is accidentally~\cite{Beenakker2008} the same as the Fano factor for transport through diffusive semiconductor or metallic wires, has been confirmed experimentally in large aspect ratio (width $\gg$ length) graphene samples~\cite{Danneau2008}. However, it  is not universally applicable to all graphene nanoribbons with different types of edges or contacts with metallic electrodes~\cite{Danneau2008,Dragomirova2008a}. It is also genuinely quantum-mechanical feature since it requires classically forbidden evanescent wave functions that decay exponentially from metallic electrodes into the graphene sample.

An example of underlying physics revealed by the Fano factor of the shot noise, such as $F=1/3$ for diffusive  wires~\cite{Beenakker1992} or $F=1/2$ for dirty
interfaces~\cite{Schep1997,Nikoli'c2005a}, is the interplay of randomness in quantum-mechanical impurity scattering and the Pauli blocking imposed by
the Fermi statistics of transported quasiparticles. In both of these cases, the Fano factor confirms bimodal distributions~\cite{Blanter2000,Beenakker2003,Schep1997,Nikoli'c2005a}
of the transmission eigenvalues of the device. Similarly, in chaotic ballistic quantum dots stochasticity is introduced by electron scattering at its irregularly shaped boundaries. Nevertheless, the Fano factor reaches $F=1/4$~\cite{Beenakker2003,Schomerus2005,Agam2000} only in the fully developed quantum regime where electron dwell time $\tau_{\rm dwell} \gg \tau_E$ is greater than the Ehrenfest time $\tau_E$ ($\tau_E$ is roughly equal to the time  it takes for the chaotic classical dynamics to stretch an initially narrow wave packet, of the size of the Fermi wavelength, to some relevant classical length scale~\cite{Schomerus2005,Agam2000}). The shot noise of chaotic quantum dots is reduced below $F=1/4$ in the classical-to-quantum crossover regime $\tau_{\rm dwell}<\tau_E$, where it  depends sensitively on the degree of chaoticity thereby allowing one to extract its Lyapunov exponent~\cite{Agam2000}.

\subsection{Recent trends in theoretical studies  of spin-dependent shot noise}\label{sec:trends_theor}

In contrast to the wealth of information acquired on the shot noise in spin degenerate transport (only briefly touched above, reviewed in Ref.~\cite{Beenakker2003}, and extensively covered in Ref.~\cite{Blanter2000} and its ``update'' Ref.~\cite{Blanter2005}), it is only recently that the study of {\em spin-dependent} and {\em spin-resolved} shot noise in ferromagnet-normal systems  has been initiated for two-terminal~\cite{Bulka1999,Bulka2000,Tserkovnyak2001,Brito2001,Mishchenko2003,Lamacraft2004,Nagaev2006,Hatami2006,Dragomirova2007,Souza2008}
and multiterminal structures~\cite{Egues2002,Belzig2004,Cottet2004,Dragomirova2008}. In these devices ferromagnetic sources (for simplicity often
assumed to be half-metallic ferromagnets~\cite{Mishchenko2003,Lamacraft2004,Nagaev2006,Hatami2006}) inject spin-polarized  charge current
into a paramagnetic central region where spin-dependent interactions affect spin dynamics in the course of transport of electrons to which spins are attached.

For example, Mishchenko~\cite{Mishchenko2003} analyzed shot noise in diffusive spin valves for parallel and antiparallel
magnetizations of their ferromagnetic electrodes, finding setups with significant increase of the Fano factor (when compared to
conventional diffusive wires with $F=1/3$) caused by generic spin-flip scattering. Lamacraft~\cite{Lamacraft2004} calculated
the effect of spin-orbit coupling, magnetic impurities, and precession in an external magnetic field on the noise in the experimentally
relevant cases of diffusive wires and lateral semiconductor dots, finding possible {\em dramatic enhancements}\footnote{Note that throughout the article ``enhancement'' of noise-to-current ratio is measured with respect to the reference value determined by transport processes in the absence of spin-dependent interactions. For example, in the case of two-terminal diffusive wires, the reference value of the Fano factor is the standard $F=1/3$ and ``enhancement'' of spin-dependent shot noise is considered to be any value $F>1/3$.} of the Fano factor. Through the specific values of the Fano factor enhancement, different types of spin-flip mechanisms leave distinctive signatures on the shot noise of injected spin-polarized charge currents, thereby making it possible to extract the spin relaxation times due to different microscopic mechanisms from electrical noise measurements in open mesoscopic systems~\cite{Lamacraft2004}. In Ref.~\cite{Nagaev2006}, Nagaev and Glazman studied finite frequency shot noise in spin valves which originates from random spin flips due to SO-dependent scattering and magnetic impurities. Although the latter mechanism does not contribute to the average current, and its effect on the noise is smaller than that of SO scattering, it can be distinguished by a unique low-frequency  noise dispersion that results from impurity-spin reorientations. By studying the shot noise of spin-polarized current injected into a diffusive ferromagnetic wires attached to two half-metallic ferromagnetic electrodes via tunnel contacts, Hatami and Zareyan~\cite{Hatami2006} demonstrated how the enhanced shot noise (most conspicuous when the electrodes are perfectly polarized in opposite directions) can probe the intrinsic density of states and the extrinsic impurity scattering  contributions to the current polarization of the wire.

Similar studies for the multiterminal spin valve-type devices by Zareyan and Belzig~\cite{Zareyan2005} also encountered enhanced (when compared to the values in identical devices but with paramagnetic electrodes) shot noise and cross-correlations measured between currents in two different ferromagnetic terminals. The enhancement depends on the relative orientation of the magnetization of electrodes, the degree of spin polarization of the terminals, and the strength of the spin-flip scattering in the normal central region. This makes it possible to determine the spin-flip scattering rate by changing the polarization of the ferromagnetic terminals~\cite{Zareyan2005}.

\begin{figure}
\centerline{\psfig{file=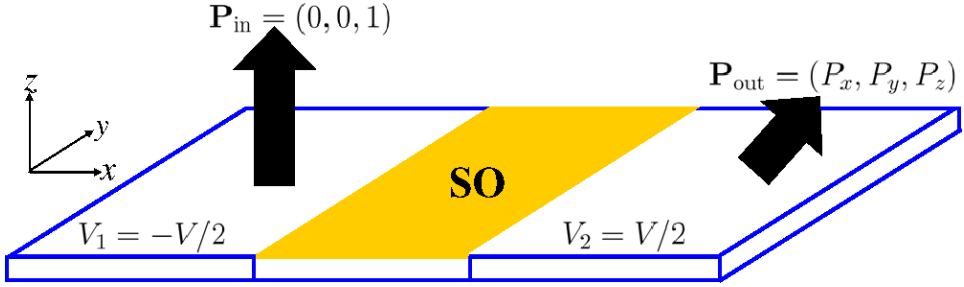,scale=0.5,angle=-90} }
\caption{Generic two-terminal low-dimensional semiconductor nanostructure for the study of spin transport and spin decoherence. Fully spin-polarized current (comprised of pure states $|{\bf P}_{\rm in}|=1$) is injected from the left lead 1 and detected in the paramagnetic right lead 2 which accepts both spin species. The central 2DEG region contains the Rashba SO coupling due to structural inversion asymmetry and can be in the ballistic or in the diffusive transport regime. The source and the drain electrodes are modeled as ideal (with no spin or charge interactions) multichannel semi-infinite leads. The detected current will have its spin polarization vector rotated by coherent spin precession in the 2DEG region, where the effective magnetic field ${\bf B}_{\rm int}({\bf p})$  is along the $y$-axis, as well as shrunk $|{\bf P}_{\rm out}|<1$ due to processes (such as scattering off static impurities or interfaces in the presence of SO coupling) which lead to the loss of spin coherence.}
\label{fig:setup}
\end{figure}

These results emphasize how enhanced Fano factor of the shot noise of spin-polarized electrons is capable of unearthing additional mechanisms of current fluctuations, as well as temporal correlations between carriers of opposite spin, which are not visible in the conventional noise of  unpolarized currents. This is due to the fact that any spin flip converts spin-$\uparrow$ subsystem particle into a spin-$\downarrow$ subsystem particle, where the two subsystems differ when spin degeneracy if lifted. The
non-conservation of the number of particles in each subsystem as the origin of an additional source of current fluctuations is analogous to more familiar example of
fluctuations of electromagnetic radiation in random optical media due to non-conservation of the number of photons~\cite{Beenakker1998}. Microscopically, spin-flips
are either instantaneous events generated by the collision of electrons with magnetic impurities and SO-dependent scattering off static disorder~\cite{Nagaev2006}, or
continuous spin precession~\cite{Lamacraft2004,Dragomirova2007,Dragomirova2008} during electron free propagation in magnetic fields imposed externally or effectively
generated by the intrinsic SO couplings~\cite{Winkler2003,Fabian2007} associated with electronic band structure. Unlike the external magnetic field, the ``internal'' magnetic field ${\bf B}_{\rm int}({\bf p})$ corresponding to the intrinsic SO couplings is momentum dependent, does not break the time-reversal invariance~\cite{Ballentine1998}, and it is capable of spin-splitting the energy bands.

In Ref.~\cite{Dragomirova2007} we addressed two key problems for the shot noise in two-terminal nanostructures with intrinsic SO coupling illustrated by the
device setup in Fig.~\ref{fig:setup}: What is the connection between the Fano factor and the degree of quantum coherence $|{\bf P}_{\rm out}|$ of transported spins? How
does the shot noise depend on the Bloch polarization vector ${\bf P}_{\rm in}$ of injected spins and its direction with respect to ${\bf B}_{\rm int}({\bf p})$?
As illustrated by Fig.~\ref{fig:setup}, the spin polarization vector of the detected current  ${\bf P}_{\rm out}$ ~\cite{Nikoli'c2005} is rotated by coherent precession, and can be shrunk $0 \le |{\bf P}_{\rm out}|<1$ by the D'yakonov-Perel' (DP) spin dephasing~\cite{Fabian2007,Mal'shukov2000,Kiselev2000,Chang2004} due to random changes
in ${\bf B}_{\rm int}({\bf p})$ after electron  scatters off impurities or boundaries (note that these collisions themselves do not involve spin flip). We review in Sec.~\ref{sec:formalism} necessary extensions~\cite{Dragomirova2007} of the scattering approach to shot noise to handle the information about the spin coherence of injected states encoded by ${\bf P}_{\rm in}$, and then discuss in Sec.~\ref{sec:two_terminal_rashba_wire} how this formalisms can be used to establish a {\em one-to-one correspondence} between the  Fano factor of the charge shot noise and $|{\bf P}_{\rm out}|$ of electrons that have traversed finite-size sample containing intrinsic SO couplings and disorder.

Besides the shot noise of spin-polarized charge currents, the shot noise of spin currents~\cite{Dragomirova2008,Lorenzo2004,Sauret2004,Wang2004b,Djuric2006}, including the noise~\cite{Dragomirova2008,Wang2004b,Djuric2006} of the {\em pure} ones which are not accompanied by any net charge flux~\cite{Nagaosa2008}, has attracted considerable theoretical attention. For instance, Sauret  and Feinberg~\cite{Sauret2004} used examples of several interacting mesoscopic devices, in which the average current is not spin polarized and spin degeneracy is not lifted, to argue how shot noise of spin currents  can probe attractive or repulsive interactions between quasiparticles, as well as how it can measure the spin relaxation time $T_1$ without the need to employ external magnetic fields. Zareyan and Belzig~\cite{Zareyan2005a} analyzed three-terminal spin valve setup, consisting of a normal diffusive wire which is connected by tunnel contacts to two oppositely polarized ferromagnetic terminals in one end and to another ferromagnetic terminal on the other end, to demonstrate how the shot noise of spin current in such a device is much more sensitive than charge current  noise to spin-flip scattering rate in the normal wire. Wang {\em et al.}~\cite{Wang2004b} computed the shot noise of pure spin current generated by pumping device (without any bias voltage applied between its two electrodes) driven by time-dependent external magnetic field, pointing out to a difference between cross- (between spin currents in different electrodes) and auto-correlation (between spin currents in the same electrode) noise that are, otherwise, identical for conventional charge currents. Although the net charge current in this device is zero, there is still non-zero shot noise of charge current due to the opposite flow of spin-$\uparrow$ and spin-$\downarrow$ electrons in the course of pure spin current induction~\cite{Wang2004b}.

Section~\ref{sec:she_noise_intro} introduces the topic of the shot noise of pure spin and zero charge currents generated by the mesoscopic spin Hall effect~\cite{Nagaosa2008,Nikoli'c2005b,Nikoli'c2006} in multiterminal nanostructures with SO couplings. We discuss its computation for single-connected 2DEGs attached to four electrodes in Sec.~\ref{sec:she_noise}  and physical insights brought by this recently initiated type of analysis~\cite{Dragomirova2008}. In  Sec.~\ref{sec:qidshe}, we  apply the same type of analysis to multiply-connected (i.e., ring-shaped) devices attached to four electrodes where noise can probe spin-dependent quantum interference effects on transport.

Among the studies of the spin-dependent shot noise, a subject that has evolved into a vast subfield~\cite{Barnas2008,Bulka1999,Bulka2000,Souza2008,Cottet2004,Elste2006,Weymann2007} on its own (and outside of the scope of this review) deals with current fluctuations and temporal correlations in nanoscale conducting islands (such as quantum dots, carbon nanotubes, and magnetic molecules) attached to ferromagnetic electrodes~\cite{Barnas2008,Seneor2007}. In such devices, an interplay between Coulomb blockade and spin accumulation  takes place with shot noise offering tools to probe various aspects of its phenomenology that cannot be extracted from mean current~\cite{Barnas2008,Souza2008}. Another set of topics which involves   spin-dependent shot noise, and is better suited for a separate review~\cite{Burkard2007} in the context of spin qubits for quantum computing, is the shot noise
probing~\cite{Beenakker2003,Egues2002,Egues2005} of two-electron spin-entangled states.

\subsection{Experimental studies of spin-dependent shot noise}\label{sec:trends_exp}

Despite increasing theoretical activity on the spin-dependent shot noise in recent years, only few experiments have been performed, mostly focusing on the shot noise
in magnetic tunnel junctions (MTJ). While high magnetoresistance of MTJs makes them well-suited for fine magnetic field sensors, their low-frequency operation is
limited by the presence of a relatively large $1/f$ noise~\cite{Jiang2004}. Although the shot noise is not the most important among noise sources~\cite{Jiang2004} as the key limiting factor~\cite{Edelstein2006} for MTJ applications, it is a sensitive tool to probe properties of different types of insulating barriers~\cite{Nikoli'c2005a}
responsible for tunneling.

For example, measurement of the Fano factor of spin-dependent shot noise in MTJs with MgO insulating barrier can be employed to test the quality (i.e., presence of impurities or imperfections) of epitaxially grown crystalline MTJs---obtaining Poissonian limit $F=1$ signals pure spin-dependent direct tunneling and validates high structural quality of the MgO barrier~\cite{Guerrero2007}. In Ref.~\cite{Guerrero2006}, Guerrero {\em et al.} measured sub-Poissonian Fano factor $F<1$ in Al$_2$O$_3$ MTJs whose value was dependent on the alignment of the ferromagnetic electrodes for certain barrier conditions. This was attributed ($F \simeq 1$ for Cr-doped and $F<1$ for Cr-free insulating barrier) to sequential tunneling via impurity levels inside the tunnel barrier. On the other hand, Garzon {\em et al.}~\cite{Garzon2007} measured super-Poissonian  shot noise in small area MTJs whose  Fano factor $F>1$ depends on the magnetization state of the ferromagnetic electrodes. Although intertwined spin and charge blockade facilitated by localized states within the barrier could account for these measurements, the search for super-Poissonian shot noise in MTJs and its theoretical explanation is still in its infancy.

\section{Overview of recent analyses of the SO coupling effects on the shot noise}

\subsection{Shot noise in Rashba SO-coupled systems}

The crucial role played by the SO interactions in all-electrical control of spin in
semiconductor nanostructures~\cite{Fabian2007,Awschalom2007} has also provoked recent studies of their effects on  the shot noise.
For example, the Rashba SO coupling~\cite{Winkler2003} induced by structural inversion asymmetry of the semiconductor heterostructure
hosting the 2DEG can be tuned by a gate electrode covering the device~\cite{Nitta1997,Grundler2000}. This is envisaged as a key ingredient
of the ``second generation'' spintronic devices~\cite{Awschalom2007}, such as semiconductor-based spin transistors that manipulate propagating
coherent spin states~\cite{Fabian2007}.

In a pioneering work on the shot noise in Rashba SO coupled systems, Egues {\em et al.}~\cite{Egues2002,Egues2005}
unveiled how the Rashba SO coupling present in a localized region  of one of the incoming leads of a clean four-terminal
beam splitter can modulate the Fano factor of the shot noise of injected spin-polarized electrons. This suggest a direct way to measure
the Rashba coupling constant via noise, as well as the degree of polarization along different directions. Using the same device one can probe injected
two-electron spin-entangled states~\cite{Burkard2007} where Rashba coupling coherently rotates spin states to modulate the noise signal~\cite{Egues2005}.

It is well known that shot noise in ballistic chaotic cavities is suppressed when electrons
follow classical deterministic trajectories during dwell time $\tau_{\rm dwell}$ shorter than the Ehrenfest time $\tau_E$~\cite{Schomerus2005,Agam2000}.
In such cases, the electron wave packet entering the quantum dot is either fully transmitted or
fully reflected, so that shot noise remains zero ($F=0$) due to absence of any quantum-mechanically-induced randomness~\cite{Beenakker2003,Beenakker1991}. However,
Ossipov {\em et al.}~\cite{Ossipov2006} demonstrated, using an example of a ballistic dot in a shape of a stadium billiard with  the Rashba SO coupling,
that SO  interactions can be solely responsible for the non-zero Fano factor  in the regime $\tau_{\rm dwell}<\tau_E$.
This arises due to the transfer of quantum mechanical uncertainty in the spin of the electron to its position via the SO coupling,
which causes a breakdown of deterministic classical dynamics.

Shangguan and Wang pointed out in Ref.~\cite{Shangguan2007} that both auto-correlation and cross-correlation noise are
required to characterize fluctuations of spin current in double quantum dots with the Rashba SO interaction. Moreover,
the sign of cross-correlation noise can be tuned by either the gate voltage  or intra-dot coupling. L\"{u} and Guo~\cite{Lu2007}
investigated the shot noise in quantum dot in the Kondo regime where the Rashba coupling is present in the dot only and the dot is
embedded in one arm of the Aharonov-Bohm ring. When bias voltage is applied across the two leads attached to the ring, spin-polarized
current flows through the dot whose shot noise is greatly affected by changing the Rashba coupling.

Several other studies~\cite{Li2005b,He2007,An2007} have computed the spin-resolved shot noise for ballistic charge transport through Datta-Das spin
field-effect transistor (spin-FET)~\cite{Fabian2007}, composed of two ferromagnetic electrodes sandwiching a semiconductor quasi-one-dimensional wire with the
Rashba SO coupling, to find the oscillatory noise behavior as the function of different parameters of the spin-FET.

\subsection{Shot noise in mesoscopic spin Hall systems with intrinsic Rashba SO coupling}\label{sec:she_noise_intro}

\begin{figure}
\centerline{\psfig{file=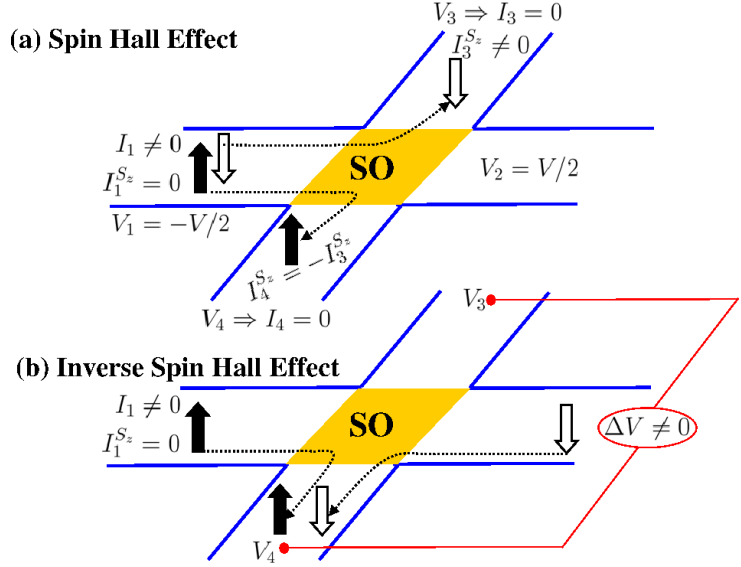,scale=0.4,angle=-90}}
\caption{Basic phenomenology of the direct and inverse SHE in multiterminal nanostructures: (a) conventional
(unpolarized) charge current flowing longitudinally through the sample experiences
transverse deflection of opposite spins in opposite direction due to SO coupling induced ``forces''.
This generates pure spin current in the transverse direction or spin accumulation (when transverse electrodes
are removed) of opposite sign at the lateral sample edges; (b) pure spin current flowing through the same sample
governed by SO interactions will induce transverse charge current or voltage drop $\Delta V=V_3 - V_4$ (when transverse leads are
removed). Note that to ensure purity ($I_3=I_4 \equiv 0$) of transverse spin Hall current in (a)
one has to apply proper voltages $V_3$ and $V_4$.}
\label{fig:she_inverse}
\end{figure}

The recently discovered spin Hall effect (SHE)~\cite{Engel2007a} in paramagnetic
semiconductor~\cite{Kato2004b,Sih2005a,Brune2008} and metallic~\cite{Valenzuela2006,Saitoh2006,Kimura2007,Seki2008} systems holds great promise to
revolutionize electrical generation, control, and detection of nonequilibrium spin populations in
the envisioned ``second generation'' spintronic devices~\cite{Awschalom2007}. The SHE actually denotes a {\em collection}~\cite{Engel2007a}
of phenomena manifesting as transverse (with respect to injected unpolarized charge current) separation of spin-$\uparrow$ and spin-$\downarrow$ states,
which then comprise either a pure spin current or accumulate at the lateral sample boundaries. Its Onsager reciprocal phenomenon---the inverse
SHE~\cite{Hirsch1999,Hankiewicz2005} where  longitudinal pure spin current generates transverse charge current or voltage between the lateral boundaries---offers one
of  the most efficient schemes to detect elusive pure  spin currents~\cite{Nagaosa2008} by converting
them into electrical signal~\cite{Valenzuela2006,Saitoh2006,Kimura2007,Seki2008}. The basic phenomenology of both the direct and the inverse SHE, as manifested in multiterminal nanostructures, is illustrated in Fig.~\ref{fig:she_inverse}.

While SHE does not require external magnetic field, it essentially relies on the SO coupling effects in solids. In addition, its  magnitude can depend
on the type of microscopic  SO interaction, impurities, charge density, geometry, and dimensionality. Such a variety of SHE manifestations poses immense challenge
for attempts at a unified theoretical description of spin transport in the presence of relativistic effects, which has not been resolved
by early hopes~\cite{Nagaosa2008,Murakami2003,Sinova2004} that auxiliary spin current operator ${\hat j}^z_y$ (for $S_z$-spins transported along the $y$-axis) 
and spin conductivity $\sigma_{\rm sH}=\langle {\hat j}^z_y \rangle/E_x$ (as the linear response to longitudinal electric field $E_x$) of infinite homogeneous systems could be elevated to universally applicable and experimentally relevant quantities.

Thus, the key task emerging for theoretical analysis is to provide guidance for increasing  and controlling the spin accumulation in confined
geometries~\cite{Nikoli'c2006,Nikoli'c2005d,Zyuzin2007,Silvestrov2008} (observed SHE in semiconductors is presently rather small~\cite{Kato2004b,Sih2005a}) or outflowing
spin currents~\cite{Nikoli'c2005b,Nikoli'c2006} driven by them. In this respect, understanding of the {\em intrinsic}~\cite{Murakami2003,Sinova2004,Guo2008} (due to
SO-induced spin-split band structure) or {\em extrinsic}~\cite{Hirsch1999,Hankiewicz2008} (due to SO-dependent scattering off impurities) origin of the SHE has been one
of the central topics in interpreting experiments~\cite{Guo2008} and developing  SHE-based spintronic devices~\cite{Awschalom2007}. For example, the intrinsic SO couplings are predicted~\cite{Guo2008} to yield much larger SHE response~\cite{Brune2008}, which, moreover, can be controlled electrically by the gate electrodes covering
low-dimensional semiconductor devices~\cite{Nikoli'c2005b,Souma2005a}. The extrinsic ones are fixed and the corresponding much smaller SHE is hardly controllable (except
through charge density and mobility~\cite{Awschalom2007}).

However, measurements of standard quantities associated with transverse spin and charge transport are often unable to resolve the intrinsic vs. extrinsic
controversy~\cite{Nagaosa2008,Guo2008} or probe the crossover between these limiting regimes~\cite{Hankiewicz2008}. This long standing issue is well-known from the studies of
the anomalous Hall effect (AHE)~\cite{Sinitsyn2008} in ferromagnetic materials (SHE can be viewed as the zero magnetization limit of AHE). For example, the frequent analysis of the AHE experimental data---fitting of the Hall resistivity vs. longitudinal zero-field resistivity by a power law---is typically insufficient~\cite{Kotzler2005} to clearly
differentiate a variety of mechanisms~\cite{Hankiewicz2008,Sinitsyn2008} driven by SO coupling effects.

Here lessons from mesoscopic quantum physics might shed new light: as discussed in Sec.~\ref{sec:intro}, much more information about transport of non-interacting or interacting quasiparticles is contained in time-dependent nonequilibrium current (or voltage) fluctuations~\cite{Blanter2000} than in traditional time-averaged quantities such as conductances and conductivities. Furthermore, recent theoretical and experimental studies have suggested that shot noise in systems with spin-dependent interactions provides a sensitive probe to differentiate between magnetic impurities, spin-flip scattering, and continuous spin precession effects, as overviewed in Sec.~\ref{sec:trends_theor} and Sec.~\ref{sec:trends_exp}.

\begin{figure}
\centerline{\psfig{file=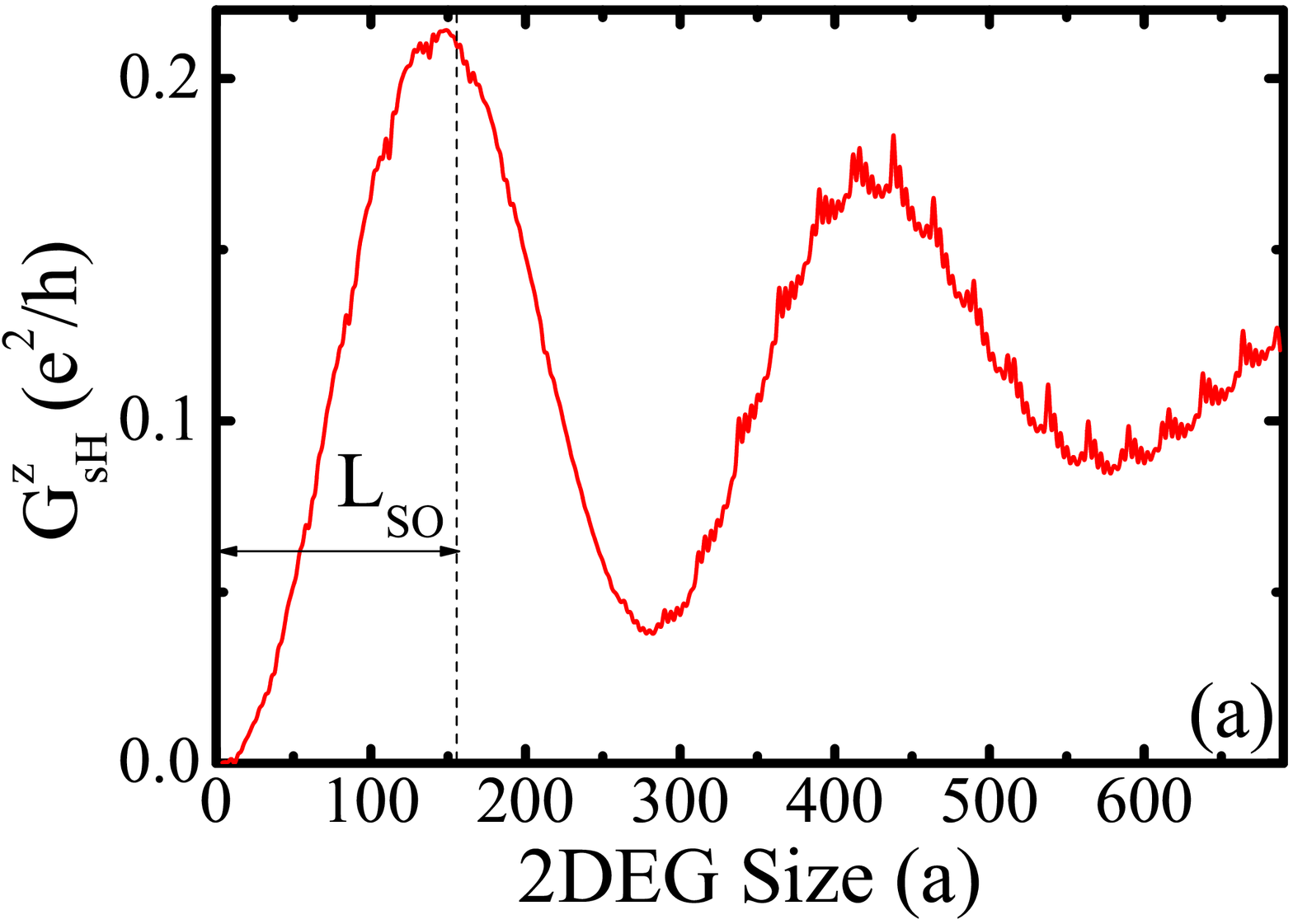,scale=0.37,angle=0} \psfig{file=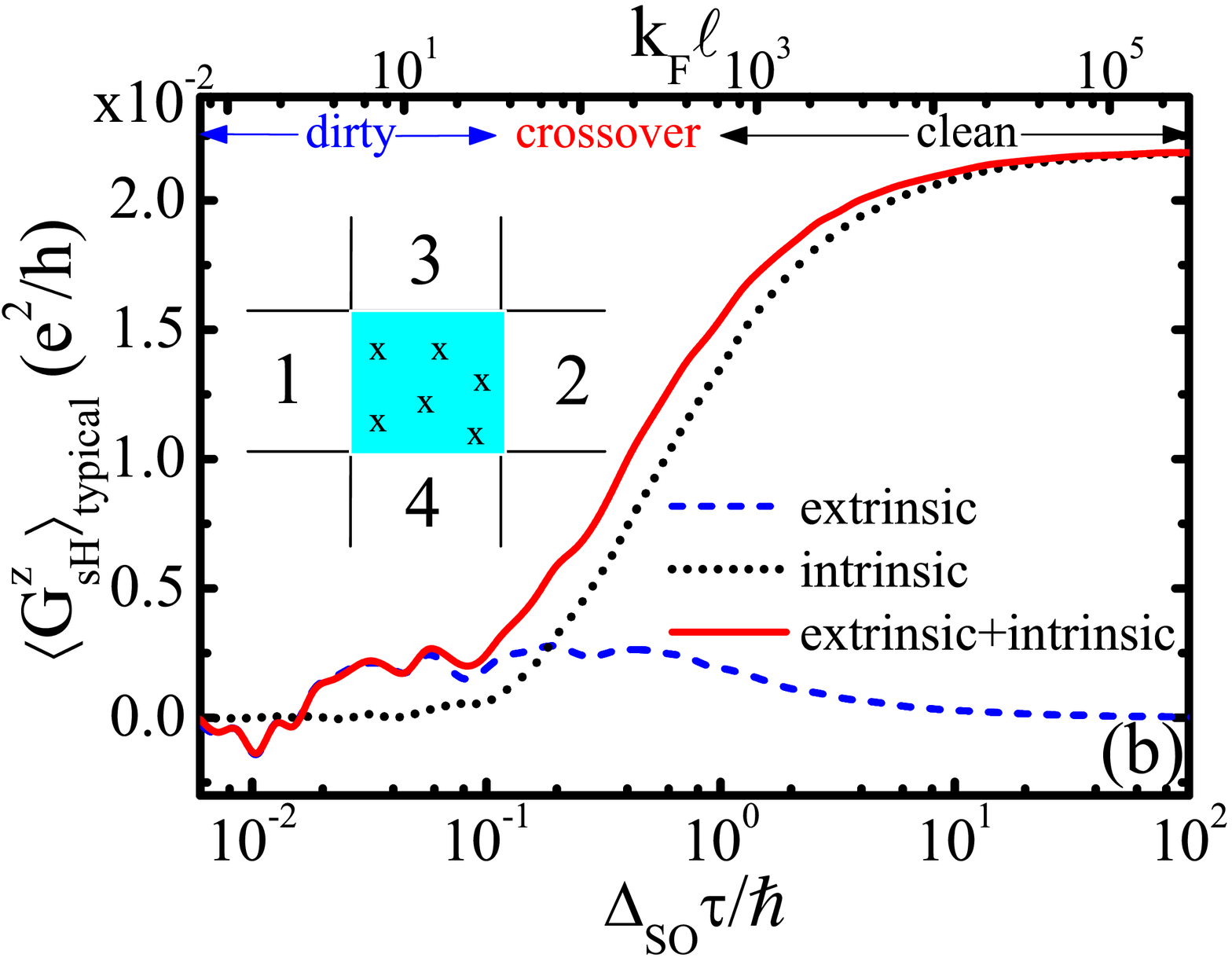,scale=0.37,angle=0}}
\caption{(a) The spin Hall conductance of a {\em clean}  square-shaped 2DEG as the function its size $L$
($a \simeq 3$ nm) with the intrinsic Rashba SO coupling setting the spin precession length
$L_{\rm SO} \approx 157a$ (along which spin precesses by an angle $\pi$). (b) The {\em disorder-averaged typical} spin Hall conductance of 2DEG nanostructures of the size $100 a \times 100 a$ which are governed by both the intrinsic Rashba SO coupling, responsible for the quasiparticle energy spin-splitting  $\Delta_{\rm SO}$, and SO scattering off impurities. The disorder strength sets the transport scattering time $\tau$ and the mean free path $\ell=v_F\tau$, while the magnitudes of the intrinsic Rashba ($L_{\rm SO} \approx 523a$) and extrinsic SO coupling $\lambda/\hbar =5.3$ \AA{}$^2$ are fixed to correspond to parameters of 2DEG in the experiment of Ref.~\cite{Sih2005a}. Panel (b) is adapted from Ref.~\cite{Nikoli'c2007}.}
\label{fig:gshe}
\end{figure}

Although seminal arguments~\cite{Sinova2004} for the intrinsic SHE in infinite homogeneous 2DEGs in the clean limit
have predicted ``universal'' SHE conductivity $\sigma_{\rm sH}=e/8\pi$, {\em a posteriori} analysis has found that for
SO couplings linear in momentum (such as the Rashba one), any scattering that leads to a stationary electric current via
deceleration of electrons by impurities or phonons will result in exact cancellation of the bulk spin Hall current in the
DC case~\cite{Engel2007a,Inoue2004,Mishchenko2004}. Such cancellation can be avoided by moving into AC domain with frequencies
exceeding the inverse spin relaxation time~\cite{Mishchenko2004} or by making sufficiently small and clean structures to support
ballistic transport (mean free path greater than the system size) across the device. The latter case gives rise to the
so-called {\em mesoscopic} SHE~\cite{Nagaosa2008,Nikoli'c2005b} which, unlike ``universal'' SHE~\cite{Sinova2004} in an infinite 2DEG
where the electric-field-driven acceleration of electron momenta and associated precession of spins~\cite{Hankiewicz2008} plays a crucial role,
was introduced~\cite{Nikoli'c2005b,Nikoli'c2006,Nikoli'c2005d} in ballistic finite-size systems attached to multiple current and voltage probes with
{\em electric field being absent} in the SO-coupled central region~\cite{Nikoli'c2005b,Hankiewicz2004a,Sheng2006b,Ren2006,Bardarson2007}.

In two-terminal SO-coupled ballistic wires, mesoscopic SHE is characterized by spin accumulation of opposite sign along opposite lateral edges~\cite{Nikoli'c2005d,Zyuzin2007,Silvestrov2008}. In four-terminal and other multiterminal~\cite{Hankiewicz2004a} nanostructures,  its description~\cite{Nikoli'c2005b} in terms of the total charge currents $I_\alpha=I_\alpha^\uparrow + I_\alpha^\downarrow$ and conserved total spin currents $I_\alpha^{S_z}=I_\alpha^\uparrow - I_\alpha^\downarrow$ (which are related to nonequilibrium spin densities within the sample~\cite{Nikoli'c2006}) outflowing through spin and charge interaction-free electrodes (ensuring terminal spin currents that do not change at different cross sections of the leads~\cite{Nikoli'c2006}) is particularly suited for spin-dependent shot noise analysis.  The SHE in four-terminal systems is quantified by the spin conductance (for labeling of the total currents and voltages in the terminals see Fig.~\ref{fig:she_inverse}):
\begin{equation}\label{eq:shecond}
G_{\rm sH}^z=I_3^{S_z}/(V_1-V_2)
\end{equation}
Unlike in three-dimensional semiconductor~\cite{Kato2004b} and metallic devices~\cite{Valenzuela2006,Saitoh2006,Kimura2007,Seki2008}, which are always disordered
and where extrinsic contribution to the SHE is therefore present or dominant, ballistic conditions for the mesoscopic SHE can be achieved in low-dimensional
semiconductor systems. For example, the very recent experiment on nanoscale H-shaped structures built on {\em high mobility} HgTe/HgCdTe quantum wells has
reported for the first time the detection of mesoscopic SHE via non-local and purely electrical measurements~\cite{Brune2008}.

The spin Hall conductance\footnote{Note that the spin conductance has a natural unit $e/4\pi=(\hbar/2e)(e^2/h)$ taking into account that spin current carries angular momenta $\hbar/2$ instead of charge $e$. Nevertheless, to simplify the noise analysis we use the same units for both the spin and the charge current, so that the unit of
$G^z_{\rm sH}$ is $e^2/h$.} in clean four-terminal 2DEG devices is shown in Fig.~\ref{fig:gshe}(a), and in disordered ones in Fig.~\ref{fig:gshe}(b). In the general cases~\cite{Sih2005a,Hankiewicz2008}, where both the extrinsic and intrinsic SO interaction effects are present, the intrinsically driven contribution to spin Hall current in finite-size devices starts to dominate~\cite{Nikoli'c2007} when the ratio of characteristic energy scales~\cite{Nagaosa2008} for the disorder and SO coupling effects satisfies $\Delta_{\rm SO}\tau/\hbar \gtrsim 10^{-1}$ ($\Delta_{\rm SO}$ is SO-induced spin-splitting of quasiparticle energies~\cite{Winkler2003,Fabian2007} and $\hbar/\tau$ is the disorder induced broadening of energy levels due to transport scattering time $\tau$).

However, spin current is not a directly measurable quantity and has to be converted into other quantities (such as spin accumulation, voltage, or charge current) to be measured by conventional techniques. Following Ref.~\cite{Dragomirova2008}, Sec.~\ref{sec:she_noise} discusses how the information stored in the shot noise of transverse spin Hall current, as well as the noise of associated transverse charge transport, can provide new tool to separate different types of SO interactions driving the SHE. We draw inspiration for this approach from the following recent intriguing theoretical findings: ({\em i}) the intrinsic aspects of AHE  have been related to
(transverse) voltage shot noise by Timm {\em et al.}~\cite{Timm2004}; ({\em ii}) Hatami and Zareyan~\cite{Hatami2006} argued that shot noise of spin-polarized current injected into a ferromagnetic diffusive wire can probe aspects of its AHE; ({\em iii}) Erlingsson and Loss~\cite{Erlingsson2005} pointed out that measurement of charge currents and their auto- and cross-correlation shot noise on a multiterminal bridge could be used to obtain the spin Hall conductance solely in terms of these {\em purely electrical} quantities (independently of the underlying microscopic SO mechanism); ({\em iv}) as discussed in Ref.~\cite{Dragomirova2007} and reviewed in Sec.~\ref{sec:two_terminal_rashba_wire}, the shot noise of spin-polarized charge current offers a sensitive electrical probe of spin precession and spin dephasing in two-terminal nanostructures, where spin precession represent crucial aspect~\cite{Hankiewicz2008} of SHEs driven by intrinsic SO couplings.

\subsection{Shot noise of quantum-interference-driven spin Hall effect in four-terminal Aharonov-Casher rings with intrinsic Rashba SO
coupling}\label{sec:qidshe_intro}

The superpositions of quantum states and thereby induced quantum interference
effects  are one of the most fundamental aspects of quantum mechanics. The interference experiments
are difficult to perform with electrons in solids which are typically coupled to a large decohering
environment~\cite{Joos2003,Schlosshauer2008}. Nevertheless, the advent of mesoscopic structures, which are smaller that the phase coherence
length $L_\phi$ ($L_\phi \lesssim 1$ $\mu$m at low temperatures $T \ll 1$ K), has made it possible to manipulate
electrons described by a single wave function throughout devices much larger than atomic or molecular scale and
to study the effects of the wave-function phase on transport properties probed by macroscopic instruments. The standard example
of mesoscopic interferometers is Aharonov-Bohm (AB) ring as a multiply-connected device with two electrodes and transmission
probability that is a periodic function of the magnetic flux penetrating the ring. Its conductance and shot noise retain the
same periodic dependence~\cite{Blanter2005}.

Since the amplitude of the oscillations diminishes in multichannel rings~\cite{Souma2004}, where topological AB phase
acquired by electrons moving along the arms of the ring is averaged over many conducting channels, recent remarkably
crafted interferometers, such as the electronic Mach-Zehnder setup~\cite{Ji2003} fabricated from the
edge channels of a 2DEG in the integer quantum Hall effect regime, operate with single (chiral edge) channel
carrying the current. The experiment of Ref.~\cite{Ji2003} also measured the shot noise with the idea that it could be
used to differentiate between phase (or thermal) averaging vs. genuine decoherence---since noise is nonlinear in the transmission
probability, its value will depend on whether averaging is performed before or after calculating its expression. On the level
of average current, decoherence cannot be distinguished easily from thermal averaging, both of which reduce the visibility of quantum
interferences encoded in the current as a function of the phase difference between the paths~\cite{Blanter2005,Marquardt2004}.

This has also motivated several theoretical proposals to use the shot noise modulation by quantum interference effects as
a probe of fundamental properties of charge carries~\cite{Feldman2007}. More recently, oscillating shot noise in
Andreev interferometers~\cite{Reulet2003} (a normal metal connected by two arms with the same superconducting
reservoir whose transport properties are sensitive to magnetic flux enclosed by the arms) as a function of magnetic flux,
or in carbon nanotube-based Fabry-Perot electronic interferometers~\cite{Herrmann2007,Wu2007} as a function of the gate
voltage, was measured  to provide an alternative probe of quantum interferences.

The AB ring represents a solid state realization of a two-slit experiment where electron entering the ring can propagate in two possible directions
(clockwise and counterclockwise). The superpositions of the corresponding quantum states are sensitive to the acquired AB topological phases~\cite{Bohm2003} in the
external magnetic field. The pursuit of fundamental spin interference effects, as well as spin transistors with  unpolarized (unlike Datta-Das spin-FET~\cite{Fabian2007}) charge currents ~\cite{Nitta1999}, has also generated considerable interest to demonstrate the Aharonov-Casher (AC) effect~\cite{Bohm2003} via transport experiments in SO-coupled semiconductor nanostructures. The electromagnetic duality (i.e., charge and spin, as well as electric and magnetic field, interchanged) entails AC effect, originally discussed in terms of a neutral magnetic dipole moment going around a charged line to acquire the AC phase~\cite{Bohm2003,Nitta2007}. Very recent vigorous experimental activity~\cite{Nitta2007,Konig2006}  has been focused on detecting the AC phase (which in one-dimensional rings is the sum of SO Berry phase and spin dynamical phase~\cite{Molnar2004,Frustaglia2004a}) difference for opposite spin states traveling clockwise and counterclockwise around the two-terminal ring with the
Rashba SO coupling. This results in oscillatory behavior of the ring conductance, more complicated than in the case of AB rings, as a function of the SO interaction strength~\cite{Souma2004,Molnar2004,Frustaglia2004a,Lucignano2007}.

\begin{figure}
\centerline{\psfig{file=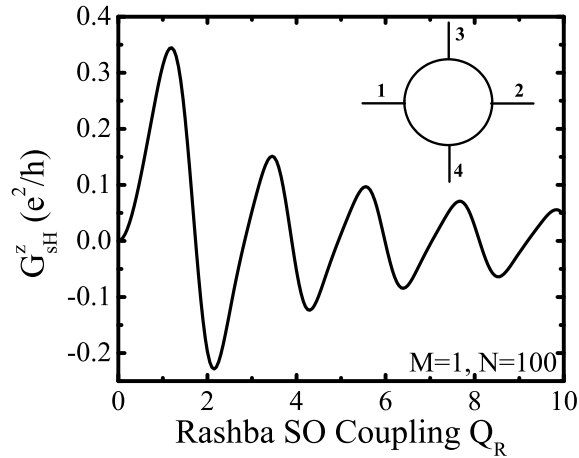,scale=0.37,angle=0}}
\caption{The spin Hall conductance $G_{\rm sH}^z=I_3^{S_z}/(V_1-V_2)$ of a one-dimensional  Aharonov-Casher ring attached to four single-channel leads as the function of the dimensionless Rashba SO coupling $Q_R$ that can be tuned by the gate electrode covering the ring. Adapted from Ref.~\cite{Souma2005a}.}
\label{fig:qidshe}
\end{figure}

Furthermore, the {\em quantum-interference-driven SHE} (QIDSHE) was predicted in Ref.~\cite{Souma2005a} to occur in a four-terminal ballistic mesoscopic rings with homogeneous  Rashba SO coupling within the ring-shaped central region. The recent studies~\cite{Tserkovnyak2007,Borunda2008} have extended the possibility of such unusual SHE (which cannot be captured by semiclassical analysis to which standard SHE can be reduced~\cite{Sinitsyn2008}) to multiterminal rings with different types of SO couplings. This AC ring-type nanostructure generates pure spin current in the transverse electrodes as a response to unpolarized charge current injected through the longitudinal leads (labeled as 1 and 2 in the inset in Fig.~\ref{fig:qidshe}). The transverse spin Hall current can be modulated between zero and large value (when compared to small extrinsically generated spin currents in dirty 2DEGs~\cite{Sih2005a}) by changing the voltage on the gate electrode covering the ring~\cite{Nitta1997,Grundler2000,Konig2006}, as demonstrated by Fig.~\ref{fig:qidshe}. This gives an unambiguous experimental signature that is particularly visible in single channel devices~\cite{Souma2005a,Souma2004}. In Sec.~\ref{sec:qidshe} we investigate if the shot noise of transverse spin Hall and zero charge current can provide additional insights into interference and dephasing effects in general multichannel  AC ring devices, or if it can offer an alternative measuring scheme to confirm QIDSHE electrically.

\section{Scattering approach to spin-resolved shot noise: Inclusion of the spin-density matrix of injected electrons}\label{sec:formalism}

At low temperatures, where small enough  conductors become phase-coherent and the
Pauli blocking renders regular injection and collection of charge carriers from the bulk electrodes,
the scattering theory of quantum transport provides the celebrated
formula~\cite{Blanter2000,Beenakker2003,Beenakker1997} for the shot noise power in terms of the transmission eigenvalues $T_n$:
\begin{equation}\label{eq:khlus}
S = \frac{4e^3V}{h} \sum_{n=1}^M T_n(1-T_n).
\end{equation}
Here $V$ is the linear response time-independent bias voltage. The physical interpretation of Eq.~(\ref{eq:khlus}) is quite transparent---in the basis of eigenchannels, which diagonalize ${\bf t} {\bf t}^\dag$ where  ${\bf t}$ is the transmission matrix of a two-terminal device, a mesoscopic structure can be viewed as a parallel circuit of  $M$ (=number of transverse propagating orbital wave functions in the leads) independent one-dimensional conductors, each characterized by the transmission probability $T_n$. To get the shot noise through disordered systems, Eq.~(\ref{eq:khlus}) has to be averaged~\cite{Blanter2000} over a proper distribution~\cite{Nikoli'c2005a} of $T_n$. However, this standard route {\em becomes inapplicable} for spin-polarized injection where one has to take into account the spin-density matrix of injected electrons~\cite{Nikoli'c2005}  and, therefore, perform the calculations in the natural basis~\cite{Blanter2000} composed of spin-polarized conducting channels of the electrodes.

We recall~\cite{Ballentine1998} that the density matrix $\hat{\rho}$ ($\hat{\rho} = \hat{\rho}^\dagger$; ${\rm Tr} \, \hat{\rho}=1)$ of spin-$\frac{1}{2}$ particles is a $2
\times 2$ matrix (${\bm 1}$ denotes the $2 \times 2$ unit matrix)
\begin{equation} \label{eq:density_matrix}
\hat{\rho} =\left( \begin{array}{cc}
        \rho^{\uparrow\uparrow} & \rho^{\uparrow\downarrow} \\
        \rho^{\downarrow\uparrow} & \rho^{\downarrow\downarrow}
         \end{array} \right)= \frac{1}{2} ({\bf 1} + {\bf P} \cdot \hat{\bm{\sigma}}),
\end{equation}
which is determined solely by measuring the three real numbers comprising the spin polarization vector  ${\bf P} = {\rm Tr} \, [ \hat{\rho} \hat{\bm{\sigma}}]$ as the expectation value of the spin operator $\hbar {\bf P}/2= \langle \hbar \hat{\bm{\sigma}}/2 \rangle$. The magnitude $|{\bf P}|$ quantifies the degree of quantum coherence or purity of a quantum state. Thus, $\hat{\rho}$ and the corresponding ${\bf P}$ provide the most general quantum-mechanical description of a spin-$\frac{1}{2}$ system, accounting for both pure (i.e., fully coherent)
$\hat{\rho}^2=\hat{\rho} \Leftrightarrow |{\bf P}|=1$ and mixed $\hat{\rho}^2 \neq \hat{\rho} \Leftrightarrow 0 \le |{\bf P}| < 1$  states~\cite{Ballentine1998}.

Below we discuss a generalization of the scattering-matrix-based formulas for the shot noise which allow one to compute spin-resolved noise as the building block of charge and spin current noise
computation, while including both the ``direction'' of injected spins and the degree of their quantum coherence. That is, all of our analytical formulas contain the spin polarization vector ${\bf P}_{\rm in}$ which determines the density matrix $\hat{\rho}_{\rm in}=({\bm 1}+{\bf P}_{\rm in} \cdot \hat{\bm \sigma})/2$ of injected spins. Moreover, in Sec.~\ref{sec:fano} we argue that the value of the Fano factor in the right electrode is directly connected to the degree of quantum coherence $|{\bf P}_{\rm out}|$ of outgoing spins, as extracted from the recently developed~\cite{Nikoli'c2005} scattering approach to their spin-density matrix $\hat{\rho}_{\rm out}=({\bm 1}+{\bf P}_{\rm out} \cdot \hat{\bm \sigma})/2$.

The analysis of the spin-dependent shot noise requires to evaluate temporal correlations between spin-resolved charge
currents $I^\uparrow_\alpha$ and $I^\downarrow_\beta$ due to the flow of spin-$\uparrow$ and spin-$\downarrow$ electrons through the terminals of a
nanostructure~\cite{Sauret2004}
\begin{equation}\label{eq:noise_time}
S_{\alpha \beta}^{\sigma \sigma^\prime} (t-t^\prime) = \frac{1}{2} \langle \delta \hat{I}_\alpha^\sigma(t)
\delta \hat{I}_\beta^{\sigma^\prime}(t^\prime) +  \delta \hat{I}_\beta^{\sigma^\prime}(t^\prime)  \delta \hat{I}_\alpha^\sigma(t) \rangle.
\end{equation}
Here $\hat{I}_\alpha^\sigma(t)$ is the quantum-mechanical operator of the spin-resolved  charge current carrying spin-$\sigma$ ($\sigma=\uparrow,\downarrow$) electrons in
lead $\alpha$. The current-fluctuation operator at time $t$ in lead $\alpha$ is
\begin{equation}
\delta \hat{I}_\alpha^\sigma(t) = \hat{I}_\alpha^\sigma(t) - \langle \hat{I}_\alpha^\sigma (t) \rangle.
\end{equation}
We use $\langle \ldots \rangle$ to denote both quantum-mechanical and statistical averaging over the states in the macroscopic reservoirs to which a mesoscopic conductor is
attached via semi-infinite interaction-free leads~\cite{Blanter2000}. The spin-resolved noise power between terminals $\alpha$ and $\beta$ is (conventionally defined~\cite{Blanter2000} as twice)
the Fourier transform of Eq.~(\ref{eq:noise_time}),
\begin{equation} \label{eq:fourier}
S_{\alpha\beta}^{\sigma \sigma^\prime} (\omega) = 2 \int d(t-t^\prime)\, \exp[-i\omega(t-t^\prime)] S_{\alpha\beta}^{\sigma \sigma^\prime} (t-t^\prime).
\end{equation}
The total noise power of charge current
\begin{equation}
I_\alpha=I_\alpha^\uparrow + I_\alpha^\downarrow,
\end{equation}
is given by
\begin{equation}
S_{\alpha\beta}^{\rm charge}(\omega) = S_{\alpha\beta}^{\uparrow \uparrow}(\omega) + S_{\alpha\beta}^{\downarrow \downarrow}(\omega) + S_{\alpha\beta}^{\uparrow
\downarrow}(\omega) + S_{\alpha\beta}^{\downarrow \uparrow}(\omega),
\end{equation}
while the total noise power of spin current
\begin{equation}
I_\alpha=I_\alpha^\uparrow - I_\alpha^\downarrow,
\end{equation}
is obtained from the spin-resolved noise powers as
\begin{equation}
S_{\alpha\beta}^{\rm spin}(\omega) = S_{\alpha\beta}^{\uparrow \uparrow}(\omega) + S_{\alpha\beta}^{\downarrow \downarrow}(\omega) - S_{\alpha\beta}^{\uparrow
\downarrow}(\omega) - S_{\alpha\beta}^{\downarrow \uparrow}(\omega).
\end{equation}
Selecting the same electrode $\alpha = \beta$ yields the {\em auto-correlation} noise, while for different electrodes $\alpha \neq \beta$ we get the {\em cross-correlation} noise.

In the  scattering theory of quantum transport, the operator of spin-resolved charge current carrying spin-$\sigma$ electrons through terminal $\alpha$ is expressed as
\begin{eqnarray}\label{eq:current_operator}
\hat{I}_{\alpha}^{\sigma}(t) & = & \frac{e}{h} \sum_{n=1}^{M} \int \!\! \int dE\,dE' \, e^{i(E-E')t/\hbar} [ \hat{a}_{\alpha n}^{\sigma \dagger}(E)\hat{a}_{\alpha
n}^{\sigma}(E') \nonumber \\
&& -\hat{b}_{\alpha n}^{\sigma \dagger}(E) \hat{b}_{\alpha n}^{\sigma}(E')].
\end{eqnarray}
The operators $\hat{a}^{\sigma \dagger}_{\alpha n}(E)$ [$\hat{a}^{\sigma}_{\alpha n}(E)$] create [annihilate] incoming electrons in lead $\alpha$ which have energy $E$,
spin-$\sigma$, and the orbital part of their wave function (i.e., ``conducting channel'') is the transverse propagating mode $|n\rangle$~\cite{Beenakker1997}. The corresponding operators
$\hat{b}^{\sigma \dagger}_{\alpha n}$, $\hat{b}_{\alpha n}^\sigma$ act on the outgoing states. Inserting $\hat{I}_{\alpha}^{\sigma}(t)$ in Eq.~(\ref{eq:noise_time}), and taking
its Fourier transform, leads to the following formula for the spin-resolved  noise power spectrum
\begin{eqnarray}\label{eq:noise_power}
S_{\alpha\beta}^{\sigma\sigma'} (\omega)  & = & \frac{e^2}{h} \int dE \, \sum_{\gamma,\gamma'} \sum_{\rho,\rho'=\uparrow,\downarrow} {\rm Tr} \,  \left [{\bf
A}_{\gamma\gamma'}^{\rho\rho'}(\alpha,\sigma,E,E+\hbar\omega) \right. \nonumber \\
&& \times \left. {\bf A}_{\gamma'\gamma}^{\rho'\rho}(\beta,\sigma',E+\hbar\omega,E) \right] \{f_{\gamma}^{\rho}(E)[1-f_{\gamma'}^{\rho'}(E+\hbar\omega)] \nonumber \\
&& + f_{\gamma'}^{\rho'}(E+\hbar\omega)[1-f_{\gamma}^{\rho}(E)]\}.
\end{eqnarray}
Here $f_\gamma^{\rho}(E)$ is the Fermi function of spin-$\rho$ electrons ($\rho=\uparrow,\downarrow$), kept at temperature $T$ and having spin-dependent chemical potential $\mu_\gamma^{\rho}$ in lead $\gamma$. The B\" uttiker's current matrix~\cite{Blanter2000} ${\bf A}_{\beta\gamma}^{\rho\rho'}(\alpha,\sigma,E,E')$, whose elements are
\begin{eqnarray} \label{eq:amatrix}
[{\bf A}_{\beta\gamma}^{\rho\rho'}(\alpha,\sigma,E,E')]_{mn}  =  \delta_{m n} \delta_{\beta \alpha} \delta_{\gamma \alpha} \delta^{\sigma \rho} \delta^{\sigma
  \rho'}  -\sum_{k} [{\bf s}_{\alpha \beta}^{\sigma \rho\dagger}(E)]_{mk} [{\bf s}_{\alpha
  \gamma}^{\sigma \rho'}(E')]_{kn},
\end{eqnarray}
is now generalized to include explicitly spin degrees of freedom through the spin-resolved scattering matrix connecting operators $\hat{a}^{\sigma}_{\alpha n}(E)$ and
$\hat{b}^{\sigma}_{\alpha n}(E)$ via
\begin{equation}\label{eq:smatrix}
\hat{b}_{\alpha n}^{\sigma}(E)=\sum_{\beta m} [{\bf s}_{\alpha \beta}^{\sigma \sigma'}]_{nm}(E) \hat{a}_{\beta m}^{\sigma'}(E).
\end{equation}
In the zero-temperature limit the thermal (Johnson-Nyquist) contribution to the noise vanishes and the Fermi function becomes a step function
$f_\gamma^{\rho}(E)=\theta(E-\mu_\gamma^{\rho})$.

\subsection{Two-terminal spin-resolved shot noise}\label{sec:two_noise}

Evaluation of Eq.~(\ref{eq:noise_power}) for  zero-temperature and zero-frequency, $S_{\alpha\beta}^{\sigma \sigma^\prime} \equiv S_{\alpha\beta}^{\sigma
\sigma^\prime}(\omega=0,T=0)$, in the right lead $\alpha=2=\beta$ of a two-terminal mesoscopic device yields the scattering theory formulae for the shot noise arising in the course of propagation of spin-polarized current through a central region with arbitrary spin-dependent interactions:
\numparts
\begin{eqnarray}
\label{eq:noise_resolved1}
S_{22}^{\uparrow\uparrow} & = & \displaystyle\frac{2e^{2}}{h}\left[  {\rm Tr} \, \left (
    {\bf t}_{21}^{\uparrow\uparrow}{\bf t}_{21}^{\uparrow\uparrow\dagger}\right)eV
    +{\rm Tr}\left({\bf t}_{21}^{\uparrow\downarrow}{\bf t }_{21}^{\uparrow\downarrow\dagger}\right)\frac{1-|{\bf P}_{\rm in}|}{1+|{\bf P}_{\rm in}|}eV \right. \nonumber \\
   \displaystyle && - \left. {\rm Tr}\, \left
     ({\bf t}_{21}^{\uparrow\downarrow}{\bf t}_{21}^{\uparrow\downarrow\dagger}{\bf t}_{21}^{\uparrow\downarrow}{\bf t}_{21}^{\uparrow\downarrow\dagger}\right )\frac{1-|{\bf
P}_{\rm in}|}{1+|{\bf P}_{\rm in}|}eV -
      {\rm Tr}\left ({\bf t}_{21}^{\uparrow\uparrow}{\bf t}_{21}^{\uparrow\uparrow\dagger}{\bf t}_{21}^{\uparrow\uparrow}{\bf
t}_{21}^{\uparrow\uparrow\dagger}\right ){eV}
\right. \nonumber \\
    \displaystyle  &  & -\left.
      2{\rm Tr}\left ({\bf t}_{21}^{\uparrow\downarrow}{\bf t}_{21}^{\uparrow\downarrow\dagger}{\bf t}_{21}^{\uparrow\uparrow}{\bf t}_{21}^{\uparrow\uparrow\dagger}\right
)\frac{1-|{\bf P}_{\rm in}|}{1+|{\bf P}_{\rm in}|}eV\right], \\
\label{eq:noise_resolved2}
     S_{22}^{\downarrow\downarrow} & = & \displaystyle\frac{2e^{2}}{h}\left[  {\rm Tr}\left (
    {\bf t}_{21}^{\downarrow\downarrow}{\bf t}_{21}^{\downarrow\downarrow\dagger}\right)\frac{1-|{\bf P}_{\rm in}|}{1+|{\bf P}_{\rm in}|}eV
    +{\rm Tr}\left({\bf t}_{21}^{\downarrow\uparrow}{\bf t }_{21}^{\downarrow\uparrow\dagger}\right)eV
    \right. \nonumber \\
    \displaystyle  &  & -\left.
     {\rm Tr}\left
     ({\bf t}_{21}^{\downarrow\downarrow}{\bf t}_{21}^{\downarrow\downarrow\dagger}{\bf t}_{21}^{\downarrow\downarrow}{\bf t}_{21}^{\downarrow\downarrow\dagger}\right
)\frac{1-|{\bf P}_{\rm in}|}{1+|{\bf P}_{\rm in}|}eV
    -  {\rm Tr}\left ({\bf t}_{21}^{\downarrow\uparrow}{\bf t}_{21}^{\downarrow\uparrow\dagger}{\bf t}_{21}^{\downarrow\uparrow}{\bf
t}_{21}^{\downarrow\uparrow\dagger}\right ){eV}
\right. \nonumber \\
\displaystyle  &  & -\left.
           2{\rm Tr}\left ({\bf t}_{21}^{\downarrow\uparrow}{\bf t}_{21}^{\downarrow\uparrow\dagger}{\bf t}_{21}^{\downarrow\downarrow}{\bf
t}_{21}^{\downarrow\downarrow\dagger}\right )\frac{1-|{\bf P}_{\rm in}|}{1+|{\bf P}_{\rm in}|}eV\right], \\
\label{eq:noise_resolved3}
S_{22}^{\uparrow\downarrow} & = & -\displaystyle\frac{2e^{2}}{h}\left[{\rm Tr}\left (
    {\bf t}_{21}^{\downarrow\uparrow}{\bf t}_{21}^{\uparrow\uparrow\dagger}{\bf t}_{21}^{\uparrow\downarrow}{\bf t}_{21}^{\downarrow\downarrow\dagger}\right)\frac{1-|{\bf
P}_{\rm in}|}{1+|{\bf P}_{\rm in}|}eV
\right. \nonumber \\
\displaystyle  &  & +\left.
    {\rm Tr}\left({\bf t}_{21}^{\downarrow\uparrow}{\bf t }_{21}^{\uparrow\uparrow\dagger}{\bf t}_{21}^{\uparrow\uparrow}{\bf t}_{21}^{\downarrow\uparrow\dagger}\right)eV
   +
    {\rm Tr} \left
     ({\bf t}_{21}^{\downarrow\downarrow}{\bf t}_{21}^{\uparrow\downarrow\dagger}{\bf t}_{21}^{\uparrow\uparrow}{\bf t}_{21}^{\downarrow\uparrow\dagger}\right) \frac{1-|{\bf
P}_{\rm in}|}{1+|{\bf P}_{\rm in}|}eV
  \right. \nonumber \\
     \displaystyle  &  & +\left.
     {\rm Tr}\left({\bf t}_{21}^{\downarrow\downarrow}{\bf t}_{21}^{\uparrow\downarrow\dagger}{\bf t}_{21}^{\uparrow\downarrow}{\bf t}_{21}^{\downarrow\downarrow\dagger}\right
)\frac{1-|{\bf P}_{\rm in}|}{1+|{\bf P}_{\rm in}|}eV
     \right],\\
\label{eq:noise_resolved4}
S_{22}^{\downarrow\uparrow} & = & -\displaystyle\frac{2e^{2}}{h}\left[{\rm Tr}\left (
    {\bf t}_{21}^{\uparrow\uparrow}{\bf t}_{21}^{\downarrow\uparrow\dagger}{\bf t}_{21}^{\downarrow\downarrow}{\bf t}_{21}^{\uparrow\downarrow\dagger}\right)\frac{1-|{\bf
P}_{\rm in}|}{1+|{\bf P}_{\rm in}|}eV
    \right. \nonumber \\
     \displaystyle  &  & +\left.
    {\rm Tr}\left({\bf t}_{21}^{\uparrow\uparrow}{\bf t }_{21}^{\downarrow\uparrow\dagger}{\bf t}_{21}^{\downarrow\uparrow}{\bf t}_{21}^{\uparrow\uparrow\dagger}\right)eV
    +
     {\rm Tr}\left
     ({\bf t}_{21}^{\uparrow\downarrow}{\bf t}_{21}^{\downarrow\downarrow\dagger}{\bf t}_{21}^{\downarrow\downarrow}{\bf t}_{21}^{\uparrow\downarrow\dagger}\right
)\frac{1-|{\bf P}_{\rm in}|}{1+|{\bf P}_{\rm in}|}eV
\right. \nonumber \\
     \displaystyle  &  & +\left.
      {\rm Tr}\left ({\bf t}_{21}^{\uparrow\downarrow}{\bf t}_{21}^{\downarrow\downarrow\dagger}{\bf t}_{21}^{\downarrow\uparrow}{\bf t}_{21}^{\uparrow\uparrow\dagger}\right
)\frac{1-|{\bf P}_{\rm in}|}{1+|{\bf P}_{\rm in}|}{eV}
     \right].
\end{eqnarray}
\endnumparts
Here the elements of the transmission matrix ${\bf t}_{21}^{\sigma \sigma'}$, which is a block of the full scattering matrix~\cite{Beenakker1997}, determine the probability $|[{\bf t}_{21}^{\sigma
\sigma^\prime}]_{nm}|^2$ for spin-$\sigma^\prime$  electron incident in lead $1$ in the orbital conducting channel $|m\rangle$ to be transmitted to lead $2$ as spin-$\sigma$ electron in channel $|n\rangle$. The direction of the spin-polarization vector of injected electrons selects the spin-quantization axis for $\uparrow$, $\downarrow$. Its magnitude quantifies the degree of spin polarization, which is introduced into Eq.~(\ref{eq:noise_power}) via the spin-dependent electrochemical potentials in the injecting (left) lead
\begin{eqnarray}\label{eq:muwithp}
\mu_1^\uparrow & = & E_F+eV, \\
\mu_1^\downarrow & = & E_F+eV \frac{1-|{\bf P}_{\rm in}|}{1+|{\bf P}_{\rm in}|}.
\end{eqnarray}
In the collecting (right) lead the electrochemical potentials for both spin-species are the same $\mu_2^\uparrow = \mu_2^\downarrow=E_F$, where $E_F$ is the Fermi energy. For instance, injection of fully spin-$\uparrow$ polarized current $|{\bf P}_{\rm in}|=1$ from the left lead (e.g., made of half-metallic ferromagnet) means that there is no voltage drop for spin-$\downarrow$ electrons $\mu_1^\downarrow=\mu_2^\downarrow=E_F$, so that they do not contribute to transport.

Equations~(\ref{eq:noise_resolved1})--(\ref{eq:noise_resolved4}), together with the expressions for mean spin-resolved currents
collected in the right lead,
\begin{numparts}\label{eq:idetect}
\begin{eqnarray}
I_{2}^{\uparrow} \equiv \langle \hat{I}_2^\uparrow(t) \rangle & = &  \left(G_{21}^{\uparrow\uparrow}+G_{21}^{\uparrow\downarrow}\frac{1-|{\bf P}_{\rm in}|}{1+|{\bf P}_{\rm
in}|}\right)V, \\
I_{2}^{\downarrow} \equiv \langle \hat{I}_2^\downarrow(t) \rangle & = &  \left(G_{21}^{\downarrow\uparrow}+G_{21}^{\downarrow\downarrow}\frac{1-|{\bf P}_{\rm in}|}{1+|{\bf
P}_{\rm in}|}\right)V,
\end{eqnarray}
\end{numparts}
define the Fano factors for parallel and antiparallel spin valve setups,
\begin{eqnarray}\label{eq:fano_spin_valve}
F_{\uparrow \rightarrow \uparrow} & = & \frac{S_{22}^{\uparrow \uparrow}(|{\bf P}_{\rm in}|=1)}{2 e I_2^\uparrow(|{\bf P}_{\rm in}|=1)}, \\
F_{\uparrow \rightarrow \downarrow} & = & \frac{S_{22}^{\downarrow \downarrow}(|{\bf P}_{\rm in}|=1)}{2 e I_2^\downarrow(|{\bf P}_{\rm in}|=1)}.
\end{eqnarray}
These equations also yield the Fano factor for a ferromagnet$|$SO-coupled-wire$|$paramagnet configuration
\begin{equation}\label{eq:fano_para}
F_{\uparrow \rightarrow \uparrow\downarrow} = \frac{S_{22}(|{\bf P}_{\rm in}|=1)}{2 e I_2(|{\bf P}_{\rm in}|=1)},
\end{equation}
where $I_2 =  I_{2}^{\uparrow} + I_{2}^{\downarrow}$ is the sum of both spin-resolved currents collected in the right paramagnetic lead. The spin-resolved two-terminal
conductances
\begin{equation}\label{eq:landauer}
G_{2 1}^{\sigma \sigma^\prime}=\frac{e^2}{h} \sum_{n,m=1}^{M} |[{\bf t}_{2 1}^{\sigma \sigma^\prime}]_{nm}|^2.
\end{equation}
are given by the usual Landauer formula.

\subsection{Four-terminal spin-resolved shot noise}\label{sec:four_noise}

Evaluation of $S_{\alpha\beta}^{\sigma \sigma^\prime} \equiv S_{\alpha\beta}^{\sigma \sigma^\prime}(\omega=0,T=0)$ at  zero-temperature and zero-frequency in the top
lead $\alpha=\beta=3$ of the four-terminal bridge in Fig.~\ref{fig:she_inverse}, typically employed in the analysis of the mesoscopic SHE~\cite{Nikoli'c2005b,Nikoli'c2006,Sheng2006b,Nikoli'c2007},   yields explicit  expressions for $S_{33}^{\sigma \sigma^\prime}$, $S_{33}^{\rm spin}$, and $S_{33}^{\rm charge}$ noise power. They are too lengthy to be written down explicitly here due to numerous terms arising from the effect of other leads on the shot noise in selected lead 3. We note that using the unitarity of the scattering matrix, $S_{33}^{\sigma \sigma^\prime}$ can be expressed solely in terms of the transmission matrix ${\bf t}_{\alpha\beta}^{\sigma \sigma'}$. The spin quantization axis for $\uparrow$ and $\downarrow$ spin states is assumed to be the $z$-axis, so that all spin currents and noises in lead 3 describe the SHE response of 2DEG~\cite{Nikoli'c2006}.

Since in two-terminal devices the spin dynamics affecting the shot noise is most pronounced when injected current is
spin-polarized~\cite{Mishchenko2003,Dragomirova2007,Hatami2006}, we also evaluate in Sec.~\ref{sec:she_noise} the noise correlators for setups where spin-polarized charge current is injected through lead 1 thereby driving the transverse charge Hall current~\cite{Bulgakov1999} through leads 3 and 4.  In this case, the magnitude $|{\bf P}_{\rm in}|$ of the spin-polarization vector enters into Eq.~(\ref{eq:noise_power}) via the spin-dependent electrochemical potentials in the injecting lead 1,  $\mu_1^\uparrow=E_F+eV$ and $\mu_1^\downarrow=E_F+eV(1-|{\bf P}_{\rm in}|)/(1+|{\bf P}_{\rm in}|)$, in complete analogy with the two-terminal spin-resolved noise formulas in Sec.~\ref{sec:two_noise}. Such setup~\cite{Bulgakov1999} is closely related to the inverse SHE where  $\mu_1^\uparrow=E_F+eV=\mu_2^\downarrow$ and $\mu_1^\downarrow=E_F=\mu_2^\uparrow$ describes injection of two counter-propagating fully spin-polarized charge currents of opposite ${\bf P}$ and, therefore, no net longitudinal charge current~\cite{Hankiewicz2005}.

\subsection{Effective SO Hamiltonian and nonequilibrium Green functions for computing the spin-resolved noise}\label{sec:negf}

The consequences of Eq.~(\ref{eq:noise_power}) can be explored by analytical means, such as the  wave function matching~\cite{Egues2005} (for one or two channel leads attached to ballistic structures~\cite{Egues2005}) or random matrix theory  applicable to ``black-box'' disordered and chaotic ballistic  structures~\cite{Lamacraft2004,Bardarson2007} which are smaller than the spin precession length $L_{\rm SO}$. However, to take into account concurrent microscopic  modeling~\cite{Hankiewicz2008} of the impurity scattering, SO effects (skew-scattering and side jump~\cite{Hankiewicz2008,Sinitsyn2008}) in the electric field of an impurity, and fast spin precession induced by strong intrinsic SO
coupling effects~\cite{Chang2004}, it is more advantageous to employ the    nonequilibrium Green function (NEGF)~\cite{Haug2007} technique via
numerically exact real$\otimes$spin space~\cite{Nikoli'c2005b,Nikoli'c2006} computation. The NEGF formalism can take as an input the microscopic
Hamiltonian of both weekly ($L \ll L_{\rm SO}$) and strongly ($L \ge L_{\rm  SO}$) SO-coupled  nanostructure of arbitrary shape and disorder 
attached to many multichannel electrodes.

In general, the central 2DEG region (such as those employed in recent SHE experiments~\cite{Sih2005a}) can be modeled by the effective mass Hamiltonian which
takes into account intrinsic and extrinsic SO coupling effects, as well as the impurity $V_{\rm dis}(x,y)$ and confining $V_{\rm conf}(y)$ potentials
\begin{eqnarray}\label{eq:hamil}
\hat{H}  & = &  \frac{\hat{p}_x^2+\hat{p}_y^2}{2m^*} + V_{\rm dis}(x,y) + V_{\rm conf}(y)  \nonumber \\
\displaystyle && + \frac{\alpha_R}{\hbar}
\left( \hat{p}_y \hat{\sigma}_x  - \hat{p}_x  \hat{\sigma}_y  \right) + \lambda \left(\hat{\bm{\sigma}} \times \hat{\bf p} \right) \cdot \nabla V_{\rm dis}(x,y).
\end{eqnarray}
Here the fourth term is the intrinsic Rashba SO coupling~\cite{Winkler2003} due to structural inversion asymmetry of the quantum well, $(\hat{\sigma}_x,\hat{\sigma}_y,\hat{\sigma}_z)$ denotes the vector of the Pauli matrices, and $\hat{\bf p}=(\hat{p}_x,\hat{p}_y)$ is the momentum operator in 2D space. The Rashba coupling is responsible for   spin splitting $\Delta_{\rm SO} = 2 \alpha_R k_F$ of quasiparticle energies at the Fermi level ($\hbar k_F$ is the Fermi momentum). The fifth  term is a relativistic correction to the Pauli equation for spin-$\frac{1}{2}$ particle where minuscule value of $\lambda$ in vacuum can be renormalized enormously by the band structure effects due to strong crystal potential (leading to, e.g.,  $\lambda/\hbar =5.3$ \AA{}$^2$ for GaAs~\cite{Winkler2003}).

The effective momentum-dependent magnetic field ${\bf B}_{\rm int}({\bf p})$ of the Rashba SO coupling lies in the plane of a 2DEG, thereby forcing the injected $z$-polarized spins to precess. This process is characterized by the spin precession length $L_{\rm SO}$, along which injected out-of-plane polarized spins precess by an angle $\pi$. The $L_{\rm SO}$ scale plays a crucial role in the mesoscopic SHE~\cite{Nikoli'c2005b}. It also plays the role of DP spin dephasing length in weakly disordered bulk SO-coupled systems~\cite{Fabian2007,Mal'shukov2000,Kiselev2000,Chang2004,Pareek2002}. This scale is inversely proportional to the Rashba coupling strength
\begin{equation}\label{eq:lso}
L_{\rm SO}= \frac{\pi \hbar^2}{2m^* \alpha_R},
\end{equation}
and can be extracted from the measurements of spin dephasing in both ballistic and diffusive systems~\cite{Chang2004}.

For NEGF computation we represent the general 2DEG Hamiltonian (\ref{eq:hamil}) in the local orbital basis~\cite{Nikoli'c2007}
\begin{eqnarray}\label{eq:tbh}
\hat{H}_{\rm TB} & = &  \sum_{{\bf m},\sigma} \varepsilon_{\bf m} \hat{c}_{{\bf
m}\sigma}^\dag\hat{c}_{{\bf m}\sigma}+\sum_{\langle {\bf
mm'}\rangle}  \sum_{\sigma\sigma'} \hat{c}_{{\bf m}\sigma}^\dag t_{\bf
mm'}^{\sigma\sigma'}\hat{c}_{{\bf m'}\sigma'} \nonumber \\
\displaystyle && -i  \lambda_{\rm SO} \sum_{{\bf m},\alpha\beta} \sum_{ij} \sum_{\nu\gamma} \epsilon_{ijz} \nu \gamma (\varepsilon_{{\bf m} + \gamma {\bf e}_j} -
\varepsilon_{{\bf m}+\nu {\bf e}_i}) \nonumber \\
\displaystyle && \times \hat{c}_{{\bf
m},\alpha}^\dag \hat{\sigma}^z_{\alpha\beta} \hat{c}_{{\bf m}+\nu{\bf e}_i+\gamma{\bf e}_j,\beta}.
\end{eqnarray}
The first term accounts for isotropic short-range spin-independent static impurity potential where $\varepsilon_{\bf m} \in [-W_{\rm dis}/2,W_{\rm dis}/2]$ is a uniform random variable. The second term is the tight-binding representation of the Rashba SO coupling whose nearest-neighbor $\langle {\bf mm'}\rangle$ hopping is a  non-trivial $2 \times 2$ Hermitian matrix  ${\bf t}_{\bf m'm}=({\bf t}_{\bf mm'})^\dagger$ in the spin space~\cite{Nikoli'c2006}
\begin{eqnarray}\label{eq:hopping}
{\bf t}_{\bf mm'}=\left\{
\begin{array}{cc}
-t_{\rm O}{\bf I}_{\rm S}-it_{\rm SO}\hat{\sigma}_y &
({\bf m}={\bf m}'+{\bf e}_x)\\
-t_{\rm O}{\bf I}_{\rm S}+it_{\rm SO}\hat{\sigma}_x &  ({\bf m}={\bf m}'+{\bf e}_y).
\end{array}\right.
\end{eqnarray}
Here ${\bf I}_{\rm S}$ is the unit $2 \times 2$ matrix in the spin space, and ${\bf e}_x$ and ${\bf e}_y$ are the unit vectors along the $x$ and $y$ axes, respectively. The strength of the SO coupling is measured by the parameter $t_{\rm SO}=\alpha_R/2a$ ($a$ is the lattice spacing), and the spin-splitting of the band structure is expressed as $\Delta_{\rm SO}=4at_{\rm SO}k_F$ in terms of $t_{\rm SO}$. A direct correspondence between  the continuous effective mass Hamiltonian Eq.~(\ref{eq:hamil})  and its lattice version Eq.~(\ref{eq:tbh}) is established  by selecting the Fermi energy of the injected electrons to be close to the bottom of the band where tight-binding dispersion reduces to the parabolic one, and by using $t_{\rm O}=\hbar^2/(2 m^* a^2)$ for the orbital hopping  which yields the effective mass $m^*$ in the continuum limit~\cite{Nikoli'c2006}.  The labels in the third term, which involves both nearest-neighbor and next-nearest-neighbor hopping, are: the dimensionless extrinsic SO scattering strength $\lambda_{\rm SO}=\lambda \hbar/(4a^2)$; $\epsilon_{ijz}$ stands for the Levi-Civita totally antisymmetric tensor with $i,j$ denoting the in-plane coordinate axes; and $\nu,\gamma$ are the dummy indices taking values $\pm 1$.

The central NEGF quantity for the computation of the transmission coefficients is the retarded Green function of the scattering region
\begin{equation}\label{eq:retarded}
{\bf G}^r=[E-{\bf H}_{\rm open}]^{-1},
\end{equation}
associated with the  matrix representation of the Hamiltonian ${\bf H}_{\rm open} = {\bf H}_{\rm TB} + \sum_{\alpha,\sigma} {\bm \Sigma}_\alpha^{r,\sigma}$ of an open system [${\bf H}_{\rm TB}$ is the matrix representation of Eq.~(\ref{eq:tbh})]. Here  non-Hermitian retarded self-energy matrices ${\bm \Sigma}_\alpha^{r,\sigma}$ introduced by the interaction with the leads determine escape rates of spin-$\sigma$ electrons into the electrodes. The block ${\bf G}_{\alpha \beta}^{r,\sigma \sigma^\prime}$ of the retarded Green function matrix, consisting of those matrix elements which
connect the layer of the sample attached to lead $\beta$ to the layer of the sample attached to lead $\alpha$, yields the spin-resolved transmission matrix
\begin{equation}\label{eq:transmission}
  {\bf t}_{\alpha \beta}^{\sigma \sigma^\prime}  =   2 \sqrt{-{\rm Im} \, {\bm \Sigma}_\alpha^{r,\sigma}} \cdot {\bf G}_{\alpha \beta}^{r,\sigma \sigma^\prime} \cdot
\sqrt{-{\rm Im} \, {\bm \Sigma}_\beta^{r,\sigma'}}.
\end{equation}
For simplicity, we assume that $\hat{\bm \Sigma}^{r,\uparrow}_\alpha=\hat{\bm \Sigma}^{r,\downarrow}_\alpha$, which experimentally corresponds to identical conditions for injection of
both spin species.

By replacing transmission matrices with Eq.~(\ref{eq:transmission}) in the spin-resolved noise expressions in Eqs.~(\ref{eq:noise_resolved1})--(\ref{eq:noise_resolved4}), or the corresponding expressions in four-terminal structures,  we arrive at NEGF formulas that can be easily evaluated in terms of ${\bf G}^r$ and ${\bm \Sigma}_\alpha^{r,\sigma}$ matrices~\cite{Dragomirova2007,Dragomirova2008}. The equivalent route to such NEGF expressions for the noise would be to start from the spin-resolved current expression in terms of NEGF, rather than the scattering formula Eq.~(\ref{eq:current_operator}), and then work through a lengthy derivation similar to the one provided in Refs.~\cite{Souza2008,Haug2007}.

\section{Shot noise in two-terminal diffusive Rashba SO-coupled wires} \label{sec:two_terminal_rashba_wire}

\begin{figure}
\centerline{\psfig{file=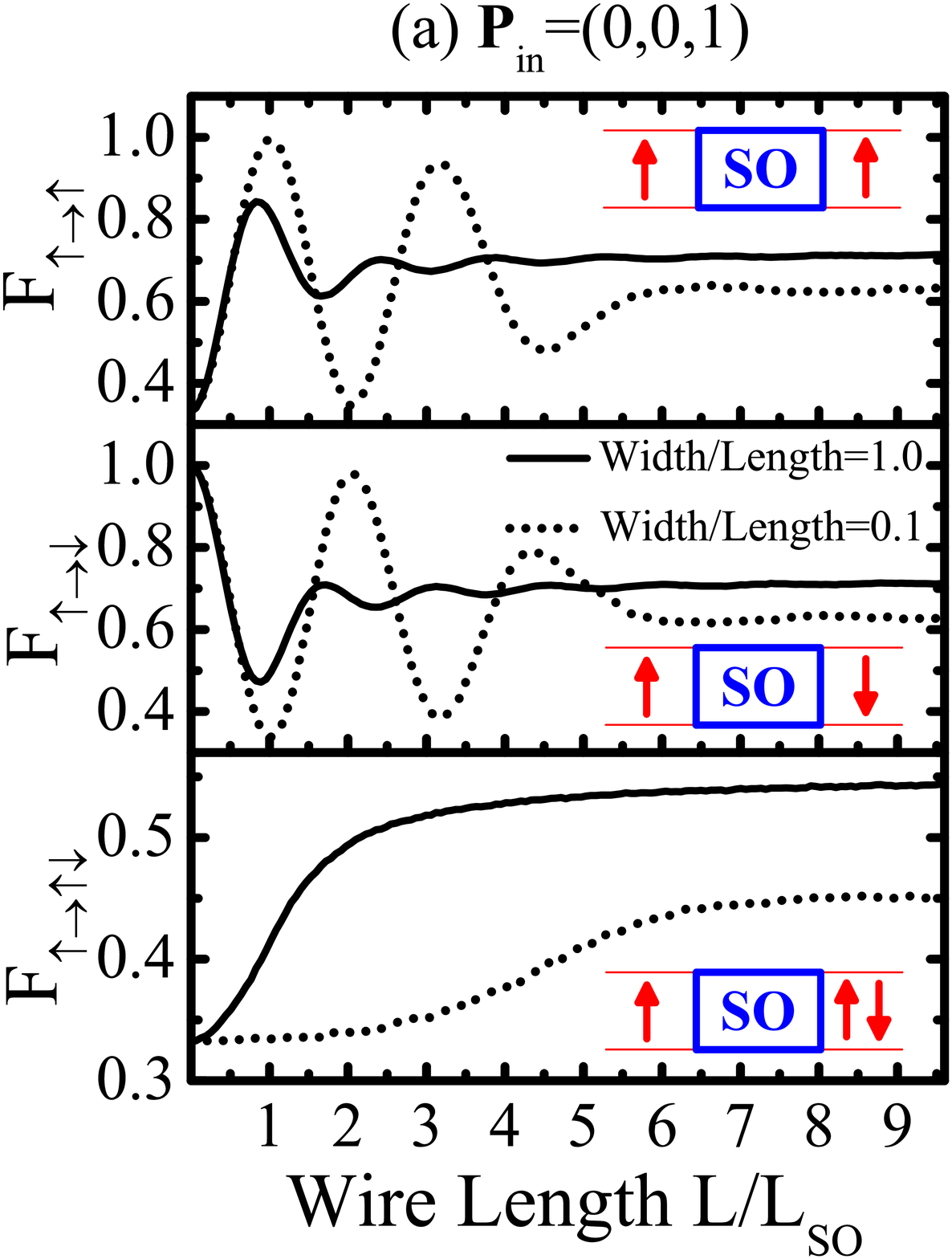,scale=0.18,angle=0} \psfig{file=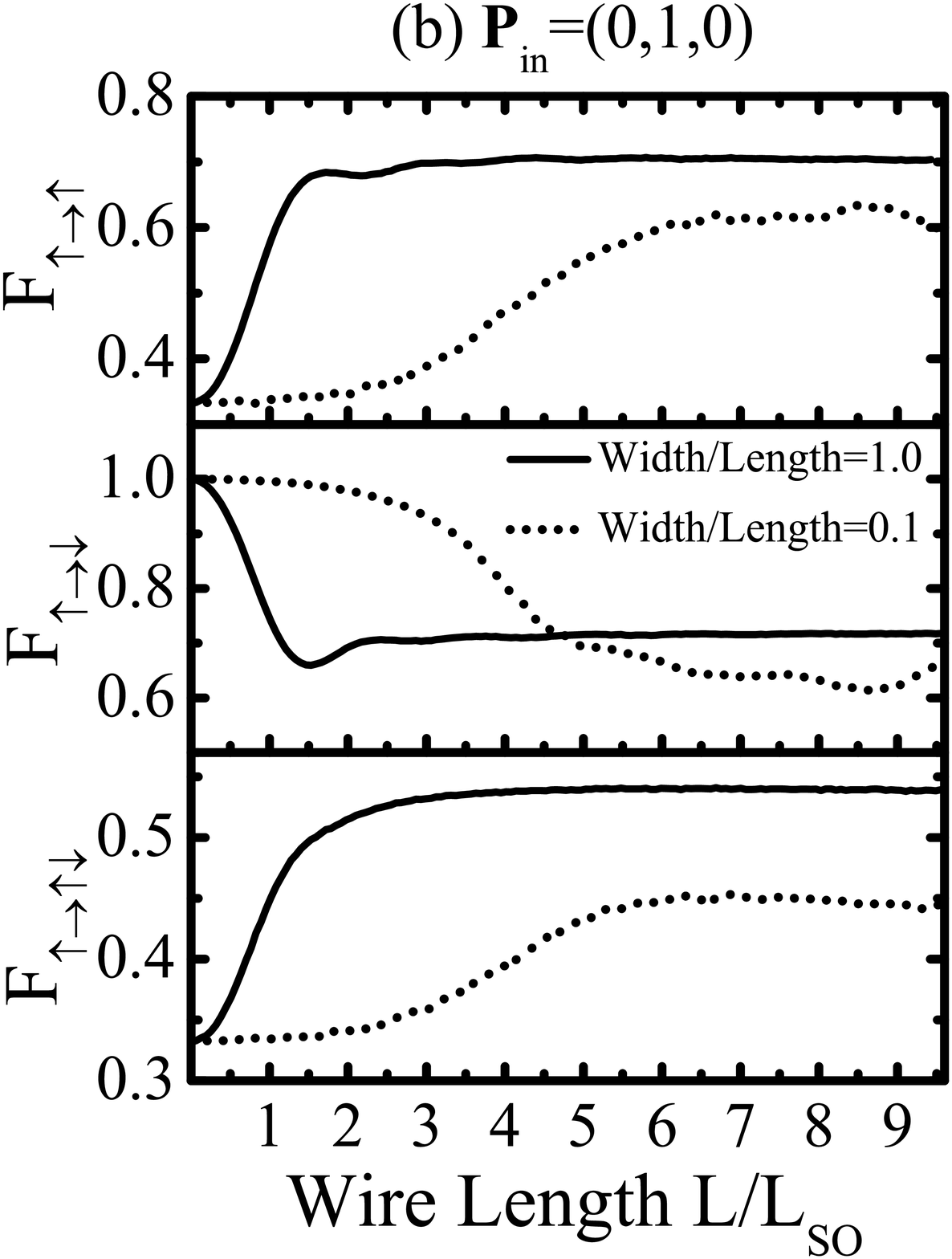,scale=0.18,angle=0} \psfig{file=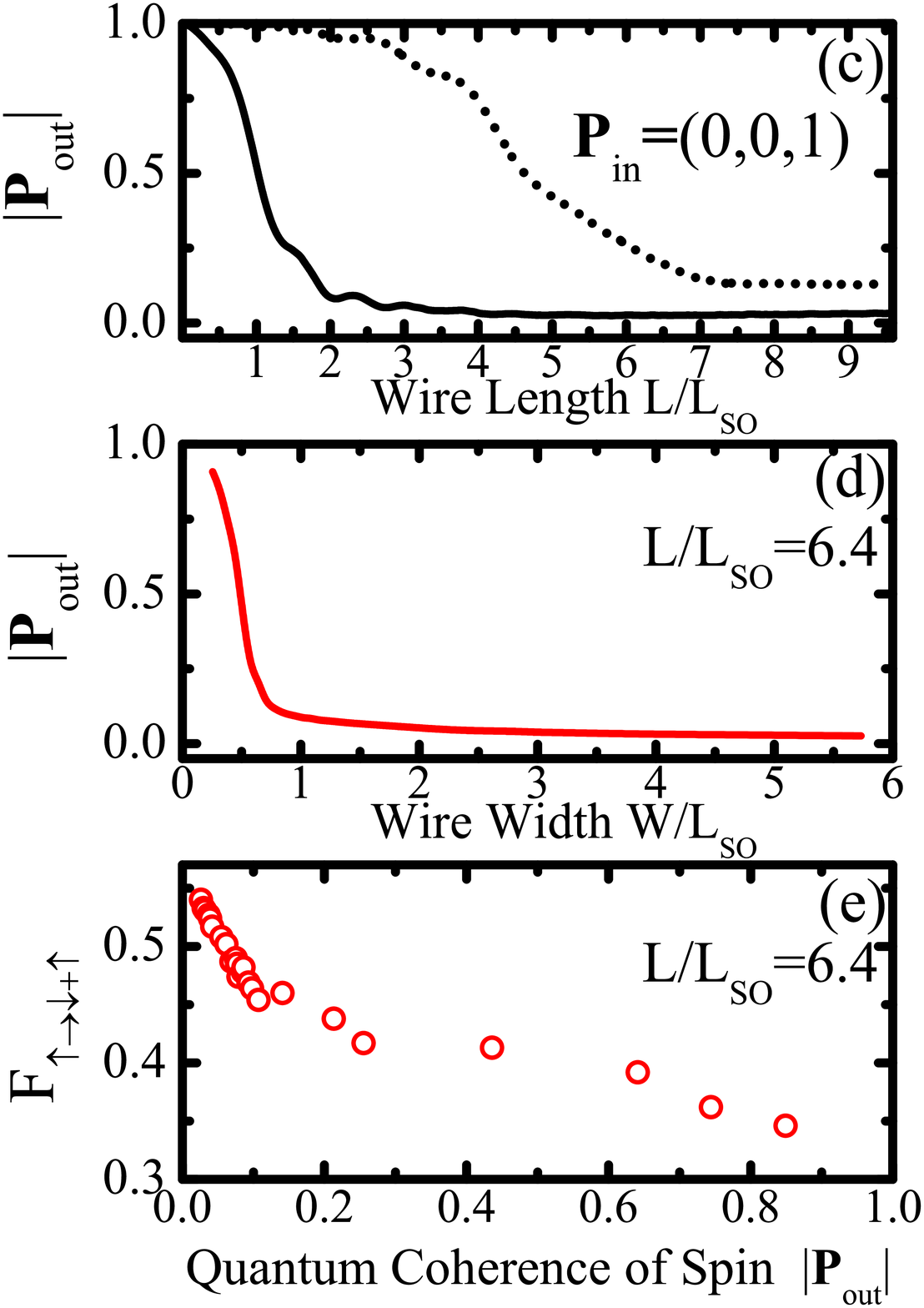,scale=0.18,angle=0}}
\caption{Panels (a) and (b) show the Fano factor vs. the spin precession length $L_{\rm SO}$ for different two-terminal setups  [Fig.~\ref{fig:setup}] where 100\% spin-$\uparrow$ polarized charge current is injected from the  source electrode (e.g., half-metallic ferromagnet) into a {\em diffusive} Rashba SO-coupled wire and spin-resolved charge currents $I_2^\uparrow$ (top), $I_2^\downarrow$ (middle), or both $I_2^\uparrow + I_2^\downarrow$ (bottom), are collected in the drain electrode. Panel (c) shows the corresponding decay of the degree of quantum-coherence of transported spin, as quantified by the magnitude $|{\bf P}_{\rm out}|$ of the spin-polarization vector of detected charge current. The injected current in the left lead is composed of fully coherent pure spin state   $|{\bf P}_{\rm in}|=1$, where ${\bf P}_{\rm in}$ points along the $z$-axis in panels (a), (c), (d), and (e) or the $y$-axis in panel (b). For fixed $L$ and $L_{\rm SO}$, the decay of $|{\bf P}_{\rm out}|$ is suppressed in narrow wires [panel (d)], which establishes a one-to-one correspondence between the Fano factor $F_{\uparrow \rightarrow \uparrow\downarrow}$ and $|{\bf P}_{\rm out}|$ in panel (e). Note that the Fano factors  attaining universal value $F_{\uparrow \rightarrow \uparrow} = F_{\uparrow \rightarrow \uparrow \downarrow}=1/3$ in the limit of zero SO coupling $L/L_{\rm SO} \rightarrow 0$ demonstrate that our wires are in the diffusive transport regime for selected disorder strength. Adapted from Ref.~\cite{Dragomirova2007}.
}\label{fig:noise}
\end{figure}

In this Section, the spin-dependent shot noise formalism introduced in Sec.~\ref{sec:formalism} is applied to diffusive SO-coupled quantum wires of different widths where the recent experiments~\cite{Holleitner2006} demonstrate how transverse confinement affects the degree of transported spin coherence~\cite{Nikoli'c2005}. The quantum wires are realized using 2DEG (in the $xy$-plane so that the unit vector ${\bf e}_z$ is orthogonal to it) with a tunable Rashba SO coupling, as described by the effective mass Hamiltonian Eq.~(\ref{eq:hamil}) and its lattice version Eq.~(\ref{eq:tbh}) [we assume that extrinsic SO scattering effects are negligible, $\lambda_{\rm SO}=0$]. The internal magnetic field ${\bf B}_{\rm int}({\bf p})=-(2 \alpha/ g\mu_B)(\hat{\bf p} \times {\bf e}_z)$  of the Rashba SO coupling is nearly parallel to the transverse $y$-axis in the case of quantum wires~\cite{Governale2004}. Therefore, the injected $z$-axis polarized spins are precessing within the wires, while the $y$-axis polarized spins are in the eigenstates of the corresponding Zeeman term and do not precess. This leads to a difference in the shot noise when changing the spin-polarization vector of the injected current in the ``polarizer-analyzer'' scheme in the top and middle panels of Figs.~\ref{fig:noise}(a) and \ref{fig:noise}(b).

Moreover, in both cases and within the asymptotic limit $L \gg L_{\rm SO}$ ($L$ is the wire length) we find that the Fano factor of the shot noise increases above the universal value $F=1/3$ (characterizing diffusive wires with zero SO coupling, $L_{\rm SO} \rightarrow \infty$) for all three measurement geometries in Fig.~\ref{fig:noise}:

\begin{enumerate}

\item spin valves with parallel magnetization of the electrodes where $\uparrow$-electrons are injected from the
left lead and $\uparrow$-electrons are collected in the right lead---a situation described by the Fano factor $F_{\uparrow \rightarrow \uparrow}$;

\item spin valves with antiparallel magnetization of the electrodes where $\uparrow$-electrons are injected through a perfect Ohmic contact and  $\downarrow$-electrons are
collected, as described by the Fano factor $F_{\uparrow \rightarrow  \downarrow}$;

\item a setup with only one spin-selective electrode where $\uparrow$-electrons are injected and both $\uparrow$- and $\downarrow$-electrons are collected in the normal
drain electrode, as described by the
Fano factor  $F_{\uparrow \rightarrow \uparrow \downarrow}$.

\end{enumerate}

The spin precession length Eq.~(\ref{eq:lso}) defined by the clean Rashba Hamiltonian can be rewritten as 
\begin{equation}\label{eq:lso_lattice}
L_{\rm SO}=\frac{a \pi t_{\rm O}}{2t_{\rm SO}},
\end{equation}
in terms of the parameters of the corresponding lattice Rashba Hamiltonian (\ref{eq:tbh}). For very small SO coupling and, therefore, large $L_{\rm SO} \rightarrow \infty$, the Fano factors $F_{\uparrow \rightarrow \uparrow}$ and $F_{\uparrow \rightarrow \uparrow\downarrow}$ start from the universal value $F=1/3$ characterizing the {\em diffusive unpolarized} transport, and then increase toward their  asymptotic values, $F_{\uparrow \rightarrow \uparrow}(L \gg L_{\rm SO}) \approx F_{\uparrow \rightarrow \downarrow}(L \gg L_{\rm SO}) \simeq 0.7$ and $F_{\uparrow \rightarrow \uparrow\downarrow}(L \gg L_{\rm SO}) \simeq 0.55$. Such enhancement of the spin-dependent shot noise is due to spin {\em decoherence} and {\em dephasing} processes~\cite{Nikoli'c2005} in SO-coupled structures that are responsible for the reduction~\cite{Joos2003,Schlosshauer2008} of the off-diagonal elements of the spin-density matrix  $\hat{\rho}_{\rm out}$ of detected current. Note that in these setups, the density matrix $\hat{\rho}_{\rm in}^2=\hat{\rho}_{\rm in}$ describes pure injected spin states  comprising fully spin-polarized current in the left lead.

However, these asymptotic Fano factor values are lowered in narrow wires where transverse confinement slows down the DP spin relaxation in the picture of semiclassical spin
diffusion~\cite{Mal'shukov2000,Kiselev2000}, or reduces the size of the ``environment'' composed of orbital conducting channels (i.e., smaller ``environment'' means smaller number of channels)
to which the spin can entangle or  which provide ``ensemble dephasing''~\cite{Schlosshauer2008} in fully quantum transport picture~\cite{Nikoli'c2005} employed to obtain $|{\bf P}_{\rm out}|$ vs. the wire width $W$  (at fixed length $L$ and the Rashba SO coupling strength) in Fig.~\ref{fig:noise}(d). The geometrical confinement effects increasing spin coherence in narrow wires~\cite{Nikoli'c2005,Mal'shukov2000,Kiselev2000,Pareek2002}  have been confirmed in very recent optical spin detection experiment~\cite{Holleitner2006}. Their utilization could be essential for the realization of all-electrical semiconductor spintronic devices~\cite{Fabian2007,Awschalom2007} where spin is envisaged to be manipulated via SO couplings while avoiding their detrimental dephasing effects~\cite{Nikoli'c2005}.

\begin{figure}
\centerline{\psfig{file=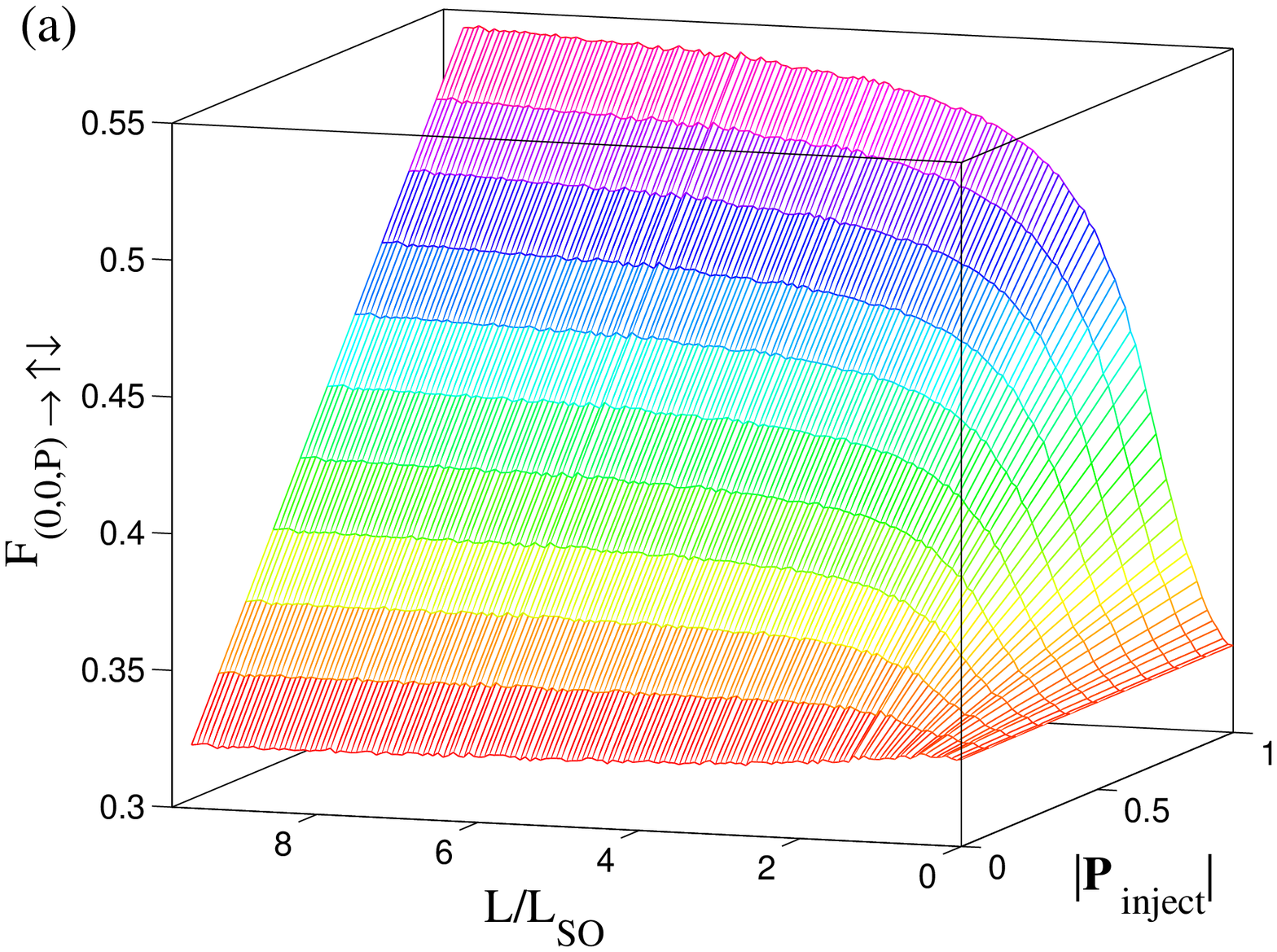,scale=0.38,angle=0}}
\centerline{\psfig{file=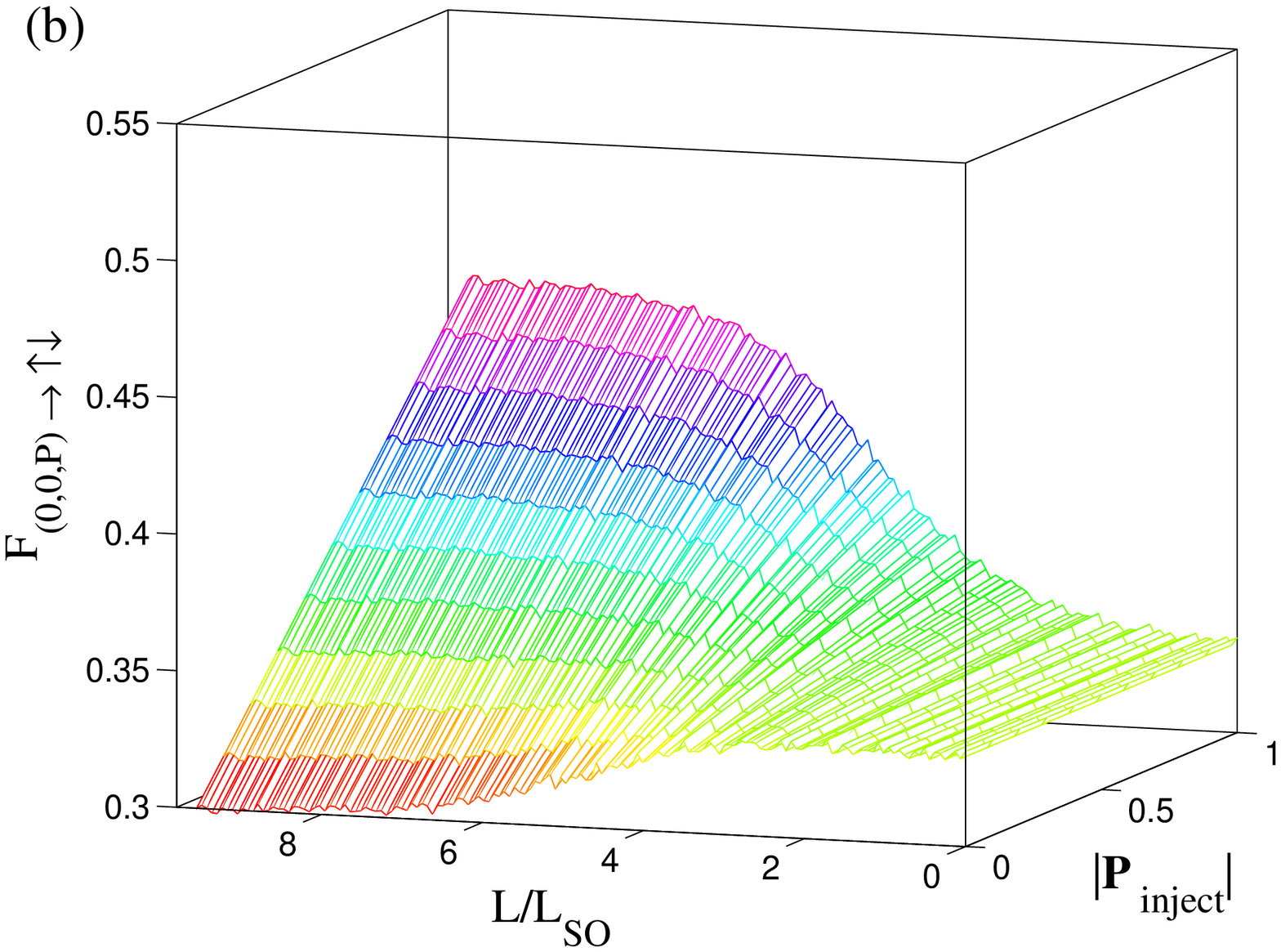,scale=0.38,angle=0}}
\caption{Fano factor as the function of SO coupling strength $L/L_{\rm SO}$ and
$|{\bf P}_{\rm in}|$ for a two-terminal device setup in Fig.~\ref{fig:setup} where partially spin-polarized current, comprised of electrons with spin-polarization vector ${\bf P}_{\rm in}=(0,0,P)$, is injected from an ideal source electrode into a diffusive Rashba SO-coupled wire and charge current of both spins
$I_2^\uparrow + I_2^\downarrow$ is collected by the spin-nonselective drain electrode. In panel (a), the
wire length $L$ and width $W$ are the same $W/L=1$, while panel (b) plots $F_{(0,0,P) \rightarrow \uparrow\downarrow}$ in narrow wires $W/L=0.1$. Note that the limiting curves extracted from the surface plots at $|{\bf P}_{\rm in}|=1$ in panels (a) and (b) are identical to solid and dotted lines, respectively, in the bottom panel of Fig.~\ref{fig:noise}(a). Adapted from Ref.~\cite{Dragomirova2007}.}\label{fig:3d}
\end{figure}

The shot noise in the antiparallel configuration reaches the full Poissonian value $F_{\uparrow \rightarrow \downarrow}(L \ll L_{\rm SO}) \simeq 1$ in the limit of small SO coupling since the probability that the spin state which has huge overlap with $|\!\! \uparrow \rangle$ can enter into the right electrode whose spin-$\uparrow$ states are empty is vanishingly small. This leads to a tunneling-type~\cite{Blanter2000,Mishchenko2003} shot noise where electrons propagate independently without being correlated by their Fermi statistics. In the asymptotic limit $L \gg L_{\rm SO}$, injected spins loose their memory on a very short length scale, so that $F_{\uparrow \rightarrow \downarrow}(L \gg L_{\rm SO})$ acquires the same asymptotic value as $F_{\uparrow \rightarrow \uparrow}(L \gg L_{\rm SO})$.

Since the present spintronic experiments are usually conducted by injecting partially spin-polarized charge currents $|{\bf P}_{\rm in}|<1$, we employ our general formulas Eq.~(\ref{eq:noise_resolved1})--(\ref{eq:noise_resolved4}) to obtain the Fano factor
\begin{equation}\label{eq:fano_p}
F_{(0,0,P) \rightarrow \uparrow\downarrow} = \frac{S_{22}[{\bf P}_{\rm in}=(0,0,P)]}{2 e I_2[{\bf P}_{\rm in}=(0,0,P)]}.
\end{equation}
This represents a generalization of $F_{\uparrow \rightarrow \uparrow \downarrow}$ to characterize the
shot noise in an experimental setup where partially polarized (along the $z$-axis) electrons are injected from the
left lead while both spin species are collected in the right lead. Figure~\ref{fig:3d} suggests that predictions for the excess shot noise $F_{(0,0,P) \rightarrow
\uparrow\downarrow} > 1/3$ should be observable even for small polarization of injected current $|{\bf P}_{\rm in}| \equiv P \gtrsim 10\%$. Note also that for conventional unpolarized current $|{\bf P}_{\rm in}|=0$, one {\em does not} observe any noise enhancement in two-terminal device geometry. In fact, the Fano factor $F_{(0,0,0) \rightarrow \uparrow\downarrow}$ in Figure~\ref{fig:3d} decreases with increasing strength of the Rashba SO coupling due to weak antilocalization correction~\cite{Beenakker1997} that reduces $S_{22}[{\bf P}_{\rm in}=(0,0,0)]$ and increases $I_2[{\bf P}_{\rm in}=(0,0,0)]$ in Eq.~(\ref{eq:fano_p}), as discussed in more detail in Sec.~\ref{sec:discussion} and its Fig.~\ref{fig:spin_resolved}.

\subsection{Fano factor as quantifier of transported spin coherence} \label{sec:fano}

To understand the evolution of quantum coherence of transported spin, we use fully quantum transport formalism of Ref.~\cite{Nikoli'c2005} which treats both the spin dynamics
and  orbital propagation of electrons to which the spins are attached phase coherently.
This allows us to obtain the spin-density matrix of charge current in the right lead in terms of the same spin-resolved transmission matrix ${\bf t}_{21}^{\sigma \sigma'}$
which determines the shot noise power $S_{22}^{\sigma \sigma^\prime}$. Note that traditional description of DP spin dephasing treats charge propagation semiclassically while the dynamics of spin attached to charges is described via quantum evolution of the
spin-density matrix~\cite{Fabian2007,Mal'shukov2000,Kiselev2000,Chang2004}.

Here we summarize principal steps, put forth by Nikoli\' c and Souma in Ref.~\cite{Nikoli'c2005} (and applied or extended in numerous recent studies of electron~\cite{Zhai2005} and hole~\cite{Brusheim2006} transport in  low-dimensional systems with SO couplings and magnetic field affecting their spins), which make it possible to define the spin-density matrix of an ensemble of phase-coherently transported spins comprising the detected current in the right lead within the framework of the scattering approach~\cite{Beenakker1997} to quantum transport. Suppose that a spin-$\uparrow$ polarized electron is injected from  the left lead through a conducting channel $|{\rm in}\rangle \equiv | m \rangle \otimes |\!\! \uparrow \rangle$. Then, a pure state emerging in the right lead after the electron has traversed the sample is described by a linear combination of the outgoing channels, 
\begin{equation}
|{\rm out} \rangle = \sum_{n \sigma} [{\bf t}_{2 1}^{\sigma \uparrow}]_{nm} |n \rangle \otimes |\sigma \rangle.
\end{equation}
Such {\em non-separable} state~\cite{Ballentine1998} encodes {\em entanglement} of spin and the ``environment'' composed of orbital conducting channels $|n \rangle$. Any entanglement to the environment is a  source of {\em spin decoherence}~\cite{Joos2003,Schlosshauer2008}. That is, the spin-density matrix obtained by tracing the full density matrix $|{\rm out} \rangle \langle {\rm out}|$ of the pure state $|{\rm out} \rangle$ over the orbital transverse propagating modes $|n \rangle$ in the right lead
\begin{equation}\label{eq:rho_out}
\hat{\rho}^{m \uparrow \rightarrow {\rm out}} = \frac{1}{Z}{\rm Tr}_{\rm orbital}  |{\rm out} \rangle \langle {\rm out}| = \frac{1}{Z} \sum_{n=1}^M \langle n|{\rm out} \rangle
\langle {\rm out}| n \rangle,
\end{equation}
will have, in general, the polarization  vector magnitude $|{\bf P}^{m \uparrow \rightarrow {\rm out}}| < 1$ reduced below one (characterizing  pure spin states) when transport takes place through multichannel wires~\cite{Nikoli'c2005}. Here $Z$ is the normalization factor ensuring that ${\rm Tr}_{\rm spin} \, \hat{\rho}^{m \uparrow \rightarrow {\rm out}}=1$.

Further decrease of the observable degree of quantum coherence encoded in the off-diagonal elements~\cite{Joos2003,Schlosshauer2008} of the spin-density matrix is generated by {\em spin dephasing}~\cite{Nikoli'c2005} due to {\em averaging} over all orbital incoming channels
\begin{equation}\label{eq:spin_dephasing}
\hat{\rho}^{\uparrow}_{\rm out} = \sum_m \hat{\rho}^{m \uparrow \rightarrow {\rm out}}.
\end{equation}
Note that this type of dephasing is equivalent to ``fake decoherence'' or ``ensemble dephasing'' discussed through examples in recent monographs on quantum decoherence Refs.~\cite{Joos2003,Schlosshauer2008}. The spin dephasing, whose meaning is defined {\em precisely} through Eq.~(\ref{eq:spin_dephasing}), can be effective in reducing the off-diagonal elements of $\hat{\rho}^{\uparrow}_{\rm out}$ even if every electron in the right lead continues to be in the orbital conducting channel through which it was originally injected, so that $|{\rm out} \rangle$ state emerges in the right lead as a separable quantum state and ``true decoherence''~\cite{Joos2003} is absent. This procedure finally leads to the spin-density matrix associated with the detected charge current in the right lead~\cite{Nikoli'c2005}
\begin{eqnarray} \label{eq:rho_c}
\hat{\rho}^{\uparrow}_{\rm out} & = &  \frac{e^2/h}{G^{\uparrow \uparrow}_{21} + G^{\downarrow \uparrow}_{21}} \! \sum_{n,m=1}^M \!\!\!
\left( \begin{array}{cc}
     |[{\bf t}_{2 1}^{\uparrow \uparrow}]_{nm}|^2 &  [{\bf t}_{2 1}^{\uparrow \uparrow}]_{nm}
       [{\bf t}_{21}^{\downarrow \uparrow}]_{nm}^*  \\

      [{\bf t}^{\uparrow \uparrow}_{21}]_{nm}^* [{\bf t}^{\downarrow \uparrow}_{21}]_{nm} &
      |[{\bf t}_{21}^{\downarrow\uparrow}]_{nm}|^2
  \end{array} \right) \nonumber \\
\displaystyle  &  = & \frac{1}{2} \left( {\bm 1} + {\bf P}_{\rm out} \cdot \hat{\bm \sigma} \right).
\end{eqnarray}
From it, one can also extract the experimentally measurable spin-polarization vector ${\bf P}_{\rm out}$.

Figure~\ref{fig:noise}(c) shows that in narrow wires quantum coherence of transported spin quantified by $|{\bf P}_{\rm out}|$ remains close to one for $L \lesssim L_{\rm SO}$. In wires of fixed length, suppression of spin decoherence in Fig.~\ref{fig:noise}(d) is governed by the wire width $W$ and the spin precession length $L_{\rm SO}$, which are also invoked as characteristic length scales to explain recent experiments~\cite{Holleitner2006}. Figure~\ref{fig:noise}(e), where the Fano factor value is directly related to $|{\bf P}_{\rm out}|$, demonstrates an exciting possibility for a novel experimental tool to quantify purity of transported spin 
state via {\em electrical} means where measurement of the Fano factor $F_{\uparrow \rightarrow \uparrow \downarrow}$ does not require demanding~\cite{Fabian2007} spin selective detection in the right lead. The preservation of spin coherence also allows for spin-interference signatures to become visible in the shot noise in Fig.~\ref{fig:noise}(a) as oscillations of the Fano factor between $F_{\sigma \rightarrow \sigma'}=1/3$ and $F_{\sigma \rightarrow \sigma'}=1$ along the $L_{\rm SO}$ spatial scale.

\subsection{Discussion} \label{sec:discussion}

The phenomenological model of Ref.~\cite{Lamacraft2004}, characterized by the spin-relaxation length
$L_{\rm S}$ (which in the bulk SO coupled systems with weak disorder is identical~\cite{Fabian2007,Mal'shukov2000,Kiselev2000,Pareek2002}
to the spin precession length $L_{\rm SO}$),  finds $F_{\uparrow \rightarrow \uparrow\downarrow}(L \gg L_{\rm S})=2/3$. This in contrast to our $F_{\uparrow \rightarrow
\uparrow\downarrow}(L \gg L_{\rm SO}) \simeq 0.55$ governed by the parameters of microscopic Rashba Hamiltonian where further modification of $F_{\uparrow \rightarrow
\uparrow\downarrow}(L \gg L_{\rm S}) < 0.55$ can be induced by geometrical confinement effects acting against spin decoherence and dephasing.

As regards the spin-valve setups, the semiclassical Boltzmann-Langevin approach~\cite{Blanter2000} applied to spin-dependent shot noise in Ref.~\cite{Mishchenko2003} predicts Fano factors $F_{\uparrow \rightarrow \uparrow}(L \gg L_{\rm S}) = F_{\uparrow \rightarrow \downarrow}(L \gg L_{\rm S}) = 1/3$ for arbitrary microscopic spin relaxation processes within the normal region, while we find $F_{\uparrow \rightarrow \uparrow}(L \gg L_{\rm SO}) = F_{\uparrow \rightarrow \downarrow}(L \gg L_{\rm SO}) \simeq 0.7$ for  specific case of wide Rashba SO coupled wires. Furthermore, oscillatory behavior of the Fano factor versus $L/L_{\rm SO}$ exhibited in our Fig.~\ref{fig:noise}, especially conspicuous when
quantum coherence of (partially coherent $0<|{\bf P}_{\rm out}| < 1$) spin is increased in narrow wires, cannot emerge
from the approach of Ref.~\cite{Mishchenko2003} where spin dynamics is characterized only by $L_{\rm S}$
(being much larger than mean free path with no restrictions imposed on its relation to the system size), rather
than by the full spin-density matrix.

\begin{figure}
\centerline{\psfig{file=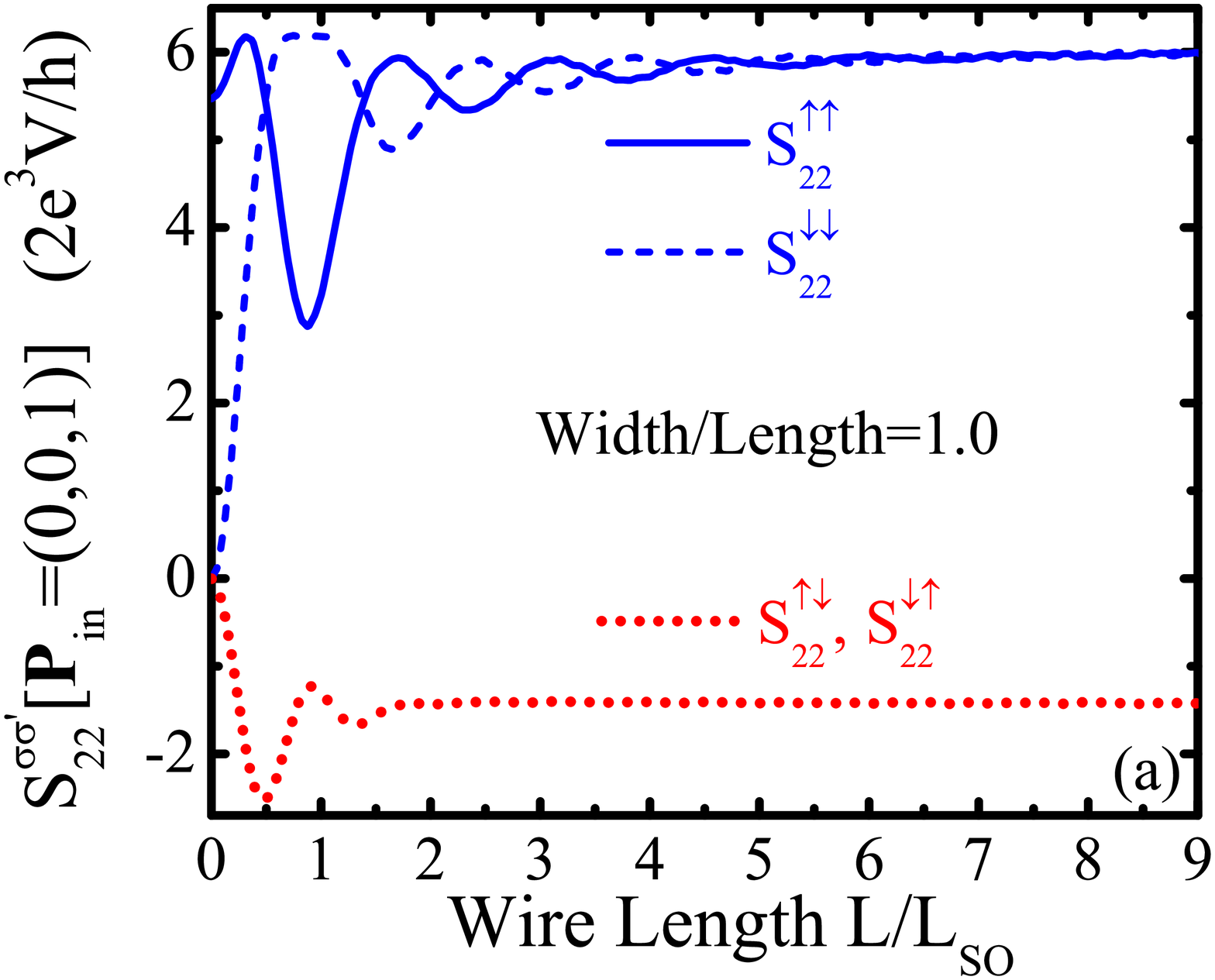,scale=0.2,angle=0} \hspace{0.5in} \psfig{file=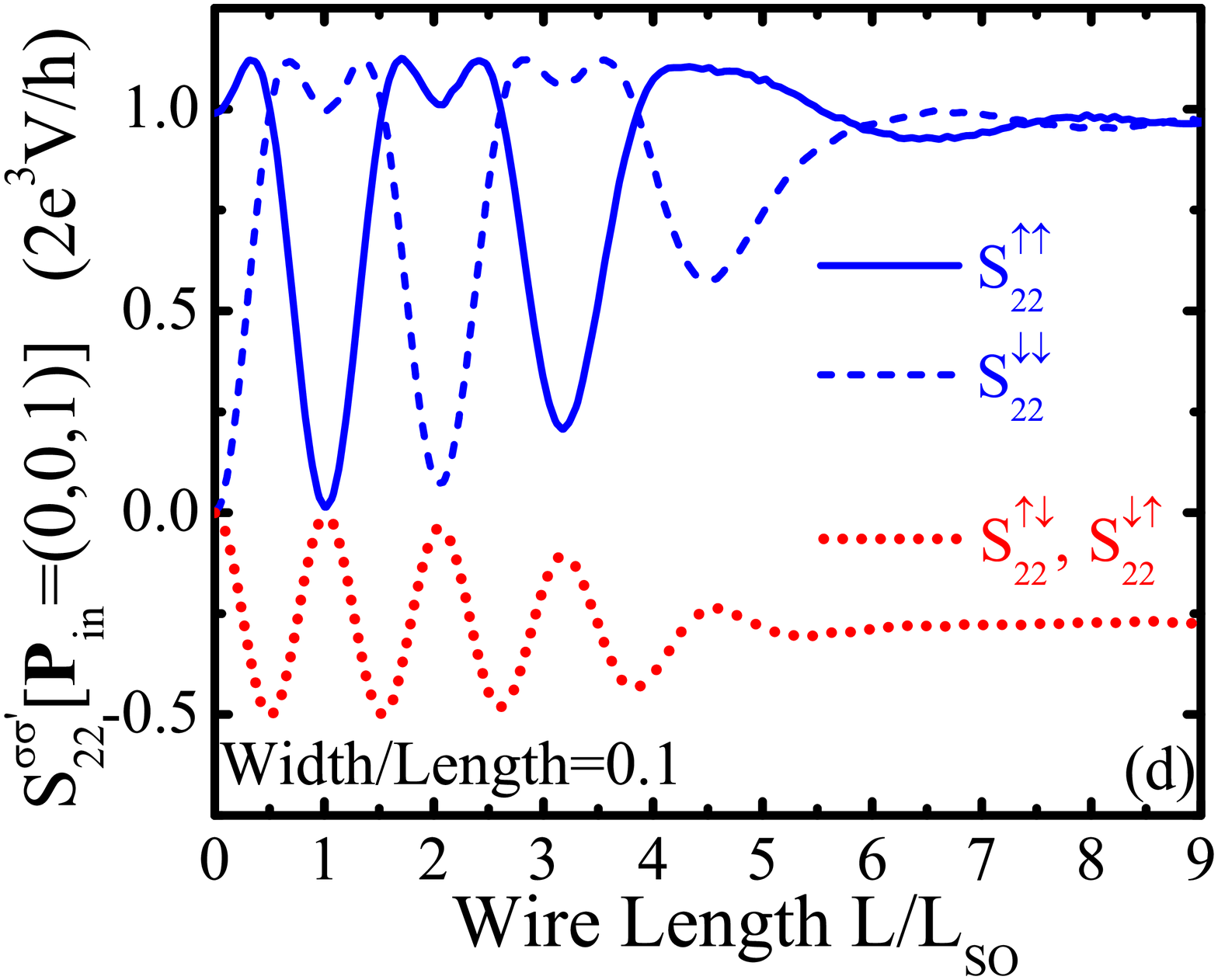,scale=0.2,angle=0}}
\centerline{\psfig{file=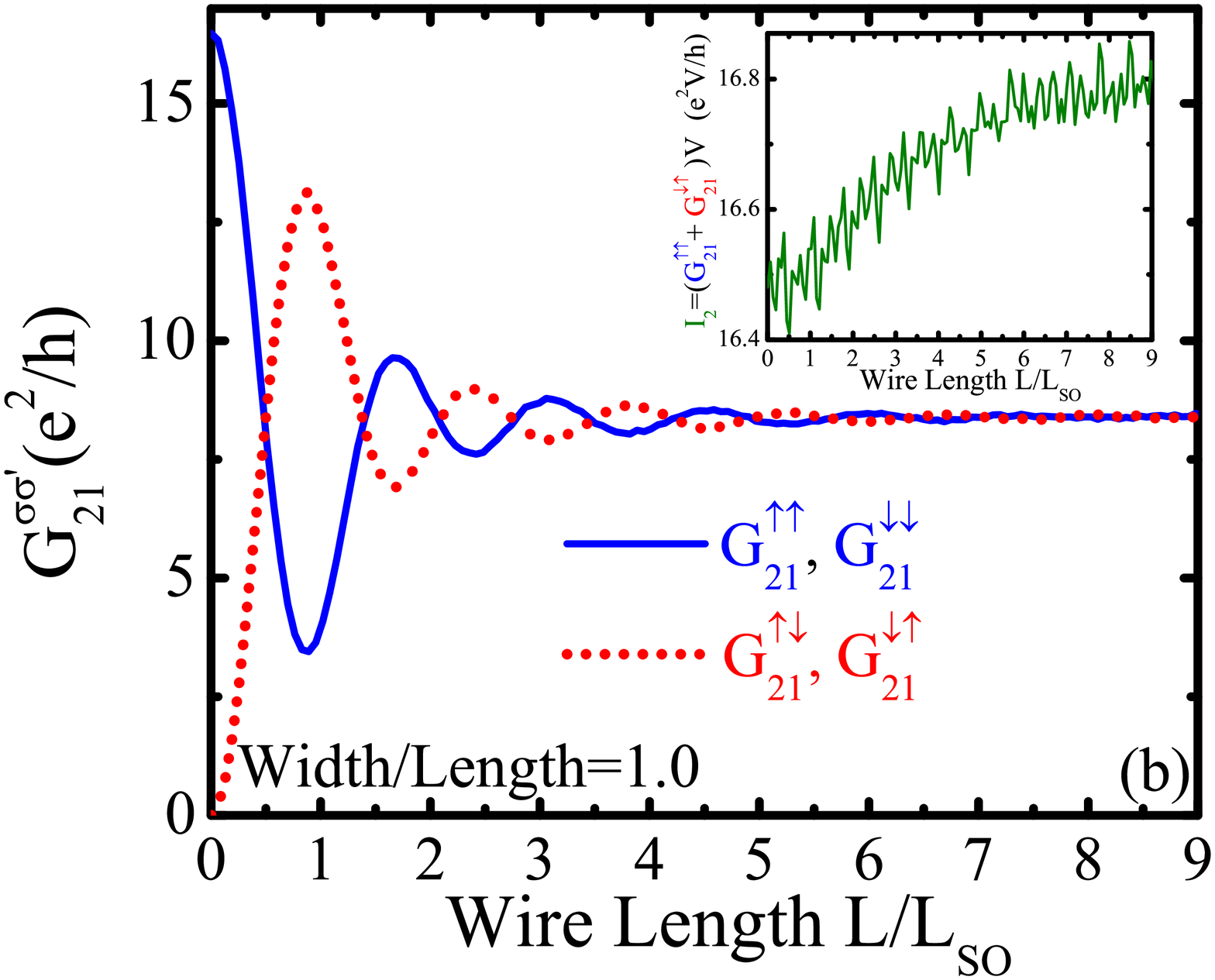,scale=0.2,angle=0} \hspace{0.5in} \psfig{file=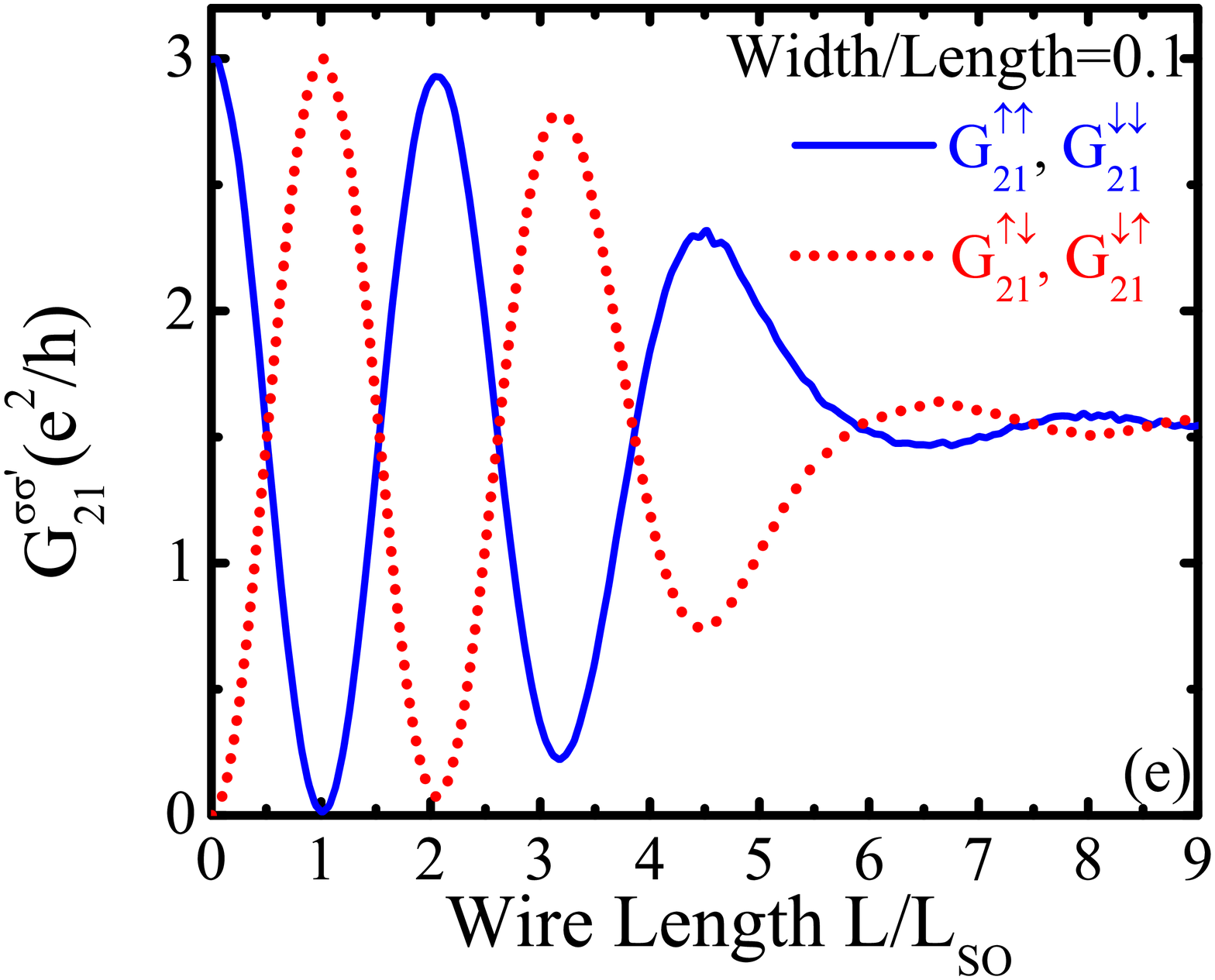,scale=0.2,angle=0}}
\centerline{\psfig{file=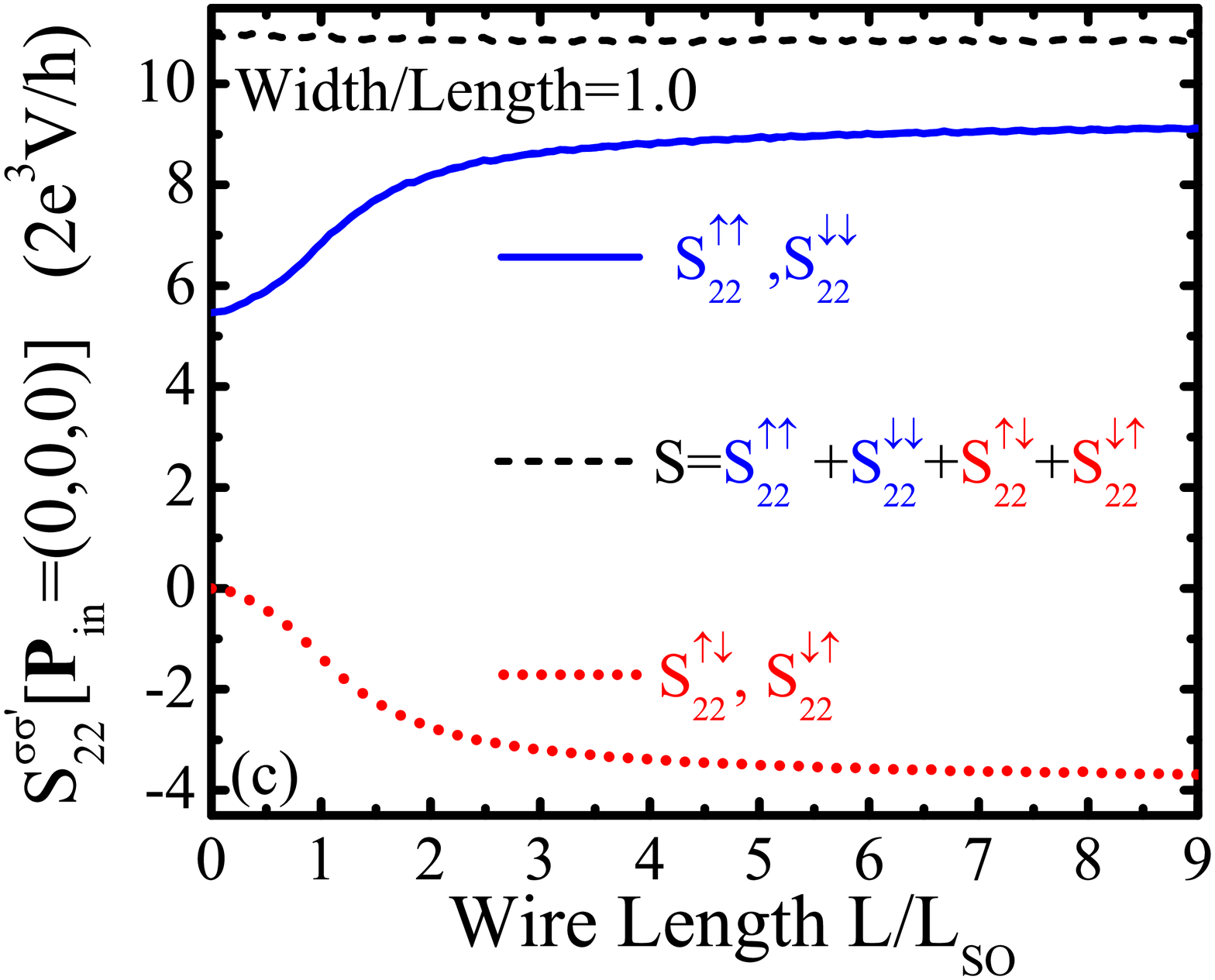,scale=0.2,angle=0} \hspace{0.5in} \psfig{file=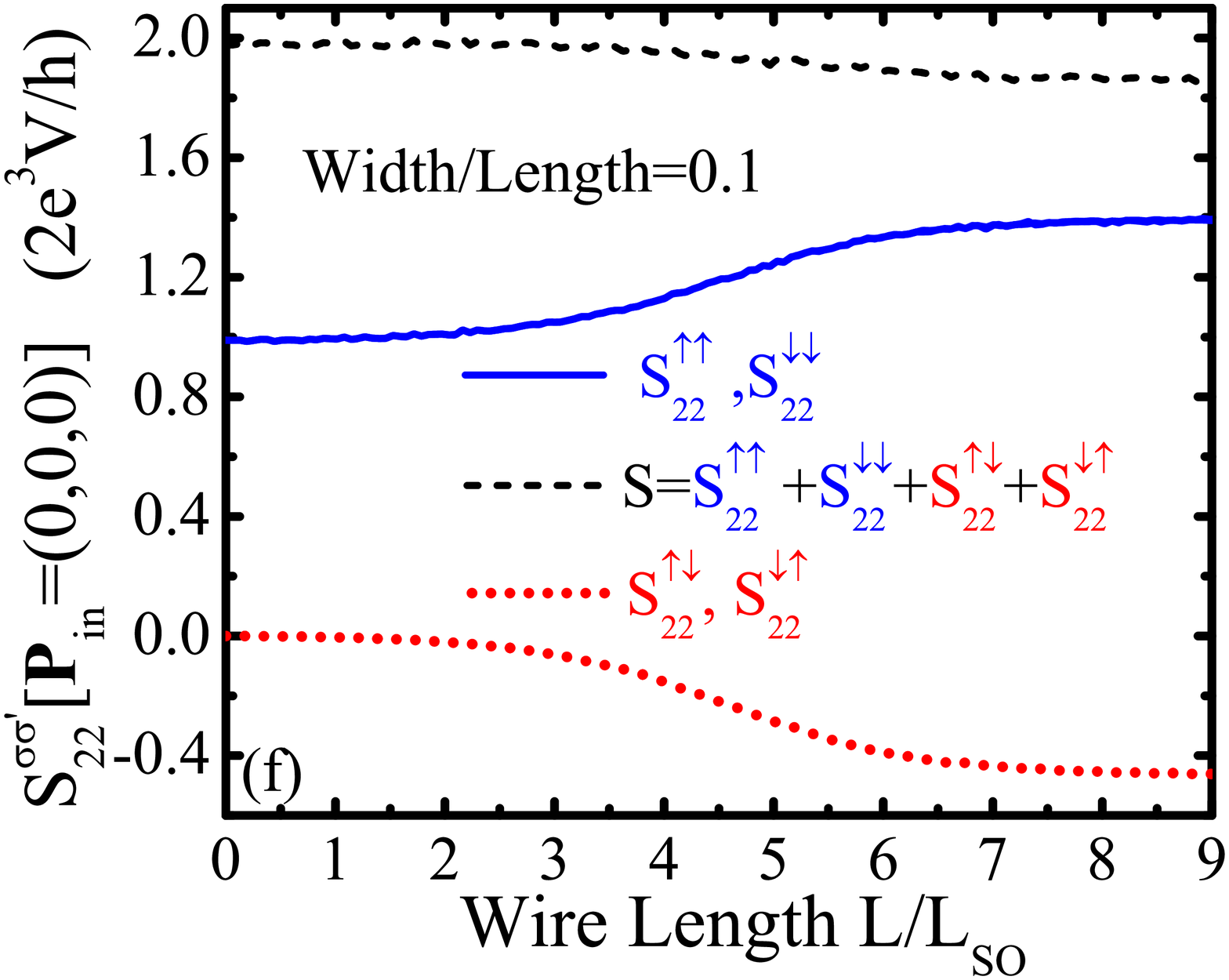,scale=0.2,angle=0}}
\caption{Zero-frequency spin-resolved shot noise power $S_{22}^{\sigma \sigma^\prime}$ [panels (a) and (d)] and spin-resolved conductances $G_{21}^{\sigma \sigma^\prime}$
[panels (b) and (e)], which define different Fano factors in Fig.~\ref{fig:noise}, for current detected in the right lead after the injection of spin-polarized (along the $z$-axis) charge current from the left lead into the diffusive wire with the Rashba SO coupling of strength $L/L_{\rm SO}$. The quantum wire is wide in panels (a)--(c) and narrow in panels (d)--(f). The inset in panel (b) shows weak antilocalization enhanced detected current in the right lead $I_2=I_2^\uparrow + I_2^\downarrow$ of a ferromagnet$|$SO-coupled-wire$|$paramagnet setup. The spin-resolved shot noise for unpolarized current injection is 
shown in panels (c) and (f), whose sums give limiting curves (for $|{\bf P}_{\rm in}|=0$) on the surface plots in Fig.~\ref{fig:3d}. Adapted from
Ref.~\cite{Dragomirova2007}.}\label{fig:spin_resolved}
\end{figure}

To elucidate the source of these apparent discrepancies, we provide in Fig.~\ref{fig:spin_resolved}
detailed picture of auto- and cross-correlations between spin-resolved charge currents. In addition, the same Fig.~\ref{fig:spin_resolved} plots the spin-resolved conductances. It is obvious that oscillations of both $S_{22}^{\sigma \sigma^\prime}$ and  $G_{21}^{\sigma \sigma^\prime}$ due to partially coherent spin
precession, visible as long as $|{\bf P}_{\rm out}| > 0$, can be captured only through fully quantum treatment of both spin dynamics and charge propagation (where spin memory
between successive scattering events is taken account~\cite{Pareek2002}). The asymptotic value $F_{\uparrow \rightarrow \uparrow}(L \gg L_{\rm SO})$ is determined by the shot
noise $S_{22}^{\uparrow \uparrow}$ that
is similar in both $L \ll L_{\rm SO}$ and $L \gg L_{\rm SO}$ limits, as well as by the value of charge current $I_2^\uparrow = G_{21}^{\uparrow\uparrow}V$ in the limit $L \gg
L_{\rm SO}$ where it becomes half of $I_2^\uparrow$
for vanishing SO coupling $L/L_{\rm SO} \rightarrow 0$ [Figs.~\ref{fig:spin_resolved}(b) and ~\ref{fig:spin_resolved}(e)]. This is due to the fact
that at the exit of the normal region with $L/L_{\rm SO} \gg 1$ charge current is unpolarized, so that one of its spin subsystems is completely reflected from the detecting
spin-selective (``analyzer'') electrode.

Figure~\ref{fig:spin_resolved}(a) also reveals that unpolarized charge current flowing out of the Rashba SO coupled
region, after injected fully spin-polarized current was completely dephased $|{\bf P}_{\rm out}|=0$ along
the Rashba wire, still displays non-trivial cross-correlations between spin-resolved currents encoded in $S_{22}^{\uparrow \downarrow} = S_{22}^{\downarrow \uparrow} \neq 0$.
They reduce $S_{22}^{\uparrow \downarrow} = S_{22}^{\downarrow \uparrow} < 0$ our Fano factor $F_{\uparrow \rightarrow \uparrow\downarrow}(L \gg L_{\rm S})= 0.55$  below $F_{\uparrow
\rightarrow \uparrow\downarrow}(L \gg L_{\rm S})=2/3$ of Ref.~\cite{Lamacraft2004} (which we recover  approximately if we characterize the shot noise in the right lead only with
$S_{22}^{\uparrow \uparrow} + S_{22}^{\downarrow \downarrow}$).

\section{Spin and charge shot noise in mesoscopic spin Hall systems}\label{sec:she_noise}

One of the principal outcomes of the analysis of spin-dependent shot noise for two-terminal nanostructures with the Rashba SO coupling in Sec.~\ref{sec:two_terminal_rashba_wire} is understanding
of how spin precession and  spin decoherence can increase the
Fano factor of the shot noise (above its value in the absence of SO
coupling) for injected current that is {\em spin-polarized}.
The analysis  of the same effects in multiterminal devices is more complicated~\cite{Sukhorukov1999} due to non-local effects where other leads contribute to the noise in a selected lead. Therefore, straightforward conclusions  about  the absence of the shot noise enhancement in the case of unpolarized current injection in two-terminal devices, found in Figs.~\ref{fig:3d} and ~\ref{fig:spin_resolved}, cannot be  extended to four-terminal devices that serve as generators of mesoscopic SHE when unpolarized charge current is injected into them. Instead, we proceed in this Section to analyze  shot noise of transverse spin Hall transport for both unpolarized and spin-polarized longitudinal charge current injection.

\subsection{Multiterminal spin Hall and charge current shot noise in ballistic 2DEG nanostructures}

\begin{figure}
\centerline{\psfig{file=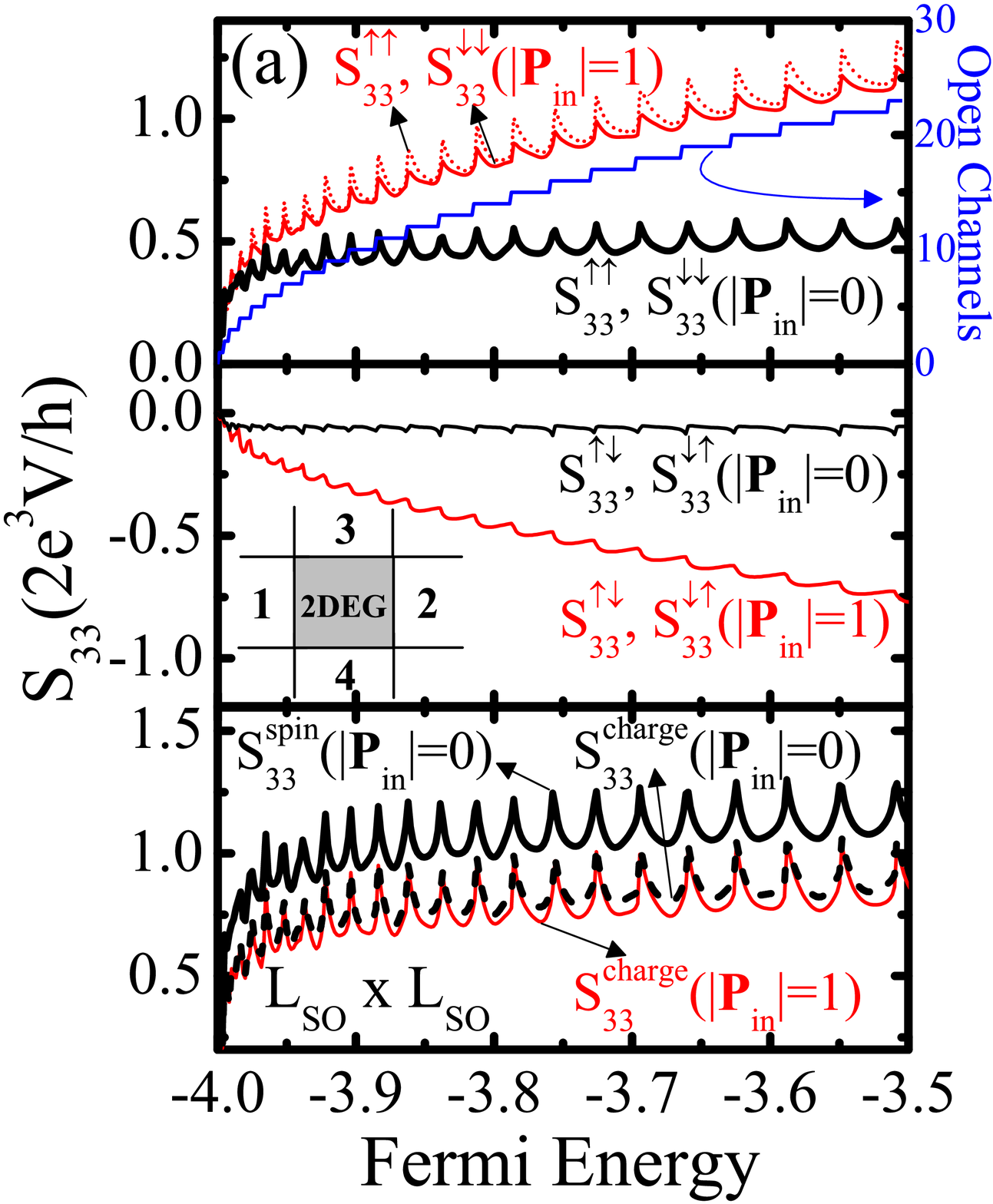,scale=0.28,angle=0} \hspace{0.5in} \psfig{file=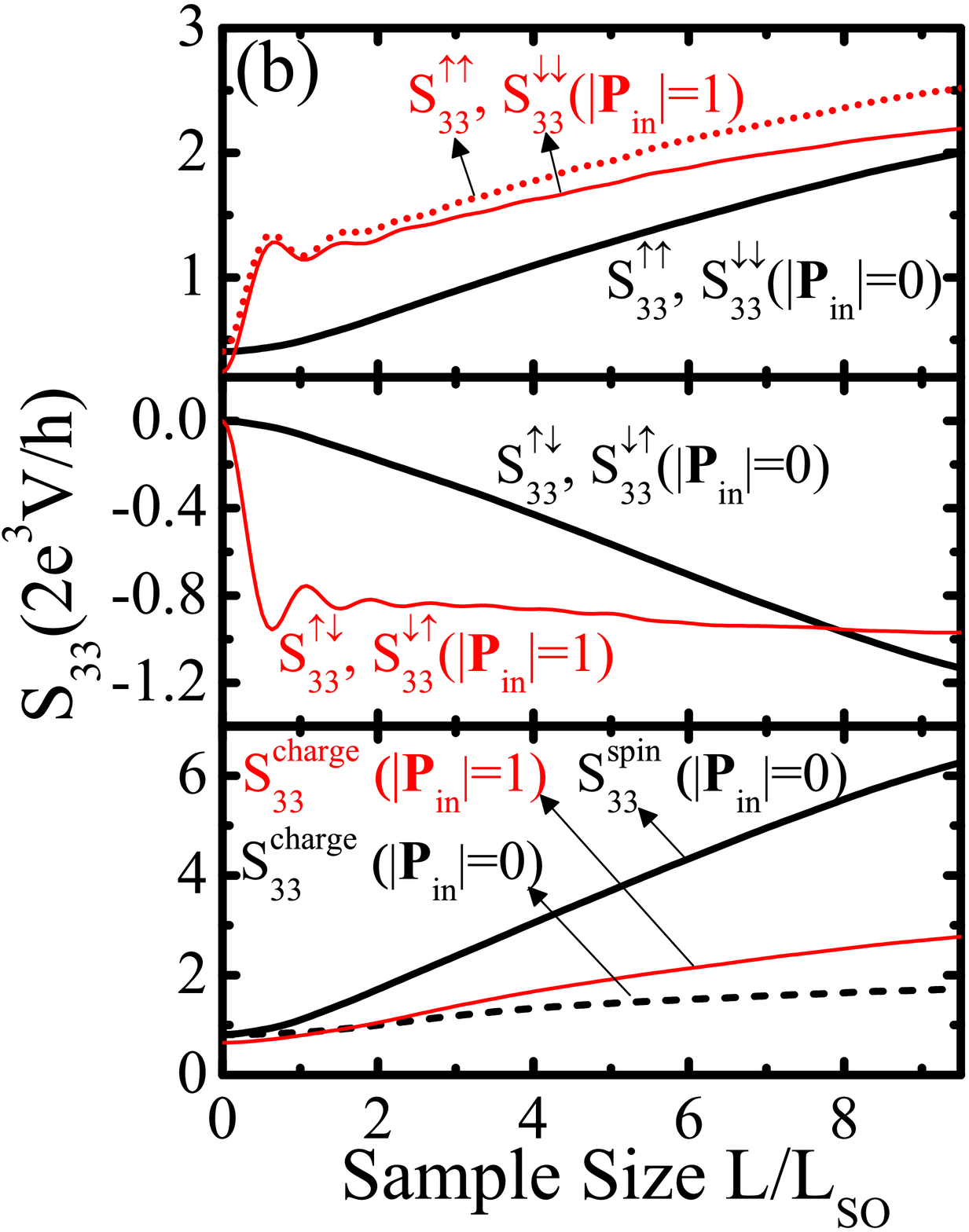,scale=0.28,angle=0}}
\caption{The ballistic spin-resolved shot noise  in the transverse electrode 3 [see Fig.~\ref{fig:she_inverse} and inset in panel (a) for electrode labeling], as well as the total shot noise of pure spin Hall current [driven by unpolarized $|{\bf P}_{\rm in}|=0$ injected charge current $I_1$] or charge Hall current [driven by spin-polarized ${\bf P}_{\rm in}=(0,0,1)$ injected $I_1$], in {\em clean} 2DEGs with the Rashba SO coupling as a function of: (a) Fermi energy $E_F$; or (b) Rashba SO coupling $\alpha_R$ measured through the spin precession length $L_{\rm SO}=\pi \hbar^2/2m^* \alpha_R$. In panel (a), the 2DEG sample is of the size $L_{\rm SO} \times L_{\rm SO}$, while in panel (b) the sample size is $300 \, {\rm nm} \times 300 \, {\rm nm}$ and $E_F$ is fixed to open 23 channels for electron injection from lead 1. Adapted from Ref.~\cite{Dragomirova2008}.}\label{fig:ballistic}
\end{figure}

In this Section and related Fig.~\ref{fig:ballistic} we assume ballistic transport [$V_{\rm dis}(x,y)=0$ or $\varepsilon_{\bf m}=0$] through 2DEG with non-zero  $L_{\rm SO}$
due to the Rashba coupling. We recall that in two-terminal ballistic structure the stream of electrons (injected from noiseless electrodes) is completely correlated by the
Pauli principle in the absence of impurity backscattering, so that the corresponding shot noise vanishes $S=0$ (except at the subband edges where new conducting channels open up)~\cite{Blanter2000}. However, in four-terminal structures in Fig.~\ref{fig:ballistic} transmission is not perfect because of the presence of the transverse leads (even if
they do not draw current~\cite{Sukhorukov1999}), so that non-zero noise appears in the absence of SO coupling. While large Rashba coupling would introduce
backscattering~\cite{Nikoli'c2005b,Sheng2006b} at the interface between the electrodes with no SO coupling and the sample, we find this effect not to be the crucial one for
noise discussion below since similar results are obtained for the bridge in Fig.~\ref{fig:she_inverse}(a) where leads 1 and 2 have the same Rashba SO coupling as in the central 2DEG sample.

The magnitude of pure spin currents flowing out of mesoscopic SHE device through ideal (with no SO coupling) electrodes is governed by the spin precession length $L_{\rm SO}$ in Eqs.~(\ref{eq:lso}) and (\ref{eq:lso_lattice}). This mesoscopic length scale (e.g., $L_{\rm SO} \sim 100$ nm in typical 2DEG experiments~\cite{Nitta1997,Grundler2000}) has been identified through intuitive physical arguments~\cite{Engel2007a} as an important parameter for spin distributions---for example, in clean systems the spin response to inhomogeneous field diverges at the wave vector $q=2/L_{\rm SO}$. In fact, the mesoscopic SHE analysis predicts~\cite{Nikoli'c2005b} via {\em numerically exact} calculations  that the optimal device size for achieving large spin polarizations and spin currents is indeed $L \simeq L_{\rm SO}$, as demonstrated  by Fig.~\ref{fig:gshe}(a). This is further confirmed by an alternative analysis of the SHE response  in disordered finite-size 2DEGs in Ref.~\cite{Moca2007}.  Therefore, we employ the 2DEG sample of the size $L_{\rm SO} \times L_{\rm SO}$ to study the dependence of the shot noise on the Fermi energy (i.e., charge density). We also assume that 2DEG is smaller than the inelastic scattering length $L_{\rm e-ph}$ because  in larger samples electron-phonon scattering would average out the ``mesoscopic'' values~\cite{Naveh1998} of the shot noise to zero~\cite{Steinbach1996}, as discussed in Sec.~\ref{sec:intro}.

The most conspicuous feature of the spin-resolved shot noise in Fig.~\ref{fig:ballistic} is the emergence of highly non-trivial temporal correlations between spin-resolved currents encoded by $S_{33}^{\uparrow \downarrow}=S_{33}^{\downarrow \uparrow}<0$ [more pronounced for polarized ${\bf P}_{\rm in}=(0,0,1)$ injection]. This stems from spin flips in the form of continuous spin precession of the $z$-axis oriented spins in the effective momentum-dependent magnetic field ${\bf B}_{\rm int}({\bf p})$ of the Rashba SO coupling. Such cross-correlations can be manipulated by changing the Fermi energy in the case of polarized injection ($|{\bf P}_{\rm in}|=1)$ or Rashba coupling in the case of unpolarized longitudinal current ($|{\bf P}_{\rm in}|=0$), thereby imprinting signatures of the intrinsic SO coupling on  experimentally measurable charge current noise $S_{33}^{\rm charge}$.

Another feature specific to mesoscopic manifestations of SHE, which is also exhibited by the SHE conductance $G_{\rm sH}^z=I_3^{S_z}/(V_1-V_2)$~\cite{Nikoli'c2005b,Sheng2006b}, is the appearance of sharp peaks in Fig.~\ref{fig:ballistic}(a) in the vicinity of subband edges. At these energies new conducting channels in the leads become available for transport [top panel of Fig.~\ref{fig:ballistic}(a)]. Although this multiterminal noise property of ballistic conductors persists even in the absence of SO coupling, additional features of this type can arise at the energies of bound states in the cross device geometry whose mixing with propagating states via SO coupling introduces
resonances in the transmission~\cite{Bulgakov1999}.

We emphasize~\cite{Nikoli'c2005b,Nikoli'c2007} that achieving pure ($I_3=I_4=0$) spin Hall current $I_3^{S_z}=-I_4^{S_z}$, akin to SHE in
infinite systems~\cite{Sinova2004,Guo2008}, requires to apply~\cite{Nikoli'c2005b,Erlingsson2005} voltages $\mu_3=\mu_4=0$ to transverse leads
of the clean bridge biased with $\mu_1=eV/2$ and $\mu_2=-eV/2$. Despite zero charge current $I_3=0$ in this case, we find non-zero fluctuations around
the zero average value (found also in some other pure spin current induction setups~\cite{Wang2004b}), as quantified by $S_{33}^{\rm charge}(|{\bf P}_{\rm in}|=1)$ in Fig.~\ref{fig:ballistic}. The noise power increases in the same setup, at fixed $E_F$ and with fast spin dynamics in samples $L/L_{\rm SO} \gtrsim 2$,  by switching from unpolarized to polarized
injection of longitudinal current $I_1$ responsible for non-zero transverse charge Hall current~\cite{Bulgakov1999}.

\subsection{Multiterminal spin Hall and charge current shot noise in diffusive 2DEG nanostructures}

\begin{figure}
\centerline{\psfig{file=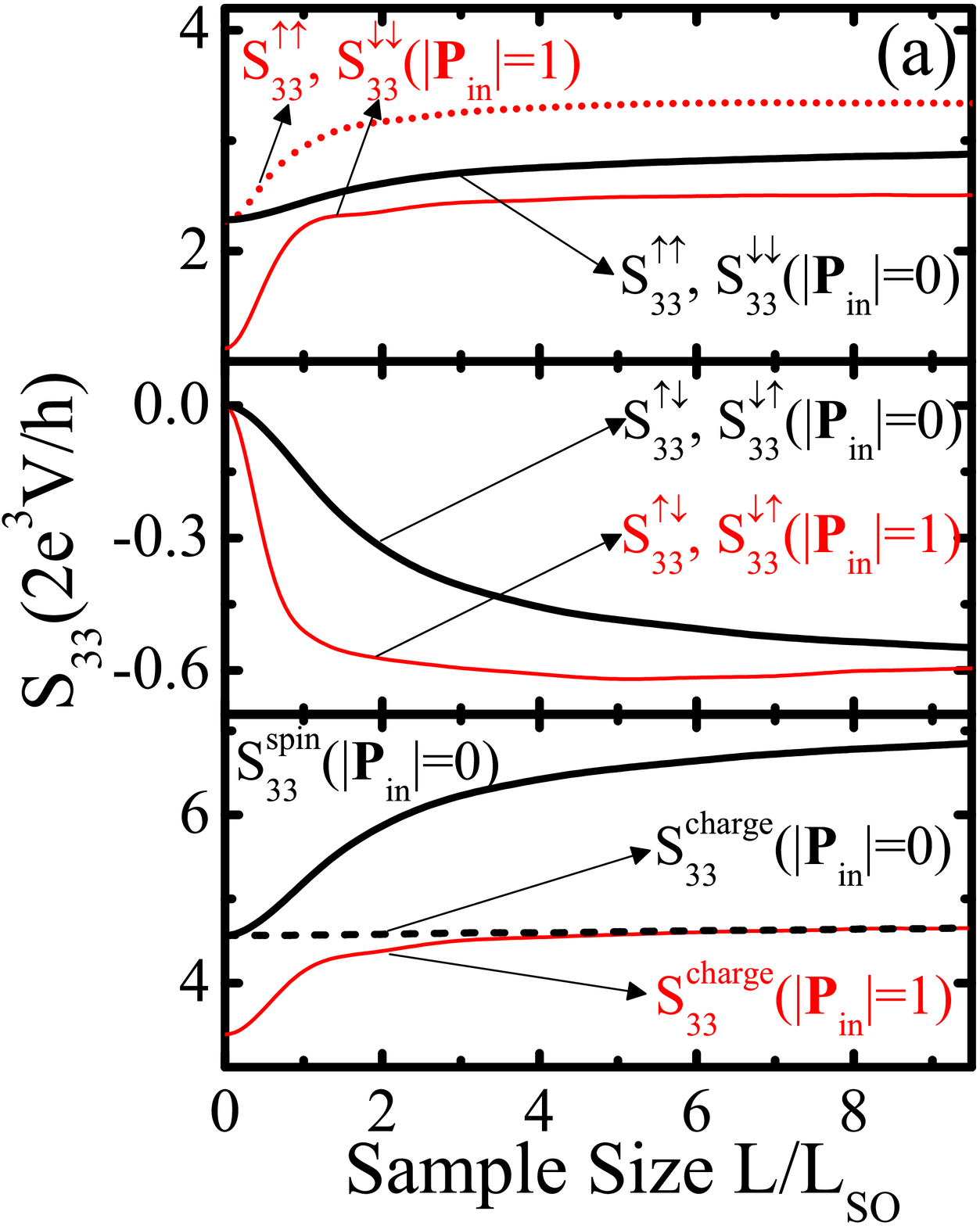,scale=0.24,angle=0} \psfig{file=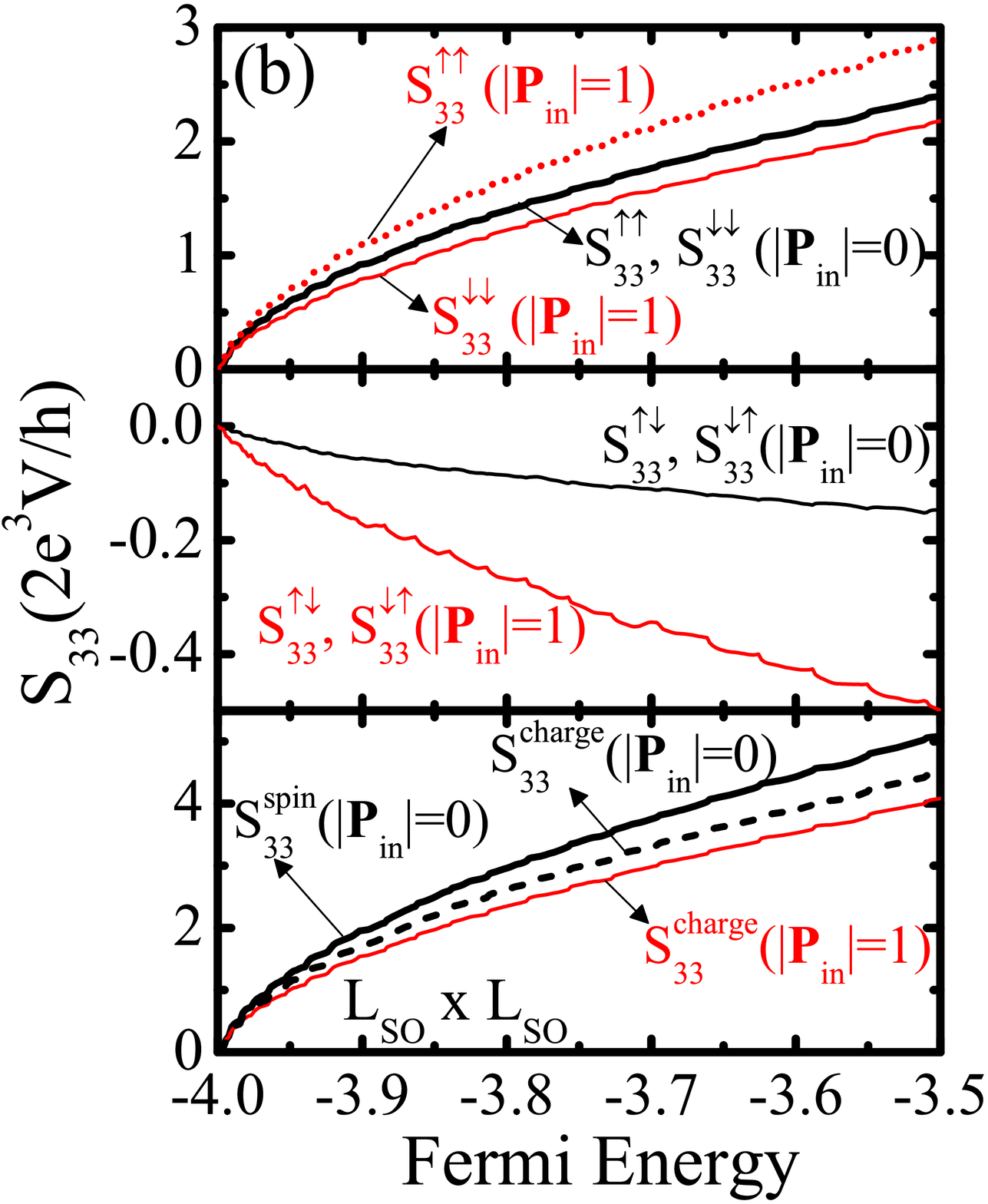,scale=0.24,angle=0} \psfig{file=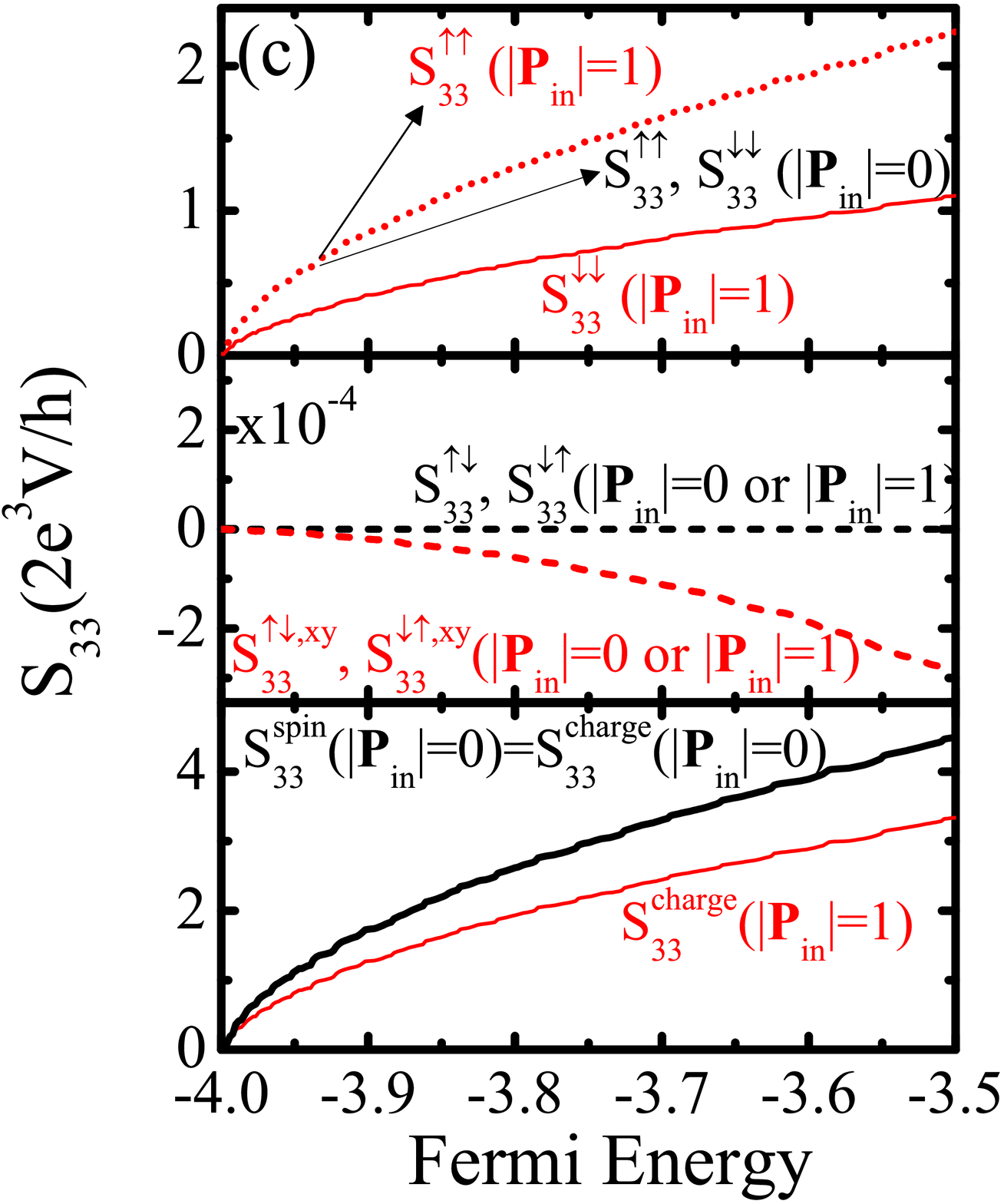,scale=0.24,angle=0}}
\caption{The diffusive spin-resolved shot noise  in the transverse electrode 3, as well as the total shot noise of pure spin Hall current [for unpolarized $|{\bf P}_{\rm in}|=0$ injection of $I_1$] or charge Hall current [for spin-polarized ${\bf P}_{\rm in}=(0,0,1)$ injection of $I_1$], in {\em disordered} four-terminal 2DEGs with the Rashba SO coupling [panels (a) and (b)] or extrinsic SO scattering of strength $\lambda/\hbar =5.3$ \AA{}$^2$ [panel (c)]. In panel (b) the 2DEG sample is of the size $L_{\rm SO} \times L_{\rm SO}$, while in panels (a) and (c) the sample size is fixed at $300 \, {\rm nm} \times 300 \, {\rm nm}$. The Fermi energy in panel (a) is set to allow for 23 open conducting channels [see Fig.~\ref{fig:ballistic}(a)]. Adapted from Ref.~\cite{Dragomirova2008}.}\label{fig:diffusive}
\end{figure}

To bring a multiterminal SHE bridge into the diffusive transport regime, we introduce disorder into the 2DEG through the on-site potential $\varepsilon_{\bf m} \in [-W_{\rm dis}/2,W_{\rm dis}/2]$ in Hamiltonian (\ref{eq:tbh}) and tune its strength $W_{\rm dis} = 1.1 t_{\rm O}$ (mean free path $\ell \approx 29 a$) to {\em ensure} that the shot noise in lead 1 attains the universal value $F_{11}=S_{11}^{\rm charge}/2eI_1=1/3$ characterizing diffusion in multiterminal devices~\cite{Sukhorukov1999}. In the absence of the SO coupling, the noise in the other three leads does not display any universal features ($F_{11}=1/3$ is expected to be independent of the impurity distribution, band structure, and shape of the conductor~\cite{Sukhorukov1999}) because of nonlocal effects---that is, other leads contribute to the noise in electrode $\alpha \neq 1$, thereby making possible arbitrarily large values of $F_{\alpha \alpha}$~\cite{Sukhorukov1999}.

In the presence of disorder, one can expect both extrinsic and intrinsic contributions to $I_3^{S_z}$. Their importance (as in the case of experimentally explored SHE
systems based on  2DEGs~\cite{Sih2005a}) is governed~\cite{Nikoli'c2007} by the ratio of the characteristic energy or length scales, $\Delta_{\rm SO} \tau/\hbar= \pi \ell/L_{\rm SO}$. For simplicity, we analyze separately 2DEGs with dominant intrinsic [$\alpha_R \neq 0$, $\lambda=0$ in Figs.~\ref{fig:diffusive}(a) and ~\ref{fig:diffusive}(b)] and extrinsic [$\alpha_R=0$, $\lambda/\hbar =5.3$ \AA{}$^2$ in Fig.~\ref{fig:diffusive}(c)] regimes of SHE. The most important insight brought about by Fig.~\ref{fig:diffusive} is substantial difference between the shot noise in the intrinsic [Fig.~\ref{fig:diffusive}(a) and ~\ref{fig:diffusive}(b)] and extrinsic [Fig.~\ref{fig:diffusive}(c)] regimes, where the former exhibits non-zero cross-correlations $S_{33}^{\uparrow \downarrow}=S_{33}^{\downarrow \uparrow}<0$ akin to its ballistic counterpart in Fig.~\ref{fig:ballistic}, but smaller. The shot noise of the extrinsic SHE device in Fig.~\ref{fig:diffusive}(c) has no temporal correlations of this type for the $z$-axis spins, while exhibiting orders of magnitude smaller cross-correlation noise, $S_{33}^{\uparrow \downarrow,x}$ and  $S_{33}^{\uparrow \downarrow,y}$ for the $x$- or $y$-spins, respectively, which carry no spin current $I_3^{S_x}=I_3^{S_y} \equiv 0$ when $\lambda \neq 0$. We also find that hypothetical (i.e., experimentally not accessible) increase of $\lambda$ would give orders of magnitude smaller noise change (in fact, decrease) compared to significant spin  $S_{33}^{\rm spin}(|{\bf P}_{\rm in}|=0)$ or charge $S_{33}^{\rm charge}(|{\bf P}_{\rm in}|=1)$ shot noise enhancement with increasing of the intrinsic $\alpha_R$ that can be experimentally controlled~\cite{Nitta1997,Grundler2000}.

\begin{figure}
\centerline{\psfig{file=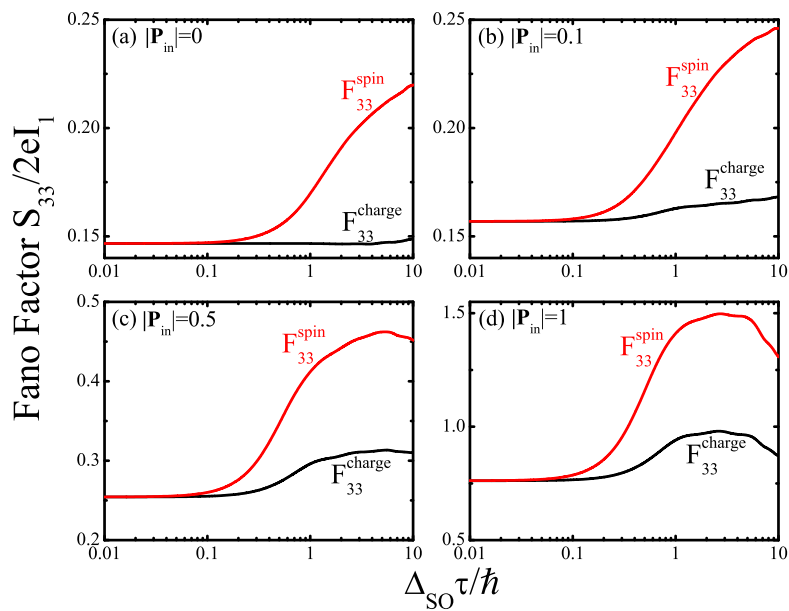,scale=0.5,angle=0}}
\caption{Fano factors $F_{33}^{\rm charge}=S_{33}^{\rm charge}/2eI_1$ and $F_{33}^{\rm spin}=S_{33}^{\rm spin}/2eI_1$ of the total spin and charge shot noise in the transverse electrode 3 of a four-terminal 2DEG bridge with variable Rashba SO coupling setting the ratio $\Delta_{\rm SO} \tau/\hbar=\pi \ell/L_{\rm SO}$ of the spin precession length and the mean free path. The sample size is fixed at $300 \, {\rm nm} \times 300 \, {\rm nm}$ and the mean free path is tuned to $\ell \simeq 86$ nm to ensure diffusive transport characterized~\cite{Sukhorukov1999} by the Fano factor $F_{11}^{\rm charge}=S_{11}^{\rm charge}/2eI_1=1/3$ in lead 1. The Fermi energy is selected to allow for injection of unpolarized (a) or polarized (b)--(d) charge current from lead 1 through 23 open conducting channels.}\label{fig:fano_she}
\end{figure}

Note that due to $I_3=0$ in the SHE setup ($|{\bf P}_{\rm in}|=0$), we plot raw noise values in Figs.~\ref{fig:ballistic} and ~\ref{fig:diffusive}, rather than normalizing them to $2eI_3$ or $2eI_3^{S_z}$ to get conventionally defined Fano factors. A useful Fano factor can actually be defined if we normalize noise to the injected current in lead 1:
\begin{equation}\label{eq:fanoshe}
F_{33}^{\rm charge}=\frac{S_{33}^{\rm charge}}{2eI_1},  \ F_{33}^{\rm spin}=\frac{S_{33}^{\rm spin}}{2eI_1}.
\end{equation}
We plot such Fano factors in Fig.~\ref{fig:fano_she} for zero charge current $I_3$ when injected current $I_1$ is unpolarized ($|{\bf P}_{\rm in}|=0$), as well as for non-zero charge Hall currents generated when partially ($|{\bf P}_{\rm in}|=0.1$ and $|{\bf P}_{\rm in}|=0.5$) or fully polarized  ($|{\bf P}_{\rm in}|=1$) current is injected through lead 1. Their dependence on the ratio of characteristic energy or length scales demonstrates that around $\Delta_{\rm SO} \tau/\hbar = \pi \ell/L_{\rm SO} \sim 10^{-1}$, for which the intrinsic Rashba SO coupling becomes dominant SHE mechanisms in disordered 2DEGs in Fig.~\ref{fig:gshe}(b), both the spin and the charge Fano factors {\em start to increase} above their reference values set in the limit of zero SO coupling ($L_{\rm SO} \rightarrow \infty$). Thus, Fig.~\ref{fig:fano_she} as one of our principal results in Sec.~\ref{sec:she_noise}, suggests that by measuring the enhanced charge Fano factor in transverse lead 3 at given polarization of the injected current one could confirm in an {\em unambiguous} fashion the dominance of the intrinsic SO mechanisms in the induction of (much more difficult to measure) spin Hall current in the same lead.

The Fano factor reference value in Fig.~\ref{fig:fano_she}, defined at negligible SO coupling strength $\pi \ell/L_{\rm SO} \rightarrow 0$, depends on $|{\bf P}_{\rm in}|$ even in this limit. At first sight, this might seem surprising since this feature is absent in Fig.~\ref{fig:3d} for two-terminal devices. In fact, this is a consequence of the four-terminal device geometry where noise power in a given lead depends on contributions from all other leads. For example, one can view four-terminal device as an eight-terminal one with fully polarized electrodes, so that injection of spin-polarized current through lead 1 corresponds to one of these eight leads being totally (for $|{\bf P}_{\rm in}|=1$) or partially blocked (for $0<|{\bf P}_{\rm in}| < 1$). This decreases both the noise power in lead 3 (as shown in Fig.~\ref{fig:diffusive}) and the current $I_1$ normalizing the noise to define the Fano factor.

\section{Quantum interference effects on the shot noise of spin Hall and charge currents in four-terminal Aharonov-Casher rings}\label{sec:qidshe}

\begin{figure}
\centerline{\psfig{file=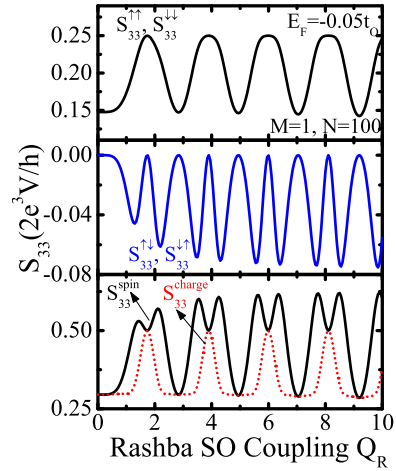,scale=0.24,angle=0}}
\caption{The ballistic spin-resolved shot noise in the transverse electrode 3 of a four-terminal 1D Aharonov-Casher ring (top and middle panels), as well as the total shot noise of transverse pure spin Hall current $I_3^{S_z}$ and zero charge current $I_3=0$ (bottom panel), driven by unpolarized $|{\bf P}_{\rm in}|=0$ injected charge current $I_1$. The spin conductance of the QIDSHE as the function of the Rashba SO coupling within the 1D ring is shown in Fig.~\ref{fig:qidshe} for the same device parameters ($E_F$ is the Fermi energy of injected electrons through single $M=1$ channel leads attached to 1D ring discretized using $N=100$ sites).}\label{fig:qidshe_1dnoise}
\end{figure}

The stationary states of the system one-dimensional (1D) ring + two 1D leads can be found exactly
by matching the wave functions in the leads to the eigenstates of the clean ($V_{\rm dis}=0$ and $\lambda=0$)
ring Hamiltonian Eq.~(\ref{eq:hamil}), and then computing the charge conductance from the Landauer transmission
formula~\cite{Molnar2004}. However, attaching two extra leads in the transverse direction, as well the finite
width of the ring and/or presence of disorder within the ring region, requires to switch from wave functions to
NEGF formalism discussed in Sec.~\ref{sec:negf} in order to compute numerically exact transmission  matrices  ${\bf t}_{\alpha \beta}^{\sigma \sigma^\prime}$
connecting the four leads $\alpha,\beta=1,\ldots,4$. The computation of the retarded Green function matrix
${\bf G}^r$ in Eq.~(\ref{eq:retarded}) can  be done efficiently using the lattice-type Hamiltonian akin to
Eq.~(\ref{eq:tbh}) (assuming $\lambda_{\rm SO}=0$), which was introduced in  Ref.~\cite{Souma2004} as a set of
$M$ concentric chains composed of $N$ lattice sites spaced at a distance $a$. Besides the usual energy scales introduced by
such Hamiltonian---the orbital hopping $t_{\rm O}$ and the Rashba hopping $t_{\rm SO}$ between nearest neighbor sites discussed
in Sec.~\ref{sec:negf}---it is advantageous to employ a dimensionless parameter $Q_{R} \equiv (t_{\rm SO}/t_{\rm O})N/\pi$ to measure the strength of the SO coupling
within the ring region~\cite{Souma2004,Frustaglia2004a}.

The charge conductance of a two-terminal 1D AC ring~\cite{Souma2004,Molnar2004,Frustaglia2004a} becomes zero at specific values  of $Q_{R}^{\rm min}$ for which  {\em destructive} spin-interference of opposite spins traveling in opposite directions around the ring takes place. For example, in a simplified
treatment~\cite{Frustaglia2004a}  $G=\frac{e^2}{h}[1 - \cos \frac{\Phi_{\rm AC}^\uparrow-\Phi_{\rm AC}^\downarrow}{2}]$ (the complete analytical solution is given in Ref.~\cite{Molnar2004}), where  $\Phi_{\rm AC}^\sigma=\pi(1+\sigma\sqrt{Q_{R}^2+1})$ is the AC phase acquired by a spin-$\uparrow$ or spin-$\downarrow$ quantum state, so that charge conductance minima $G(Q_{R}^{\rm min})=0$ are at $Q_{R}^{\rm min} \simeq\sqrt{n^2-1}$ ($n=2,4,6,\ldots$). However, adding two transverse leads onto the same 1D ring lifts the minima of the longitudinal conductance to $G_L(Q_R^{\rm min})=I_2/(V_1-V_2) \simeq e^2/h$  due to the contribution from incoherent (indirect) paths, $1 \to 3 \to 2$ and $1 \to 4 \to 2$, which do not exhibit destructive spin interference effects that characterize coherent (direct) paths from terminal 1 to 2 (see inset in Fig.~\ref{fig:qidshe} for labeling of the terminals). In these cases, electron goes into the macroscopic reservoirs, where the phase of its wave function is lost, before reaching the second longitudinal electrode. Nevertheless, $G_{\rm sH}^z$  vanishes at $Q_{R}^{\rm min}$, while the amplitude of its quasiperiodic oscillations  gradually decreases at large $Q_{R}$ because of the reflection at the ring-lead interface, as shown in Fig.~\ref{fig:qidshe}.

\begin{figure}
\centerline{\psfig{file=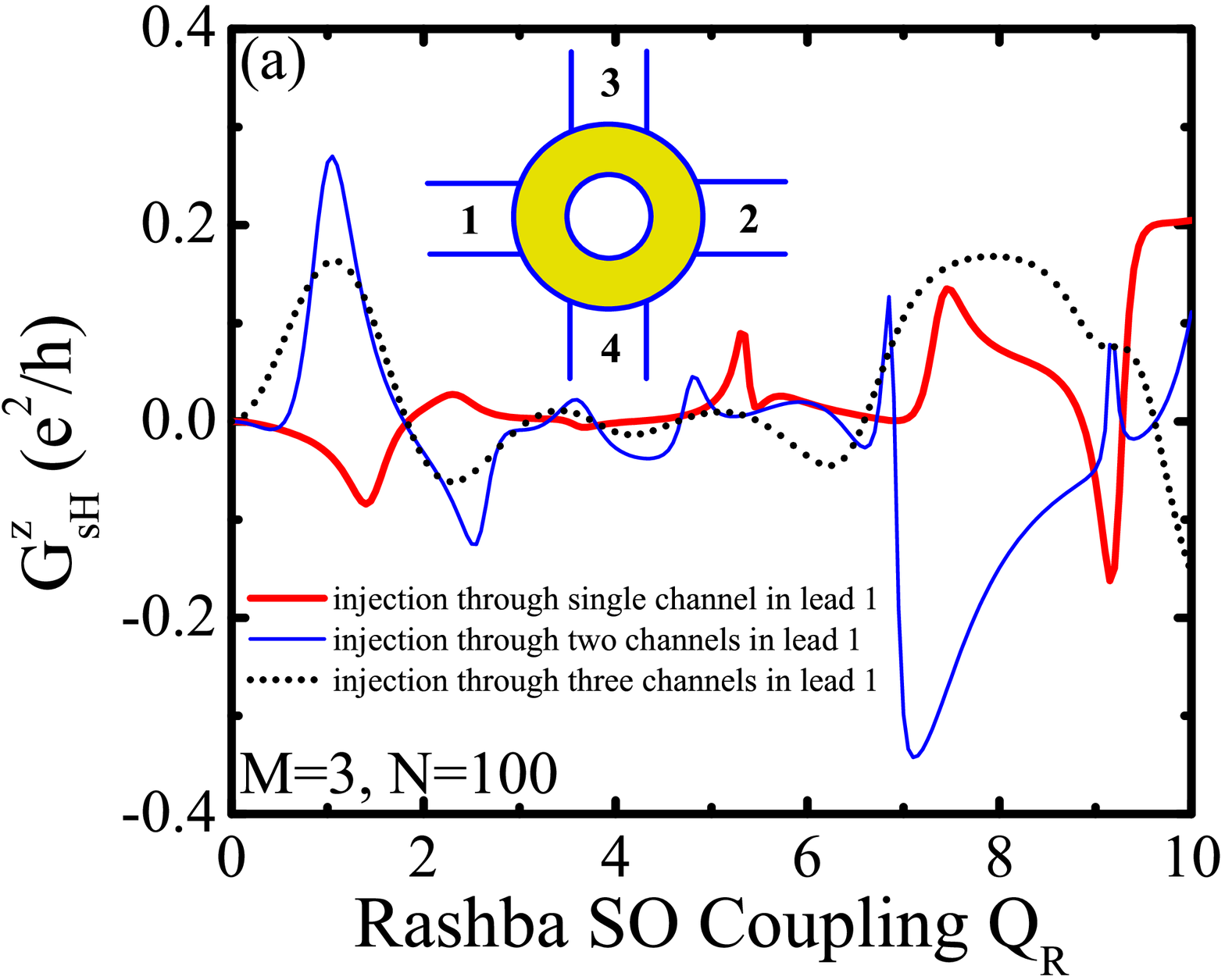,scale=0.35,angle=0} \hspace{0.1in} \psfig{file=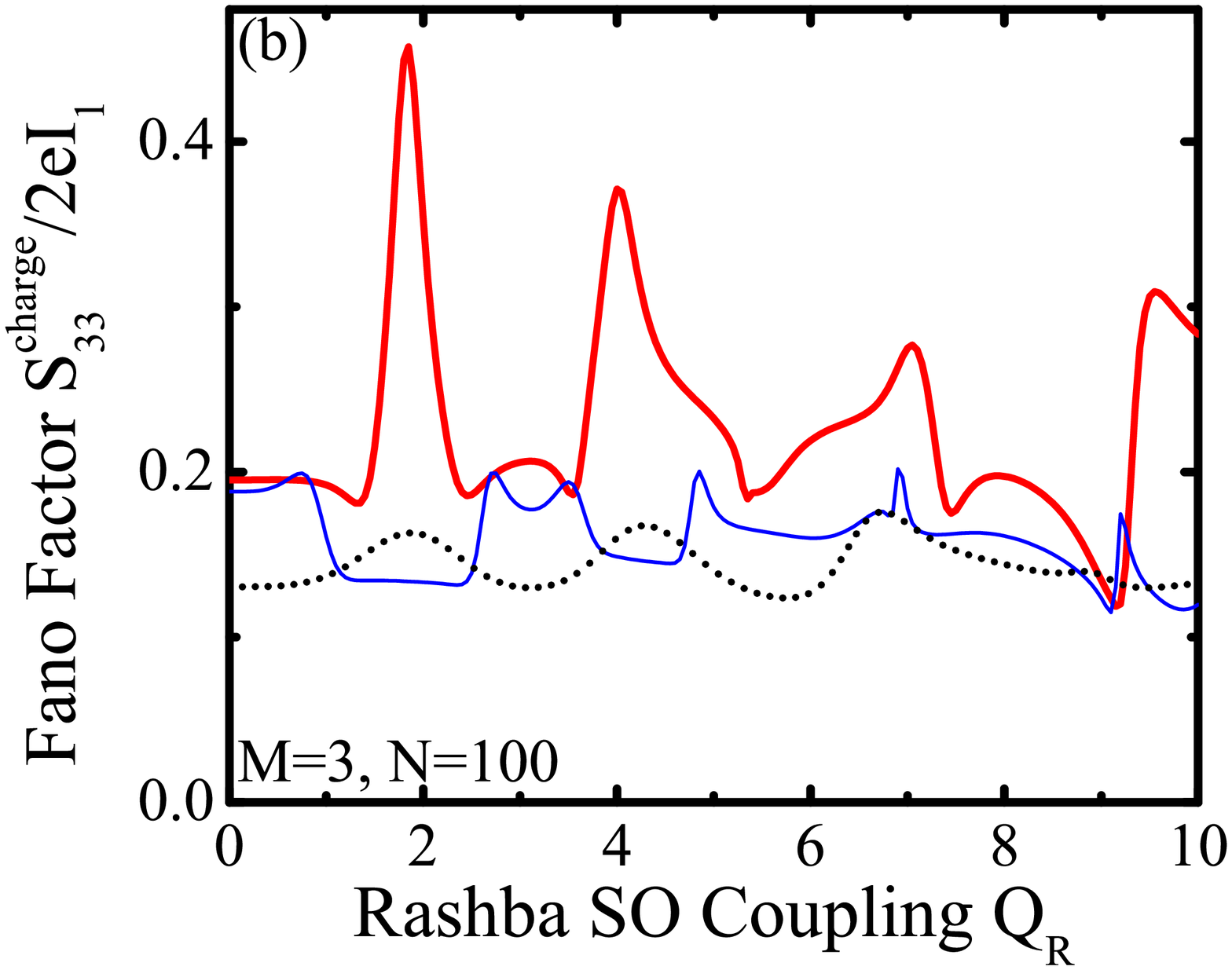,scale=0.35,angle=0}}
\caption{(a) The spin Hall conductance  of a four-terminal 2D AC ring attached to multichannel ($M=3$) electrodes as the function of the dimensionless Rashba coupling $Q_{R}$ within the ring. (b) Fano factor characterizing the shot noise of zero charge current $I_3=0$, normalized to current in lead 1. The Fermi energy $E_F$ is tuned to inject electrons from lead 1 through a single ($E_F=-2.7t_{\rm O}$; solid line), two ($E_F=-1.7t_{\rm O}$; thin solid line), and three ($E_F=-0.17t_{\rm O}$; doted line) conducting channels. The ring is modeled~\cite{Souma2004} using three coupled concentric circles, each discretized with $N=100$ sites.}
\label{fig:qidshe_2d}
\end{figure}

This type of four-terminal spin interferometer and its two-terminal counterpart~\cite{Souma2004}, based on tunable Rashba SO coupling, share the same limitations with other types of interferometers discussed in Sec.~\ref{sec:qidshe_intro}. The visibility of quantum interference effects encoded into the quasi-periodic oscillations of $G_{\rm sH}^z$($Q_{R})$ in Fig.~\ref{fig:qidshe} is reduced by ``dephasing'' when accumulated AC phases are averaged over many Feynman paths through 2D rings with $M>1$ [see Fig.~\ref{fig:qidshe_2d}(a)].

Here we examine if the shot noise of pure spin Hall current $I_3^{S_z}$ and zero charge current $I_3$ contains any additional insights, beyond the spin Hall conductance  $G_{\rm sH}^z$ in Fig.~\ref{fig:qidshe}, about the spin interference effects in four-terminal AC rings. The result for spin $S_{33}^{\rm spin}(|{\bf P}_{\rm in}|=0)$ and
charge $S_{33}^{\rm charge}(|{\bf P}_{\rm in}|=0)$ noise power as the function of the Rashba coupling strength $Q_{R}$ is plotted in Fig.~\ref{fig:qidshe_1dnoise}. In contrast to na\"ive expectations~\cite{Blanter2005} about the shot noise in interferometers simply retracing the oscillations of $G_{\rm sH}^z$ in Fig.~\ref{fig:qidshe}, we see more complicated pattern for the total spin noise, as well as  maxima of the total charge noise (for zero time-averaged value of the charge current $I_3=0$) at around zeros of $G_{\rm sH}^z$. Furthermore, the spin interference effects are inducing highly non-trivial oscillatory pattern in the cross-correlation noise $S_{33}^{\uparrow \downarrow}=S_{33}^{\downarrow \uparrow}<0$ encoding temporal correlations between $I_3^\uparrow$ and $I_3^\downarrow$ spin-resolved charge currents. Unlike the same type of temporal correlations for two-terminal simply-connected devices in Fig.~\ref{fig:spin_resolved}, they appear only for finite values of $Q_{R}$.

\begin{figure}
\centerline{\psfig{file=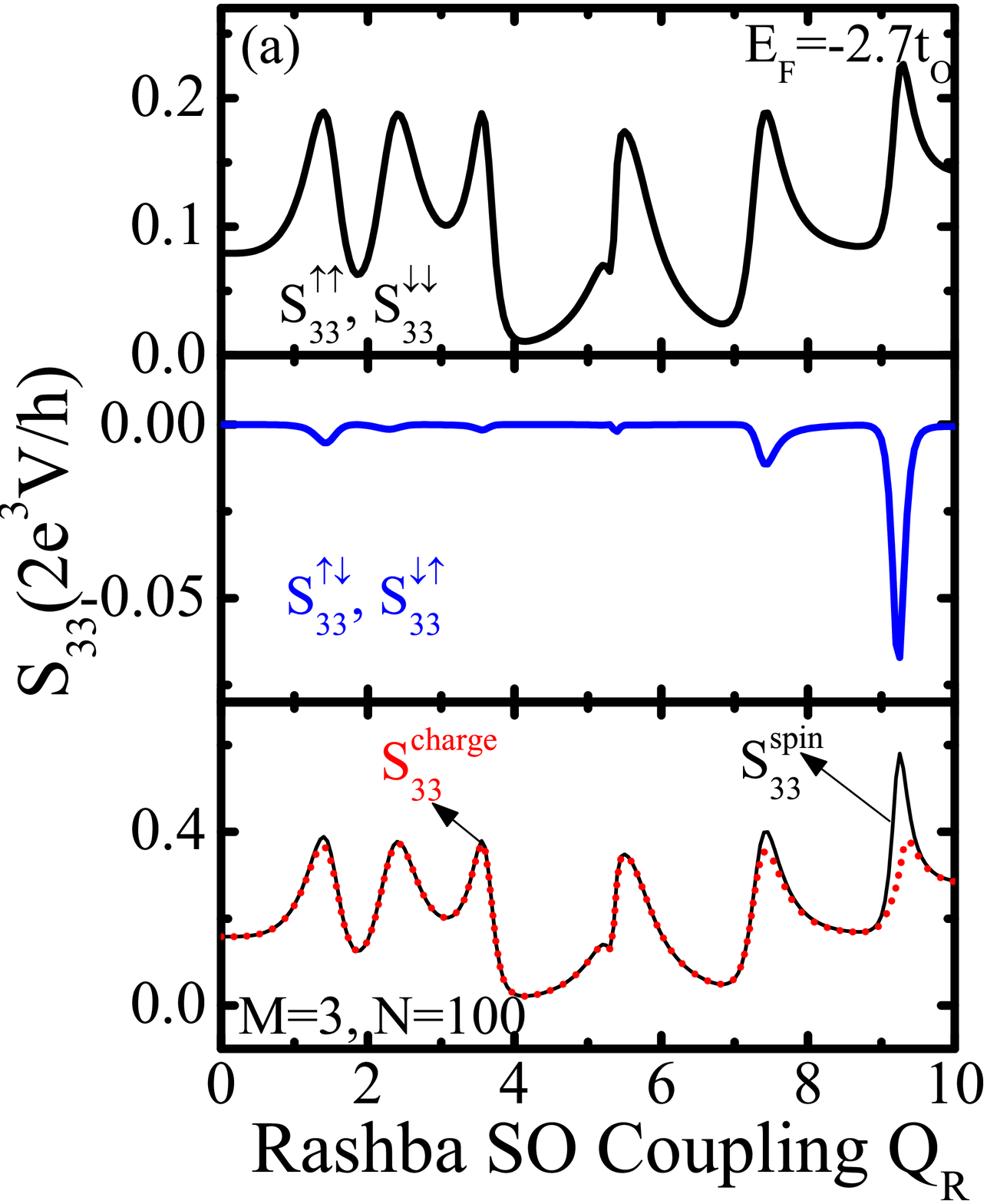,scale=0.24,angle=0} \psfig{file=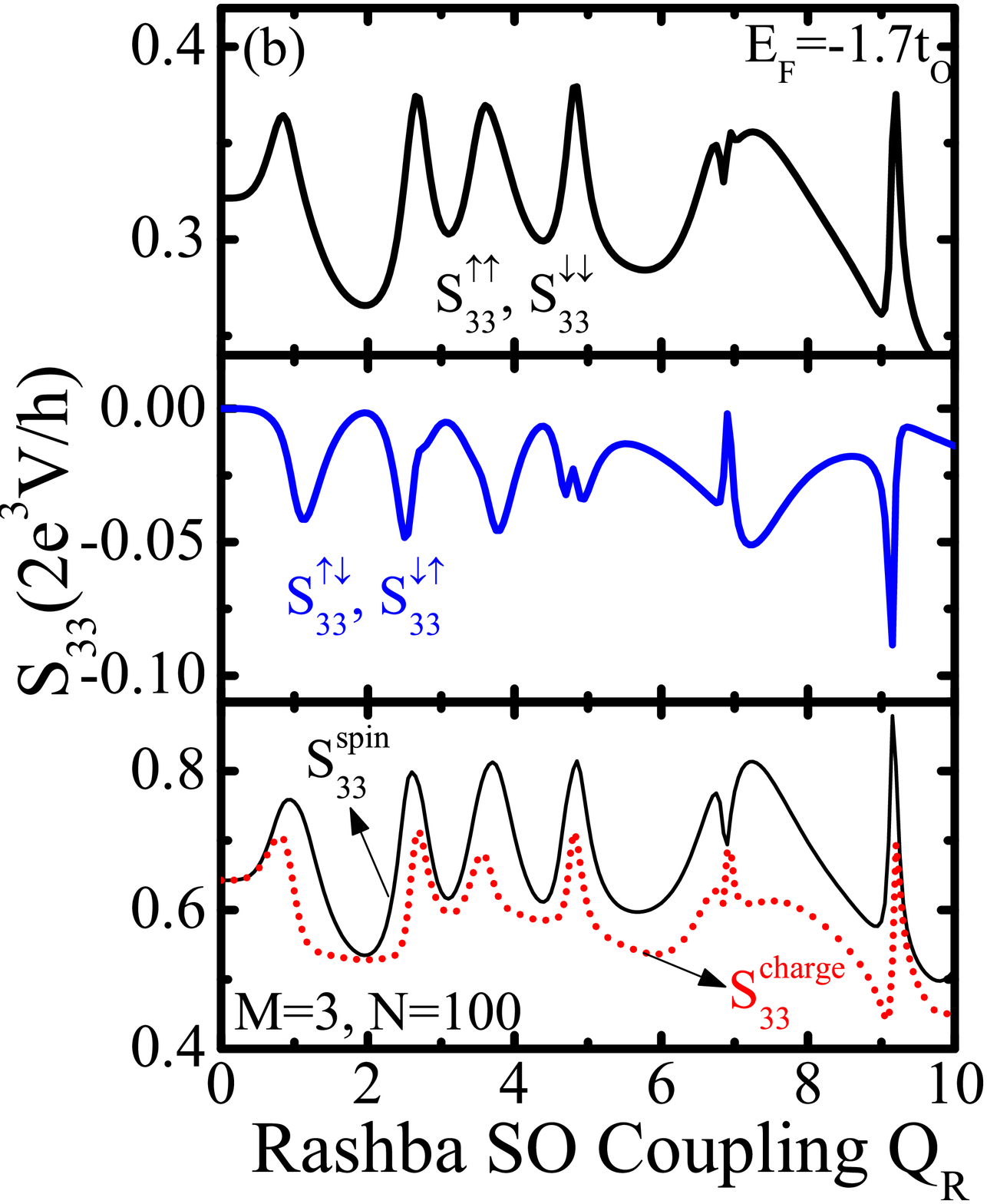,scale=0.24,angle=0} \psfig{file=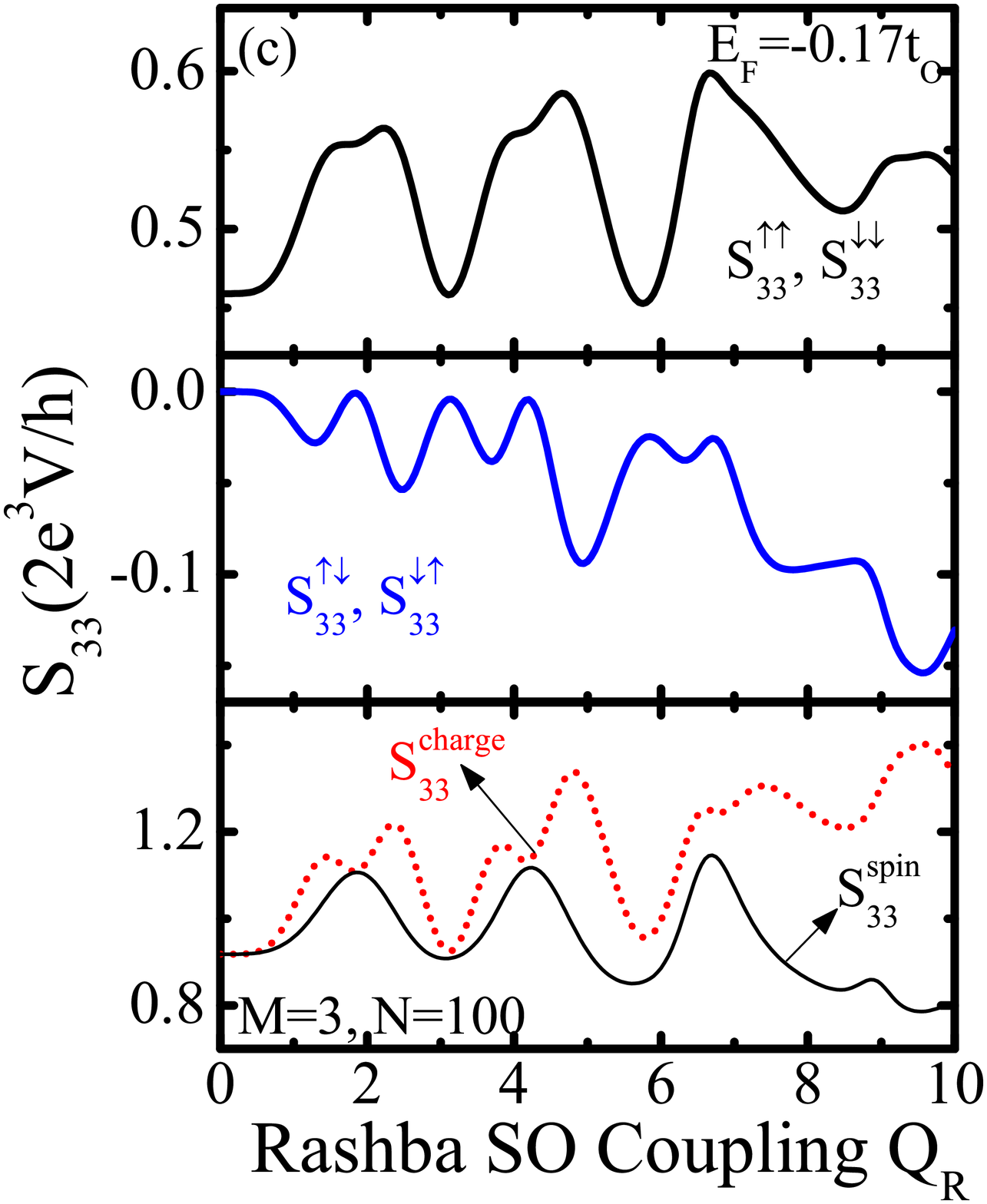,scale=0.24,angle=0}}
\caption{The ballistic spin-resolved shot noise in the transverse electrode 3 of a four-terminal 2D Aharonov-Casher ring (top and middle panels), as well as the total shot noise of transverse pure spin Hall current $I_3^{S_z}$ and zero charge current $I_3=0$ (bottom panel), driven by unpolarized $|{\bf P}_{\rm in}|=0$ injected charge current $I_1$. The spin conductance of the QIDSHE and the Fano factor of zero transverse charge current as the function of the Rashba SO coupling within the 2D ring is shown in Fig.~\ref{fig:qidshe_2d} for the same device parameters ($M=3$, $N=100$) and the same number of open channels [one in (a), two in (b), and three in (c)] used to inject electrons from lead 1 by tuning their Fermi energy $E_F$.}\label{fig:qidshe_2dnoise}
\end{figure}

As discussed in Sec.~\ref{sec:qidshe_intro}, the key issue for experimental~\cite{Ji2003} realization of solid-state interferometers is confinement of electrons to single channel transport to avoid averaging of the phase of their wave function when many channels contribute to measured transport properties~\cite{Marquardt2004}. In the case of QIDSHE,  Ref.~\cite{Souma2005a} examined the effect of 2D transport within the ring and electron injection through 1D leads (assuming that, e.g., point contact has been introduced between the lead and the ring), finding that oscillations of $G_{\rm sH}^z$ due to spin interference effects are still clearly visible. However, a more realistic device amenable to nanofabrication~\cite{Konig2006} is ballistic 2D ring attached to few-channel leads where electrons can be injected through one or more conducting channels of lead 1 by tuning their Fermi energy. The spin Hall conductance of such device is shown in Fig.~\ref{fig:qidshe_2d}, where we still find non-zero QIDSHE, but with greatly distorted oscillations of $G_{\rm sH}^z$ even when electrons are injected through a single transverse propagating mode of lead 1. Nevertheless, the Fano factor of transverse zero charge current, defined as $S_{33}^{\rm charge}/2eI_1$ where $I_1$ is used for normalization taking into account that $I_3=0$, displays much more regular oscillations with maxima appearing at similar values as in the case of the noise in strictly 1D structures of Fig.~\ref{fig:qidshe_1dnoise}. Analogously to Fig.~\ref{fig:qidshe_1dnoise}, we show the origin of these Fano factors in terms of the spin-resolved shot noise contributions to it plotted in Fig.~\ref{fig:qidshe_2dnoise}.

\section{Concluding remarks} \label{sec:conclusion}

The number of theoretical studies on spin-dependent shot noise has grown at an accelerated pace in recent years, carried by a wave of
interest in spintronics and spin-based quantum computing, as well as by fundamental interest to unravel new tools for probing
spin dynamics and electron-electron in nanostructures. In particular, akin to earlier studies of the shot noise in
spin-degenerate mesoscopic devices, the results on spin-dependent shot noise have divulged how random time-dependent  current fluctuations
encode the signatures of interactions of transported spin with magnetic impurities, SO couplings, and other internal and external magnetic
fields. These unique signatures are not visible when measuring the time-averaged currents and conductances.

In contrast to theoretical endeavors, large-scale experimental effort on spin-dependent shot noise in semiconductor spintronic devices is
still lacking. The recent spin-dependent shot noise measurements have mostly been focused on magnetic tunnel junctions. While shot noise does not
impose the most important limiting factor (when compared to debilitating $1/f$ noise) for MTJ applications, it does offer a sensitive tool to probe
microscopic features of their imperfect insulating barriers or Coulomb interaction effects in spin-polarized tunneling.

Here we focused on reviewing, as well as  extending, recent results~\cite{Dragomirova2007,Dragomirova2008} on the effect of the Rashba SO coupling on: ({\em i})
the shot noise of spin-polarized current injected into two-terminal diffusive quasi-1DEG-based nanowires; ({\em ii}) the shot noise of pure spin
and charge currents generated by the mesoscopic SHE in four-terminal ballistic and diffusive 2DEG nanostructures; and ({\em iii})
the shot noise associated with quantum-interference-driven SHE in four-terminal Aharonov-Casher rings realized using 2DEG.

To study these problems requires to extend~\cite{Dragomirova2007} the conventional scattering theory formulae for spin-degenerate noise~\cite{Blanter2000}
to spin-resolved shot noise, while taking as an input the degree of quantum coherence of injected spins $|{\bf P}_{\rm in}|$ and the direction of the spin-polarization vector ${\bf P}_{\rm in}$ with respect to relevant internal and external magnetic fields within the sample. The application of this formalism to two-terminal
multichannel diffusive quantum wires with the Rashba SO coupling shows how decoherence and dephasing of spin dynamics are essential to observe enhancement of charge shot noise in spin-polarized transport. That is, in narrow wires, where loss of spin coherence is suppressed and $|{\bf P}_{\rm out}|$ decays
much slower  than in the bulk systems, increase of the Fano factor (above $F=1/3$
of spin-degenerate diffusive transport~\cite{Beenakker1992}) in the strong SO coupling regime ($L \gg L_{\rm SO}$
inducing fast spin dynamics within the sample) is reduced when compared to wide wires. This occurs despite the fact
that {\em partially coherent}  spin state continues to ``flip'', but through (partially coherent~\cite{Nikoli'c2005})
spin precession $0 <|{\bf P}_{\rm out}|<1$. To obtain the Fano factor of charge currents comprised of partially coherent spins requires
to treat both charge propagation and spin dynamics quantum-mechanically, as suggested by the spin-resolved shot noises and conductances in
Fig.~\ref{fig:spin_resolved} (which cannot be reproduced by semiclassical approaches to spin-dependent shot noise where spin dynamics is captured
only through phenomenological spin-flip diffusion length~\cite{Mishchenko2003}). A remarkable one-to-one correspondence between  the values of
$F_{\uparrow \rightarrow \uparrow \downarrow}$ and the degree of quantum coherence $|{\bf P}_{\rm out}|$ predicted in Fig.~\ref{fig:noise}(e) offers
an exciting possibility to measure the coherence properties of transported spin as a magnetic degree of freedom in a {\em purely charge transport experiment} 
on open SO-coupled systems. This  offers an all-electrical alternative to usually employed optical tools to probe
transport of spin coherence in semiconductors~\cite{Holleitner2006}.

While enhancement of the shot noise due to fluctuations involving spin-flips (i.e., continuous spin precession) in two-terminal Rashba SO-coupled devices is absent when the injected current is unpolarized, this conclusion cannot be trivially extended to multiterminal devices due to non-local effects where other leads contribute to the noise in a selected lead. In fact, in multiterminal SO-coupled device exhibiting mesoscopic SHE, where intrinsic SO mechanisms relying on {\em precessing spins} can dominate over the extrinsic ones, we find a possibility for a significant enhancement of the shot noise of transverse spin and charge transport even when the current injected through longitudinal leads is unpolarized. This is related  to the fact that extrinsic SO scattering off impurities in 2D has no measurable effect on the shot noise. Therefore, experiments observing shot noise enhancement in the transverse electrodes upon changing the voltage of the gate electrode~\cite{Nitta1997,Grundler2000} covering 2DEG could unambiguously resolve the dominance of the intrinsic contribution to the spin Hall or the charge Hall effect (and related inverse SHE) in multiterminal nanostructures.  The {\em central result} of this novel approach to long-standing ``intrinsic vs. extrinsic controversy''~\cite{Nagaosa2008,Engel2007a,Sinitsyn2008} surrounding experimental tests of the origin of SHE (and related AHE) is shown in Fig.~\ref{fig:fano_she}---the spin and charge current Fano factors in the transverse electrode starts to increase in the same region of intrinsic SO coupling strength in which intrinsic mechanisms begin to dominate~\cite{Nikoli'c2007} SHE manifestations in Fig.~\ref{fig:gshe}(b). The specific Fano factor values that can be measured {\em electrically} are set by the polarization of injected current. Thus, by detecting the increase of the Fano factor of transverse charge current while increasing the Rashba SO coupling via the gate voltage will confirm that intrinsic mechanisms (i.e., spin precession associated with them) dominate the induction of the SHE~\cite{Sih2005a} in the same device.

Finally, in four-terminal Aharonov-Casher rings we find that both the spin and charge shot noise of the spin Hall transport in the transverse electrodes
oscillate as the strength of the Rashba SO coupling is modified by the gate electrode covering the ring to tune constructive and destructive spin interference effects. However, the pattern of these oscillations is much more different from the oscillations of time-average quantities, such as the spin Hall conductance or longitudinal charge conductance. This is related to complicated (when compared to the same quantities studies in simply-connected SO-coupled nanostructures in Sec.~\ref{sec:discussion})  oscillatory pattern of the spin-resolved shot noise measuring temporal correlations between currents of spin-$\uparrow$ and spin-$\downarrow$ electrons. Despite the net transverse charge current being identically zero, there is still a non-zero oscillatory charge shot noise due to the opposite flow of spin-$\uparrow$ and spin-$\downarrow$ electrons in the course of pure spin Hall current induction. This noise reaches maxima at around zeros of the spin Hall conductance. Such effects could be utilized to detect features of the QIDSHE without the need for demanding direct measurement of transverse pure spin Hall currents. Also, the Fano factor defined by normalizing such noise by the non-zero injected current in the longitudinal leads, displays far more regular oscillatory pattern in multichannel AC rings despite spin Hall conductance oscillations being washed out by phase averaging (i.e., ``ensemble dephasing''~\cite{Schlosshauer2008,Souma2004}) over many channels~\cite{Blanter2005}.

\ack
The authors appreciate stimulating discussions with J. C. Egues, S. Garzon, E. R. Nowak, M. G. Vavilov and L. P. Z\^{a}rbo. This work was supported by  DOE Grant No. DE-FG02-07ER46374 through the Center for Spintronics and Biodetection at the University of Delaware.

\end{document}